\documentclass[a4paper,11pt]{article}
\pdfoutput=1
\usepackage{jheppub}
\pdfoutput=1

\usepackage{epsf}
\usepackage{epsfig}
\usepackage{subfig}
\usepackage{mathtools}
\usepackage{hhline}
\usepackage{float}
\usepackage{multirow}
\usepackage{nicefrac}
\usepackage{epstopdf}
\usepackage{slashed}
\usepackage{xcolor}
\usepackage{url}

\usepackage{braket}
\usepackage{subdepth}

\graphicspath{ {figures/} }

\newcommand{\be}{\begin{eqnarray}}
\newcommand{\ee}{\end{eqnarray}}

\newcommand{\ba}{\begin{array}}
\newcommand{\ea}{\end{array}}
\newcommand{\bee}{\begin{equation}\ba{c}}
\newcommand{\eee}{\ea\end{equation}}

\newcommand{\bi}{\begin{itemize}}
\newcommand{\ei}{\end{itemize}}

\allowdisplaybreaks

\def\nnb{\nonumber}

\def\gev{{\rm GeV}}
\def\mev{{\rm MeV}}
\def\be{\begin{equation}}
\def\ee{\end{equation}}
\def\bea{\begin{eqnarray}}
\def\eea{\end{eqnarray}}
\def\nnb{\nonumber}
\def\dps{\displaystyle}

\def\bbuildrel#1_#2^#3{\mathrel{\mathop{\kern 0pt#1}\limits_{#2}^{#3}}}
\def\slash#1{\setbox0=\hbox{$#1$}#1\hskip-\wd0\dimen0=5pt\advance
       \dimen0 by-\ht0\advance\dimen0 by\dp0\lower0.5\dimen0\hbox
         to\wd0{\hss\sl/\/\hss}}
\def\gev{{\rm GeV}}
\def\mev{{\rm MeV}}

\def\be{\begin{equation}}
\def\ee{\end{equation}}
\def\beq{\begin{eqnarray}}
\def\eeq{\end{eqnarray}}

\def\slash#1{#1 \hskip-0.45em /}

\def\DB0{\partial B_0}

\newcommand{\lsim}
{\;\raisebox{-.3em}{$\stackrel{\displaystyle <}{\sim}$}\;}

\def\Cl2{\mbox{Cl}_2}

\def\slash#1{#1 \hskip-0.45em /}

\newcommand{\spp}{\vphantom{$\Big($}}

\usepackage{color}

\definecolor{Brown}{rgb}{0.5,0.25,0}

\title{Phenomenology of inclusive $\bar{B}\to X_s \ell^+ \ell^-$ \\
	for the Belle~II era}
\author{Tobias Huber$^1$,}
\author{Tobias Hurth$^2$,}
\author{Jack Jenkins$^3$,}
\author{Enrico Lunghi$^3$,}
\author{Qin Qin$^{4}$,}
\author{K.~Keri Vos$^5$}
\affiliation{
$^1$Theoretische Physik 1, Naturwissenschaftlich-Technische Fakult\"at, Universit\"at Siegen, \\ 
Walter-Flex-Stra{\ss}e 3, D-57068 Siegen, Germany\\
$^2$PRISMA+ Cluster of Excellence and Institute for Physics (THEP), \\ Johannes Gutenberg University, \\
Staudingerweg 9, D-55099 Mainz, Germany\\
$^3$Physics Department, Indiana University, \\ 727 E. Third St., Bloomington, IN 47405, USA \\
$^4$School of Physics, Huazhong University of Science and Technology, \\ Luoyulu 1037, Wuhan 430074, China \\
$^5$Physics Department T31, Technische Universit\"at M\"unchen, \\ James Franck-Stra{\ss}e 1, D-85748 Garching, Germany
}

\emailAdd{huber@physik.uni-siegen.de}
\emailAdd{tobias.hurth@cern.ch}
\emailAdd{jackjenk@iu.edu}
\emailAdd{elunghi@indiana.edu}
\emailAdd{qqin@hust.edu.cn}
\emailAdd{keri.vos@tum.de}

\abstract{With the first data being recorded at Belle~II, we are at the brink of a new era in quark flavour physics. The many exciting new opportunities for Belle~II include a full angular analysis of inclusive ${\bar B \to X_{s} \, \ell^+\ell^-}$ which has the potential to reveal new physics, in particular by its interplay with the exclusive $b \to s \ell^+\ell^-$ counterparts studied extensively at LHCb. In this paper, we present fully updated Standard Model predictions for all angular observables necessary for this endeavour. These predictions are tailored to Belle~II and include an elaborate study of the treatment of collinear photons which become crucial when aiming for the highest precision. In addition, we present a phenomenological study of the potential for Belle~II to reveal possible new physics in the inclusive decay channel, both in an independent manner and in combination with exclusive modes.

}

\preprint{
\begin{minipage}{3cm}
\small
\flushright
SI-HEP-2020-13\\
P3H-20-030\\
MITP/20-37\\
TUM/1268-20
\end{minipage}}

\begin{document}

\maketitle

%%%%%%%%%%%%%%%%%%%%%%%%%%%%%%%%%%%%%%%%%%%%%%%%%%%%%%%%%%%%%%%%%%%%%%%%%%%%%%%%%%%%%%%%%%%%%%%%%%%%%%%%%%%%%%%%%%%%%%%%%%%%%%%%%%%%%%%%%%%%%%%%%%%%%%%%%%%%%%%%%%%%%%
%%%%%%%%%%%%%%%%%%%%%%%%%%%%%%%%%%%%%%%%%%%%%%%%%%%%%%%%%%%%%%%%%%%%%%%%%%%%%%%%%%%%%%%%%%%%%%%%%%%%%%%%%%%%%%%%%%%%%%%%%%%%%%%%%%%%%%%%%%%%%%%%%%%%%%%%%%%%%%%%%%%%%%

\section{Introduction}
\label{sec:introduction}

Many of the yet unanswered questions of particle physics are related to the Yukawa sector of the Standard Model (SM). In the past decades, flavour physics experiments at electron-positron~\cite{Bevan:2014iga} and hadron machines have already revealed much of our current understanding of the quark sector. With Run-2 data from the LHC being analysed and Belle~II having the first dozens of inverse femtobarns on tape, the quark sector of the SM is currently being investigated to unprecedented precision, possibly revealing and quantifying the remaining mysteries in this sector.
Flavour-changing neutral current (FCNC) decays of heavy quarks are among the prime candidates to further scrutinize the quark flavour sector of the SM and to search for physics beyond it. While exclusive decays of $B$ and $B_s$ mesons such as $\bar{B} \to K^{(*)} \ell^+\ell^-$ have played a major role in the experimental programs~\cite{Aaltonen:2011ja,Lees:2015ymt,Wehle:2016yoi,Sirunyan:2017dhj,Aaij:2017vbb,Aaboud:2018krd,Abdesselam:2019wac,Aaij:2019wad,Aaij:2020nrf} and have revealed certain interesting tensions between experimental data and SM predictions~\cite{Beaujean:2013soa,Hiller:2014yaa,Descotes-Genon:2015uva,Bordone:2016gaq,DAmico:2017mtc,Altmannshofer:2017yso,Ciuchini:2017mik,Capdevila:2017bsm,Alguero:2019ptt,Hurth:2020rzx}, inclusive channels such as $\bar{B}\to X_s \ell^+\ell^-$ will be analysed at Belle~II, where a full angular analysis is expected to become feasible for the first time~\cite{Kou:2018nap}. Taken together, the experiments at hadron and electron-positron machines have a huge potential in tackling fundamental questions of particle physics and searching for new phenomena.

On the theoretical side, the description of inclusive $\bar{B}\to X_s \ell^+\ell^-$ is already very much advanced. The short-distance partonic rate is known to NLO~\cite{Misiak:1992bc,Buras:1994dj} and NNLO~\cite{Bobeth:1999mk,Gambino:2003zm,Gorbahn:2004my,Asatryan:2001zw,Asatrian:2001de,Asatryan:2002iy,Ghinculov:2002pe,Asatrian:2002va,Asatrian:2003yk,Ghinculov:2003bx,Ghinculov:2003qd,Greub:2008cy,Bobeth:2003at,deBoer:2017way} in QCD, and to NLO in QED~\cite{Huber:2005ig,Huber:2007vv,Huber:2015sra}. Recently, also CKM suppressed contributions from multi-particle final states at leading power have become available analytically~\cite{Huber:2018gii}. In addition, local power-corrections that scale as $1/m_b^2$~\cite{Falk:1993dh,Ali:1996bm,Chen:1997dj,Buchalla:1998mt} and $1/m_b^3$~\cite{Bauer:1999kf,Ligeti:2007sn}~have been analysed. Other long-distance effects stem from intermediate charmonium resonances --~most prominently $J/\psi$ and $\psi(2S)$~-- which show up as large peaks in the dilepton invariant mass spectrum. Their effect in the low- and high-$q^2$ regions\footnote{$q^2$ denotes the dilepton invariant mass squared.} is treated via the Kr\"uger-Sehgal (KS) approach~\cite{Kruger:1996cv,Kruger:1996dt}, which has been refined and improved in several respects in~\cite{Huber:2019iqf}.

In addition, there are the so-called resolved contributions, which describe nonlocal power corrections arising from operators in the effective field theory other than the ones proportional to $(\bar s \Gamma_1 b)(\bar \ell \Gamma_2 \ell)$. In the low-$q^2$ region, the resolved contributions can be systematically computed using soft-collinear effective theory
(SCET) at subleading power~\cite{Hurth:2017xzf,Benzke:2017woq,Benzke:2020htm}, while in the high-$q^2$ region the dominating terms
(nonfactorizable $c \bar c$ contributions) can be re-expanded in local operators and treated along the lines of~\cite{Voloshin:1996gw,Buchalla:1997ky}. 

Over the years, additional observables have been proposed besides the traditionally studied decay rate and forward-backward asymmetry. In~\cite{Lee:2006gs} the full set of independent angular observables was identified. Furthermore, it was proposed in~\cite{Ligeti:2007sn} to normalise the ${\bar B \to X_{s} \, \ell^+\ell^-}$ rate in the high-$q^2$ region to the inclusive semi-leptonic $\bar B^0 \to X_u \ell \nu$ rate {\emph{with the same dilepton mass cut}} in order to tame the ${\cal O}(30-40\%)$ uncertainty coming from poorly known HQET matrix elements at orders $1/m_b^2$ and $1/m_b^3$. This behavior was indeed confirmed in subsequent phenomenological analyses~\cite{Huber:2007vv,Huber:2015sra,Huber:2019iqf}, including the present work.

Since it will still take some time until a fully inclusive measurement using the recoil technique will become feasible at Belle~II, one has to rely on the sum-over-exclusive method which requires a cut on the hadronic invariant mass $M_X$ to remove $b \to c (\to s\ell\nu)\ell\nu$ charged-current semi-leptonic and other sources of background at Belle~II. The effect of an $M_X$ cut in ${\bar B \to X_{s} \, \ell^+\ell^-}$, including the sensitivity to sub-leading shape functions, was analysed in~\cite{Lee:2005pk,Lee:2005pwa,Lee:2008xc}, with certain problems about the SCET scaling of the virtual photon in the low-$q^2$ region indicated in~\cite{Bell:2010mg,Hurth:2017xzf,Benzke:2017woq}. In the present work our predictions are given without a hadronic mass cut, leaving such a study for future work.

The novelties of the present article are still manifold and tailored to the Belle~II era. First, we update the SM predictions of all angular observables, integrated over two bins in the low-$q^2$ region. For selected observables, the high-$q^2$ integrated results are also provided. Depending on the observable and the $q^2$-bin, the updated central values  differ by several percent from those of the previous numerical analysis in~\cite{Huber:2015sra}. The main reasons for this behaviour can be traced back to updated input parameters and the more sophisticated treatment of non-perturbative effects, coming for instance from resonances treated via the Kr\"uger-Sehgal approach as developed in~\cite{Huber:2019iqf}. To probe effects of lepton-flavour violation, we give predictions for $R_{X_s}$, the inclusive analogue of $R_{K^{(*)}}$, for the first time. Second, we perform a new Monte Carlo study on the treatment of collinear photon radiation tailored to the treatment of collinear photons at Belle~II, including the effect of bin migration from the charmonium resonances into the perturbative low-$q^2$ window. Third, we carry out a comprehensive model-independent new-physics analysis which also considers a study of the synergy and complementarity between inclusive and exclusive $b \to s \ell^+ \ell^-$ transitions with the full Belle~II data set. Thus, our new analysis paves the road for a full phenomenological study of ${\bar B \to X_{s} \, \ell^+\ell^-}$ at Belle~II.

This article is organised as follows: In section~\ref{sec:observables} we define all ${\bar B \to X_{s} \, \ell^+\ell^-}$ obs\-ervables under consideration, while section~\ref{sec:results} contains the phenomenological results of the main observables. In section~\ref{sec:collinear} we describe the treatment of collinear photons and quantify resulting corrections. In section~\ref{sec:newphysics} we carry out our comprehensive model-independent new-physics analysis. We conclude in section~\ref{sec:conclusion}. The paper is supplemented by two appendices. Appendix~\ref{sec:pheno} contains the SM predictions for the remaining observables relegated from section~\ref{sec:results}, while we collect new-physics formulas in terms of high-scale Wilson coefficients in appendix~\ref{sec:formulae}.

%%%%%%%%%%%%%%%%%%%%%%%%%%%%%%%%%%%%%%%%%%%%%%%%%%%%%%%%%%%%%%%%%%%%%%%%%%%%%%%%%%%%%%%%%%%%%%%%%%%%%%%%%%%%%%%%%%%%%%%%%%%%%%%%%%%%%%%%%%%%%%%%%%%%%%%%%%%%%%%%%%%%%%
%%%%%%%%%%%%%%%%%%%%%%%%%%%%%%%%%%%%%%%%%%%%%%%%%%%%%%%%%%%%%%%%%%%%%%%%%%%%%%%%%%%%%%%%%%%%%%%%%%%%%%%%%%%%%%%%%%%%%%%%%%%%%%%%%%%%%%%%%%%%%%%%%%%%%%%%%%%%%%%%%%%%%%

\section{Definition of the observables}
\label{sec:observables}

We start from the double-differential decay width ${d^2\Gamma}/{dq^2}/{dz}$, where $z = \cos\theta$ and $\theta$
is the angle between the three-momenta of the positively charged lepton and the decaying $B$ meson in the dilepton center-of-mass frame.
The differential decay width ${d\Gamma_{\ell\ell}}/{dq^2}$ and the unnormalized differential forward-backward asymmetry ${dA_{\rm FB}^{\ell\ell}}/{dq^2}$ for the $\ell=e$ or $\ell=\mu$ final state are then defined as
\begin{align}
\frac{d\Gamma_{\ell\ell}}{dq^2} &\equiv \int_{-1}^{+1} dz \, \frac{d^2\Gamma(\bar B \to X_s \ell\ell)}{dq^2 dz} \; , \label{eq:gamma} \\
\frac{dA_{\rm FB}^{\ell\ell}}{dq^2} &\equiv \int_{-1}^{+1} dz \, \frac{d^2\Gamma(\bar B \to X_s \ell\ell)}{dq^2 dz} \, {\rm sign}(z) \; . \label{eq:afb}
\end{align}%
The normalized forward-backward asymmetry $\overline{A}_{\rm FB}$ integrated in a region $q_m^2 < q^2 < q_M^2$ (in units of GeV$^2$)
is then given by
\begin{align}
\overline{A}_{\rm FB}[q_m^2, q_M^2]_{\ell\ell} \equiv
\frac{\int_{q_m^2}^{q_M^2}dq^2 ({dA_{\rm FB}^{\ell\ell}}/{dq^2}) }{\int_{q_m^2}^{q_M^2}dq^2 ({d\Gamma_{\ell\ell}}/{dq^2}) } \; .
\end{align}

In the absence of QED corrections the double-differential decay width ${d^2\Gamma}/{dq^2}/{dz}$ is a second order polynomial in $z$,
giving rise to three independent angular-distribution observables $H_I^{\ell\ell}(q^2)$, $I=T,A,L$~\cite{Lee:2006gs}. As pointed out in~\cite{Huber:2015sra},
QED corrections lead to a distortion of the simple polynomial $z$ dependence and result in a complicated function of $z$. It is therefore
instructive to use projections with weight functions $W_I(z)$ to define the $H_I^{\ell\ell}(q^2)$.
In the absence of QED corrections the original definitions
from~\cite{Lee:2006gs} are restored, but the use of the weight functions better captures the effects of QED radiation in the angular observables.
In addition, the weight functions will give us the flexibility to define further observables, as we will demonstrate below.
We therefore define
\begin{align}
H_I^{\ell\ell} (q^2) &= \int_{-1}^{+1} dz \, \frac{d^2\Gamma(\bar B \to X_s \ell\ell)}{dq^2dz} \, W_I(z)\; , \nonumber \\[0.3em]
H_I[q_m^2, q_M^2]_{\ell\ell} & =   \int_{q_m^2}^{q_M^2}dq^2 \, H_I^{\ell\ell} (q^2) \; . \label{eq:projectionHI}
\end{align}
Almost all weight functions $W_I(z)$ are constructed from Legendre polynomials $P_n(z)$, which are orthogonal on $z \in [-1,1]$.
Moreover, we can use Legendre polynomials with $n>2$ to define observables which vanish in the absence of QED corrections. We do this by defining
$H_3^{\ell\ell}(q^2)$ and $H_4^{\ell\ell}(q^2)$ for $n=3$ and $n=4$, respectively, to get a handle on even and odd powers of $z$. This leads to the following weight functions,
\begin{equation}
\begin{aligned}
W_T &= \frac{2}{3} \; P_0 (z) + \frac{10}{3} \; P_2 (z) \, , &\quad\quad & W_3 = P_3 (z) \, , \\
W_L &= \frac{1}{3} \; P_0 (z) - \frac{10}{3} \; P_2 (z) \, , &\quad\quad & W_4 = P_4 (z) \, , \\
W_A &= \frac{4}{3} {\rm sign} (z) \; .
\label{weights}
\end{aligned}
\end{equation}
The differential rate and unnormalized forward-backward asymmetry are related to the angular-distribution observables via
\begin{align}\label{eq:relation}
{\frac{d\Gamma_{\ell\ell}}{dq^2}} = H_T^{\ell\ell}(q^2) + H_L^{\ell\ell}(q^2)\; ,\qquad \frac{dA_{\rm FB}^{\ell\ell}}{dq^2} = {3\over4}H_A^{\ell\ell}(q^2) \; .
\end{align}
The observables ${\cal H}_I$ differ from the $H_I$ merely by a normalization which can be deduced from eqs.~(4.4) and~(4.6) of~\cite{Huber:2015sra}. To the latter paper we also refer for master formulas of all observables. Our operator basis is the same as in~\cite{Huber:2005ig}. Finally, the branching ratio is calculated via
\begin{equation}
{\cal B}[q_m^2, q_M^2]_{\ell\ell} = {\cal H}_T[q_m^2, q_M^2]_{\ell\ell} + {\cal H}_L[q_m^2, q_M^2]_{\ell\ell} \, .
\end{equation}

In the high-$q^2$ region, we also consider the ratio~\cite{Ligeti:2007sn}
\begin{equation}
\label{eq:zoltanR}
{\cal R}(s_0)_{\ell\ell} =
\int_{\hat s_0}^1 {\rm d} \hat s \, {{\rm d} {\Gamma_{\ell\ell}} \over {\rm d} \hat s}\;  \Big/ \,
\int_{\hat s_0}^1 {\rm d} \hat s \, {{\rm d} {\Gamma} (\bar B^0 \to X_u \ell \nu) \over {\rm d} \hat s}\, ,
\end{equation}
where $\hat{s} = q^2/m_{b,\rm{pole}}^2$. The ratio ${\cal R}(14.4)_{\ell\ell}$ significantly reduces the uncertainties
introduced by hadronic power corrections, which dominate the uncertainties of the high-$q^2$ $\bar B \rightarrow X_s \ell^+ \ell^-$ decay rate.

Moreover, to quantify the effects of lepton-flavour universality violation in the inclusive
$\bar{B}\to X_s\ell^+\ell^-$ decay, we define the ratio of the decay widths of the muon- to electron-modes
\begin{align}\label{eq:Rratdef}
R_{X_s}[q_m^2, q_M^2] \equiv \int_{q_m^2}^{q_M^2}dq^2 \, {d\Gamma_{\mu\mu}\over dq^2} \; \Big/
\int_{q_m^2}^{q_M^2}dq^2 \, {d\Gamma_{ee}\over dq^2} \; ,
\end{align}
analogous to the ratios $R_{K^{(*)}}$ in the exclusive channels. Besides, the corresponding ratios
for the angular observables $H_I$ ($I=T,A,L$) are also calculated. They are defined by
\begin{align}
R_{H_I}[q_m^2, q_M^2] \equiv H_I[q_m^2, q_M^2]_{\mu\mu} \Big/ H_I[q_m^2, q_M^2]_{ee} \; .
\end{align}

%%%%%%%%%%%%%%%%%%%%%%%%%%%%%%%%%%%%%%%%%%%%%%%%%%%%%%%%%%%%%%%%%%%%%%%%%%%%%%%%%%%%%%%%%%%%%%%%%%%%%%%%%%%%%%%%%%%%%%%%%%%%%%%%%%%%%%%%%%%%%%%%%%%%%%%%%%%%%%%%%%%%%%
%%%%%%%%%%%%%%%%%%%%%%%%%%%%%%%%%%%%%%%%%%%%%%%%%%%%%%%%%%%%%%%%%%%%%%%%%%%%%%%%%%%%%%%%%%%%%%%%%%%%%%%%%%%%%%%%%%%%%%%%%%%%%%%%%%%%%%%%%%%%%%%%%%%%%%%%%%%%%%%%%%%%%%

\section{Phenomenological results}
\label{sec:results}

For the updated numerical analysis we use the same input parameters as in our $\bar{B}\to X_d \ell^+\ell^-$ analysis~\cite{Huber:2019iqf}. They are presented in table~\ref{tab:inputs}. The most significant changes compared to the previous $\bar{B}\to X_s \ell^+\ell^-$ analysis~\cite{Huber:2015sra} are, on the one hand, the inclusion of the resolved photon contributions~\cite{Hurth:2017xzf,Benzke:2017woq,Benzke:2020htm}, which we discussed in detail in~\cite{Huber:2019iqf}. Moreover, we implemented the new and more sophisticated treatment of the non-perturbative effects following the Kr\"uger-Sehgal approach~\cite{Huber:2019iqf}. Finally, in the high-$q^2$ region, the HQET matrix elements $\lambda_2, \rho_1$ and the weak annihilation matrix elements $f_u^0, f_u^\pm$ and $f_s$ play a crucial role. We have updated these parameters as discussed in~\cite{Huber:2019iqf} and give their explicit values in table~\ref{tab:inputs}. Here the weak annihilation matrix elements are defined as\footnote{This equation corrects (5.5) in~\cite{Huber:2019iqf}, where the factor $4\pi^2$ was missing.}
\begin{align}
f^a_q &\equiv \frac{4\pi^2}{2 m_B} \langle B^a | Q_1^q - Q_2^q | B^a \rangle \; , &f_q\equiv (f_q^0+f_q^\pm)/2 \, ,
\end{align}
where $Q_1^q = \bar h_v \gamma_\mu (1-\gamma_5) q \; \bar q \gamma^\mu (1-\gamma_5) h_v$ and $Q_2^q = \bar h_v  (1-\gamma_5) q \; \bar q (1+\gamma_5) h_v$~\cite{Voloshin:2001xi},
and $a=0,\pm$ denotes the charge of the meson. Taking into account isospin and flavour $\mathrm{SU}(3)$ considerations, we can rewrite the weak annihilation matrix elements in terms of the valence $f_{\rm V}$ and non-valence $f_{\rm {NV}}$ ones. The observables depend on
\begin{align}
{\mathcal B}(\bar{B}\to X_s\ell^+\ell^-) &\Longrightarrow
\begin{cases}
f_s = f_\text{NV}\cr
f_u = (f_\text{V} + f_\text{NV})/2 \, , \cr
\end{cases}\\
{\cal R}(s_0,\bar{B}\to X_s\ell^+\ell^-) &\Longrightarrow
\begin{cases}
(f_s + f_u^0)/2 = f_\text{NV} \cr
f_s - f_u^0  = [\delta f]_{\mathrm{SU}(3)} \, . \cr
\end{cases}
\end{align}
The input parameters in table~\ref{tab:inputs} are obtained from  a re-analysis of \cite{Asner:2009pu,Gambino:2010jz}. For the ratio ${\cal R}(s_0,\bar{B}\to X_s\ell^+\ell^-)$, the symmetry breaking corrections play an important role. Following ref.~\cite{Ligeti:2007sn}, we estimated these effects as $[\delta f]_{\mathrm{SU}(3)} = 0.04$ and $[\delta f]_{\mathrm{SU}(2)} = 0.004$, respectively.
\begin{table}[t]
	\begin{center}
		\begin{displaymath}
		\begin{tabular}{|l|l|}
		\hline\spp
		$\alpha_s (M_z) = 0.1181 (11)$ &
		$m_e = 0.51099895 \;\mev $ \\ \spp
		$\alpha_e (M_z) =  1/127.955 $ &
		$m_\mu = 105.65837 \;\mev$ \\ \spp
		$s_W^2 \equiv \sin^2\theta_{W}^{{\overline{\rm MS}}} = 0.2312$ &
		$m_\tau = 1.77686 \;\gev$ \\ \spp
		$|V_{ts}^* V_{tb}/V_{cb}|^2 = 0.96403 (87)$~\cite{Charles:2004jd} &
		$\overline{m}_c(\overline{m}_c) = 1.275 (25) \;\gev$ \\\spp
		$|V_{ts}^* V_{tb}/V_{ub}|^2 = 123.5 (5.3) $~\cite{Charles:2004jd} &
		$m_b^{1S} = 4.691(37) \;\gev$~\cite{Amhis:2012bh,Schwanda:2013bg} \\ \spp
		$|V_{td}^* V_{tb}/V_{cb}|^2 = 0.04195 (78) $~\cite{Charles:2004jd} &
		$|V_{us}^* V_{ub}/(V_{ts}^*V_{tb})| = 0.02022(44)$~\cite{Charles:2004jd} \\ \spp
		$|V_{td}^* V_{tb}/V_{ub}|^2 = 5.38(26)$~\cite{Charles:2004jd} &
		$\arg \left[V_{us}^* V_{ub}/(V_{ts}^*V_{tb})\right] = 115.3(1.3)^{\circ} $~\cite{Charles:2004jd}\\ \spp
		$\mathcal{B}(\bar{B}\to X_c e \bar\nu)_{\rm exp}=0.1065 (16) $~\cite{Amhis:2016xyh} &
		$|V_{ud}^* V_{ub}/(V_{td}^*V_{tb})| = 0.420(10)$ \\ \spp
	   $m_B = 5.2794\;\gev$ &
      $\arg \left[V_{ud}^* V_{ub}/(V_{td}^*V_{tb})\right] = -88.3(1.4)^{\circ}$ \\ \spp
      $M_Z = 91.1876\;\gev$ & $m_{t,{\rm pole}}= 173.1 (0.9) \;\gev$\\ \spp
		$M_W = 80.379\;\gev$ &
		$C = 0.568 (7)(10)$~\cite{Alberti:2014yda} \\ \spp
		$\mu_b = 5^{+5}_{-2.5}\;\gev$ & $\mu_0 = 120^{+120}_{-60}\;\gev$ \\ \spp
		$f_\text{NV} = (0.02 \pm 0.16)\;\gev^3$ &
      $\lambda_2^{\rm eff} = 0.130 (21)\;\gev^2$ \cite{Gambino:2016jkc} \\ \spp
		$f_\text{V}-f_\text{NV} = (0.041 \pm 0.052)\;\gev^3$ &
      $\lambda_1 = - 0.267 (90)\;\gev^2$~\cite{Gambino:2016jkc} \\ \spp
		$[\delta f]_{\mathrm{SU}(3)} = (0 \pm 0.04)\;\gev^3$ &
      $\rho_1 = 0.038 (70)\;\gev^3$~\cite{Gambino:2016jkc} \\ \spp
      $[\delta f]_{\mathrm{SU}(2)} = (0 \pm 0.004)\;\gev^3$ &
       \\ \hline
		\end{tabular}
		\end{displaymath}
		
		\vspace*{-10pt}
		
		\caption{Numerical inputs used in the phenomenological analysis as in \cite{Huber:2019iqf}, taken from PDG~\cite{Tanabashi:2018oca} and CKMfitter Group~\cite{Charles:2004jd}.}
		\label{tab:inputs}
	\end{center}
\end{table}
In the remainder of this section, we present updated numerical results for the branching ratio in two bins of the low dilepton mass region $1\ {\rm GeV}^2 <  q^2  < 6\ \rm{GeV}^2$ and the	high dilepton mass region $q^2 > 14.4$ GeV$^2$. In addition, we give the ratios $R_{X_s}$, $\mathcal{R}(s_0)$ and the forward-backward asymmetry. The remaining angular observables $\mathcal{H}_T, \mathcal{H}_L,\mathcal{H}_A,\mathcal{H}_3,\mathcal{H}_4$ are relegated to Appendix~\ref{sec:pheno}. The quoted uncertainties are
obtained by varying the inputs within their ranges indicated in table~\ref{tab:inputs}, where we assume that $m_c$ and $C$ are
fully anti-correlated. Moreover, we have added a $5\%$ uncertainty due to the resolved contributions as in~\cite{Huber:2019iqf}. The total uncertainties are obtained by adding the individual ones in quadrature. Our results are summarized in table~\ref{tab:mainres}, 
including also the ratios $R_{H_I}$ ($I=T,L,A$).
%%%%%%%%%%%%%%%%%%%%%%%%%%%%%%%%%%%%%%%%%%%%%%%%%%%%%%%%%%%%%%%%%%%%%%%%%%%%%%%%%%%%%%%%%%%%%%%%%%%%%%%%%%%%%%%%%%%%%%%%%%%%%%%%%%%%%%%%%%%%%%%%%%%%%%%%%%%%%%%%%%%%%%
\renewcommand{\arraystretch}{1.2}
\begin{table}
	\begin{center}
		\begin{tabular}{|c|l|l|l|}
			\hline %{$q^2 \in [1,6] \; {\rm GeV}^2$}
			\rule{0pt}{14pt}$q^2 \rm{~range} \;[{\rm GeV}^2]$	& {$[1,3.5]$} & {$[3.5,6]$} & $[1,6]$ \\
			\hline
			\multirow{2}{*}{$\cal B$}	& $9.82\pm 0.67$ 	\multirow{2}{*}{$\times \;\;10^{-7}$} & $7.98\pm 0.67$ 	\multirow{2}{*}{$\times \;\;10^{-7}$} & $17.80\pm 1.33$ 	\multirow{2}{*}{$\times \;\;10^{-7}$} \\
			& $9.44\pm 0.63$ & $7.85\pm 0.66$ &$17.29\pm 1.28$ \\ \hline
			$R_{X_s}$ & $0.961\pm0.004$ & $0.984 \pm 0.002$ & $0.971 \pm 0.003$ \\ \hline
			\multirow{2}{*}{	${\cal H}_T$} & $2.91\pm 0.22$ 	\multirow{2}{*}{$\times \;\;10^{-7}$} & $2.51\pm 0.24$ 	\multirow{2}{*}{$\times \;\;10^{-7}$} & $5.42\pm 0.46$ 	\multirow{2}{*}{$\times \;\;10^{-7}$}\\
			& $2.08\pm 0.14$ &  $2.00\pm 0.20$ & $4.08\pm 0.34$ \\ \hline
			$R_{H_T}$ & $0.714 \pm 0.013$ & $0.798 \pm 0.013$ & $0.753 \pm 0.013$ \\ \hline
			\multirow{2}{*}{	${\cal H}_L$} & $6.92\pm 0.50$ 	\multirow{2}{*}{$\times \;\;10^{-7}$}& $5.43\pm 0.44$	\multirow{2}{*}{$\times \;\;10^{-7}$} & $12.35\pm 0.92$ 	\multirow{2}{*}{$\times \;\;10^{-7}$} \\
			& $7.37\pm 0.52$ & $5.81\pm 0.47$ & $13.18\pm 0.96$ \\ \hline
			$R_{H_L}$ & $1.065 \pm 0.006$ & $1.070 \pm 0.006$  & $1.067 \pm 0.006$ \\ \hline
			\multirow{2}{*}{	${\cal H}_A$} & $-0.95\pm 0.08$ 	\multirow{2}{*}{$\times \;\;10^{-7}$}& $0.91\pm 0.16$	\multirow{2}{*}{$\times \;\;10^{-7}$} & $-0.04\pm 0.21$ 	\multirow{2}{*}{$\times \;\;10^{-7}$} \\
			& $-1.03\pm 0.08$  &$0.85\pm 0.16$ & $-0.18\pm 0.21$ \\ \hline
			$R_{H_A}$ & $1.077 \pm 0.008$ & $0.933 \pm 0.013$ & \qquad\qquad $-$ \\ \hline
			\multirow{2}{*}{	${\cal H}_3$} & $4.14\pm 0.70$ 	\multirow{2}{*}{$\times \;\;10^{-9}$}& $5.00\pm 0.59$	\multirow{2}{*}{$\times \;\;10^{-9}$} & $9.14\pm 1.29$	\multirow{2}{*}{$\times \;\;10^{-9}$} \\
			& $1.72\pm 0.29$ & $2.08\pm 0.25$ & $3.80\pm 0.53$ \\ \hline
			\multirow{2}{*}{	${\cal H}_4$}  & $6.37\pm 0.67$	\multirow{2}{*}{$\times \;\;10^{-9}$} & $2.24\pm 0.21$	\multirow{2}{*}{$\times \;\;10^{-9}$} & $8.60\pm 0.88$ 	\multirow{2}{*}{$\times \;\;10^{-9}$} \\
			& $2.65\pm 0.28$ & $0.93\pm 0.09$ & $3.58\pm 0.36$ \\
			\hline \hline
			\rule{0pt}{14pt}$q^2 \rm{~range} \;[{\rm GeV}^2]$ & \multicolumn{3}{c|}{$>14.4$ }\\ \hline
			\multirow{2}{*}{	$\cal B$} & \multicolumn{3}{c|}{$2.04\pm 0.87$	\multirow{2}{*}{$\;\;\;\times \;\;10^{-7}$}}\\
			& \multicolumn{3}{c|}{$2.38\pm 0.87$ \multirow{2}{*}{$\;\;\;\;\;\;\;\;\;\;\;\;\;\;\;\;$}}\\\hline
			$R_{X_s}$  & \multicolumn{3}{c|}{$1.17\pm 0.08$} \\ \hline
			\multirow{2}{*}{	$\mathcal{R}(s_0)$ 	}	& \multicolumn{3}{c|}{$21.53 \pm 2.35$\multirow{2}{*}{$\;\;\;\times \;\;10^{-4}$}}\\
			&	 \multicolumn{3}{c|}{$25.33\pm 1.93$\multirow{2}{*}{$\;\;\;\;\;\;\;\;\;\;\;\;\;\;\;$}}\\\hline

		\end{tabular}
		\caption{Summary of the numerical results for the different observables. If applicable, the first entry in each row is for electrons, the second for muons.
		}
	\label{tab:mainres}
	\end{center}
\end{table}
\renewcommand{\arraystretch}{1} 
%%%%%%%%%%%%%%%%%%%%%%%%%%%%%%%%%%%%%%%%%%%%%%%%%%%%%%%%%%%%%%%%%%%%%%%%%%%%%%%%%%%%%%%%%%%%%%%%%%%%%%%%%%%%%%%%%%%%%%%%%%%%%%%%%%%%%%%%%%%%%%%%%%%%%%%%%%%%%%%%%%%%%%
\subsection{Branching ratio, low-$q^2$ region}\label{sec:brlowq2}
We give the results for the branching ratios integrated over two bins in the low-$q^2$ region $1~\text{GeV}^2 < q^2 < 6~\text{GeV}^2$. As is customary, we present our results for both electron and muon final states separately.
For the low-$q^2$ region, we neglect $1/m_b^3$ corrections.
\begin{align}
\dps {\cal B}[1,3.5]_{ee} =&\   (9.82\pm 0.34_{\text{scale}}\pm 0.10_{m_t}\pm 0.21_{C,m_c}\pm 0.11_{m_b}\pm 0.04_{\alpha _s}\pm 0.009_{\text{CKM}}\nonumber \\
& \hspace*{27pt} \pm 0.15_{\text{BR}_{\text{sl}}}\pm 0.06_{\lambda _2} \pm 0.49_\text{resolved})\cdot 10^{-7} = (9.82\pm 0.67) \cdot 10^{-7} \; .
\end{align}
\begin{align}
\dps {\cal B}[3.5,6]_{ee} =&\   (7.98\pm 0.47_{\text{scale}}\pm 0.09_{m_t}\pm 0.19_{C,m_c}\pm 0.09_{m_b}\pm 0.06_{\alpha _s}\pm 0.01_{\text{CKM}}\nonumber \\
& \hspace*{27pt} \pm 0.12_{\text{BR}_{\text{sl}}}\pm 0.06_{\lambda _2} \pm 0.40_\text{resolved})\cdot 10^{-7} = (7.98\pm 0.67) \cdot 10^{-7} \; .
\end{align}
\begin{align}\label{eq:BRee}
\dps {\cal B}[1,6]_{ee} =&\  (17.80\pm 0.80_{\text{scale}}\pm 0.19_{m_t}\pm 0.39_{C,m_c}\pm 0.20_{m_b}\pm 0.10_{\alpha _s}\pm 0.02_{\text{CKM}}\nonumber \\
& \hspace*{32.5pt} \pm 0.27_{\text{BR}_{\text{sl}}}\pm 0.12_{\lambda _2} \pm 0.89_\text{resolved})\cdot 10^{-7} = (17.80\pm 1.33) \cdot 10^{-7} \; .
\end{align}
\begin{align}
\dps {\cal B}[1,3.5]_{\mu\mu} =&\   (9.44\pm 0.30_{\text{scale}}\pm 0.10_{m_t}\pm 0.20_{C,m_c}\pm 0.11_{m_b}\pm 0.04_{\alpha _s}\pm 0.009_{\text{CKM}}\nonumber \\
& \hspace*{27pt} \pm 0.14_{\text{BR}_{\text{sl}}}\pm 0.06_{\lambda _2} \pm 0.47_\text{resolved})\cdot 10^{-7} = (9.44\pm 0.63) \cdot 10^{-7} \; .
\end{align}
\begin{align}
\dps {\cal B}[3.5,6]_{\mu\mu} =&\ (7.85\pm 0.45_{\text{scale}}\pm 0.09_{m_t}\pm 0.19_{C,m_c}\pm 0.10_{m_b}\pm 0.06_{\alpha _s}\pm 0.01_{\text{CKM}}\nonumber \\
& \hspace*{27pt} \pm 0.12_{\text{BR}_{\text{sl}}}\pm 0.06_{\lambda _2} \pm 0.39_\text{resolved})\cdot 10^{-7} = (7.85\pm 0.66) \cdot 10^{-7} \; .
\end{align}
\begin{align}\label{eq:BRmumu}
\dps {\cal B}[1,6]_{\mu\mu} =&\ (17.29\pm 0.76_{\text{scale}}\pm 0.19_{m_t}\pm 0.39_{C,m_c}\pm 0.20_{m_b}\pm 0.09_{\alpha _s}\pm 0.02_{\text{CKM}}\nonumber \\
& \hspace*{32.5pt} \pm 0.26_{\text{BR}_{\text{sl}}}\pm 0.12_{\lambda _2} \pm 0.86_\text{resolved})\cdot 10^{-7} = (17.29\pm 1.28) \cdot 10^{-7} \; .
\end{align}
These new results are $6$--$7\%$ larger compared to the previous numerical analysis \cite{Huber:2015sra}, and have an increased uncertainty. The difference in the central value can partially be traced back to changes in the input parameters (mainly CKM factors and the value of the semileptonic branching ratio). The remaining shift~--~and in fact the dominant one~--~stems from the more sophisticated analysis of the non-perturbative effects by updating the Kr\"uger-Sehgal analysis along the lines of~\cite{Huber:2019iqf}. In addition, in the low-$q^2$ region we do not implement the $1/m_c^2$ effects as in~\cite{Buchalla:1997ky} any more, but add in quadrature a $5\%$ uncertainty for the resolved contributions~\cite{Hurth:2017xzf,Benzke:2017woq,Benzke:2020htm}, a procedure that was already applied in~\cite{Huber:2019iqf}. It shifts the central value only marginally, but is entirely responsible for the increase in uncertainty.

%%%%%%%%%%%%%%%%%%%%%%%%%%%%%%%%%%%%%%%%%%%%%%%%%%%%%%%%%%%%%%%%%%%%%%%%%%%%%%%%%%%%%%%%%%%%%%%%%%%%%%%%%%%%%%%%%%%%%%%%%%%%%%%%%%%%%%%%%%%%%%%%%%%%%%%%%%%%%%%%%%%%%%

\subsection{Branching ratio, high-$q^2$ region}\label{sec:brhighq2}
In the high-$q^2$ region, $q^2 > 14.4~\text{GeV}^2$, we find
\begin{align}
\dps {\cal B}[>14.4]_{ee} =&\ ( 2.04\pm 0.28_{\text{scale}}\pm 0.02_{m_t}\pm 0.03_{C,m_c}\pm 0.19_{m_b}\pm 0.002_{\text{CKM}} \pm 0.03_{\text{BR}_{\text{sl}}}\nonumber \\[0.5em]
   & \hspace*{25pt} \pm 0.006_{\alpha _s} \pm 0.13_{\lambda _2}\pm 0.57_{\rho _1}\pm 0.54_{f_{u,s}}) \cdot 10^{-7} = ( 2.04 \pm 0.87) \cdot 10^{-7} \; , \\
\dps {\cal B}[>14.4]_{\mu\mu} =&\ ( 2.38 \pm 0.27_{\text{scale}}\pm 0.03_{m_t}\pm 0.04_{C,m_c}\pm 0.21_{m_b}\pm 0.002_{\text{CKM}} \pm 0.04_{\text{BR}_{\text{sl}}}\nonumber \\[0.5em]
   & \hspace{25pt} \pm 0.006_{\alpha _s} \pm 0.12_{\lambda _2}\pm 0.57_{\rho _1}\pm 0.54_{f_{u,s}}) \cdot 10^{-7}
   = (2.38\pm 0.87) \cdot 10^{-7} \; .
\end{align}
Here the power corrections proportional to $\lambda_{1,2}, \rho_1, f_u^{0,\pm}, f_s$, expanded to linear power in these parameters, are also included. We only quote the combined uncertainty of the weak annihilation parameters $f_{u,s}$ due to their correlation. Compared to the previous analysis in ref.~\cite{Huber:2015sra}, we find an increased uncertainty caused by the power-corrections $\rho_1$ and $f_{u,s}$.

%%%%%%%%%%%%%%%%%%%%%%%%%%%%%%%%%%%%%%%%%%%%%%%%%%%%%%%%%%%%%%%%%%%%%%%%%%%%%%%%%%%%%%%%%%%%%%%%%%%%%%%%%%%%%%%%%%%%%%%%%%%%%%%%%%%%%%%%%%%%%%%%%%%%%%%%%%%%%%%%%%%%%%

\subsection{The ratio ${R}_{X_s}$}\label{sec:ratioRXs}
With our updated results, we can now also consider the lepton-universality ratio $R_{X_s}$ for the inclusive decays, defined in eq.~\eqref{eq:Rratdef}.
We discuss this ratio again in more detail in section~\ref{sec:ratioRXSNP}, where we study the constraints on new physics.
For the SM, we find the following predictions,
\begin{align}
\dps R_{X_s}[1,3.5] =&\ 0.961 \pm 0.004_{\text{scale}} \pm {3\times10^{-5}}_{m_t} \pm 0.0002_{C,m_c} \pm 0.0004_{m_b}  \nonumber \\
&\hspace*{25pt} \pm  {4\times10^{-5}}_{\alpha_s}  \pm {8\times10^{-5}}_{\lambda _1}\pm {7\times10^{-5}}_{\lambda _2}  = 0.961 \pm 0.004\; ,  \\[0.5em]
\dps R_{X_s}[3.5,6] =&\ 0.984 \pm 0.001_{\text{scale}} \pm {4\times10^{-5}}_{m_t} \pm 0.0002_{C,m_c} \pm 0.0005_{m_b}  \nonumber \\
&\hspace*{25pt} \pm  {2\times10^{-5}}_{\alpha_s}  \pm {3\times10^{-5}}_{\lambda _1}\pm {4\times10^{-5}}_{\lambda _2}  = 0.984 \pm 0.002\; , \\[0.5em]
\dps R_{X_s}[1,6] =&\ 0.971 \pm 0.003_{\text{scale}} \pm {7\times10^{-6}}_{m_t} \pm 0.0002_{C,m_c} \pm 0.0004_{m_b}  \nonumber \\
&\hspace*{25pt} \pm  {3\times10^{-6}}_{\alpha_s}  \pm {6\times10^{-5}}_{\lambda _1}\pm {7\times10^{-5}}_{\lambda _2}  = 0.971 \pm 0.003\; , 
\\[0.5em]
\dps R_{X_s}[>14.4] =&\ 1.17 \pm 0.03_{\text{scale}} \pm 0.0003_{m_t} \pm 0.002_{C,m_c} \pm 0.006_{m_b}  \nonumber \\
&\hspace*{25pt} \pm  0.0009_{\alpha_s}  \pm 0.01_{\lambda _2}\pm 0.04_{\rho _1}\pm 0.06_{f_{u,s}}  = 1.17 \pm 0.08\; .
\end{align}

%%%%%%%%%%%%%%%%%%%%%%%%%%%%%%%%%%%%%%%%%%%%%%%%%%%%%%%%%%%%%%%%%%%%%%%%%%%%%%%%%%%%%%%%%%%%%%%%%%%%%%%%%%%%%%%%%%%%%%%%%%%%%%%%%%%%%%%%%%%%%%%%%%%%%%%%%%%%%%%%%%%%%%

\subsection{The ratio ${\cal R}(s_0)$}\label{sec:ratioRhighq2}
In order to reduce the large uncertainties from power corrections in the high-$q^2$ region, we compute the ratio $\mathcal{R}(s_0)_{\ell\ell}$ from eq.~\eqref{eq:zoltanR}. We find
\begin{align}
  {\cal R}(14.4)_{ee} =&\ ( 21.53 \pm 0.54_{\text{scale}} \pm 0.25_{m_t} \pm 0.15_{C,m_c} \pm 0.09_{m_b} \pm 0.06_{\alpha
   _s} \pm 0.92_{\text{CKM}} \nonumber \\[0.5em]
   & \hspace*{25pt} \pm 0.11_{\lambda _2} \pm 1.38_{\rho_1} \pm 1.54_{f_{u,s}} )  \times 10^{-4}
                       =( 21.53 \pm 2.35 )  \times 10^{-4} \; ,
                       \label{eq:RZO1}\\[1.0em]
  {\cal R}(14.4)_{\mu\mu} =&\ ( 25.33  \pm 0.27_{\text{scale}} \pm 0.29_{m_t} \pm 0.14_{C,m_c} \pm 0.03_{m_b} \pm 0.07_{\alpha
   _s} \pm 1.09_{\text{CKM}} \nonumber \\[0.5em]
   & \hspace*{25pt} \pm 0.04_{\lambda _2} \pm 0.83_{\rho_1} \pm 1.29_{f_{u,s}} )  \times 10^{-4}
                       = ( 25.33 \pm 1.93 )  \times 10^{-4} \; .
\label{eq:RZO2}
\end{align}
Even though this ratio is much less sensitive to power corrections, the latter contributes significantly to the uncertainty. However, note that the uncertainty has been reduced to about $10\%$, which is smaller than in previous analysis although we include $30\%$ $\mathrm{SU}(3)$ breaking effects in the weak annihilation parameters. This reveals once more the robustness of this ratio.

%%%%%%%%%%%%%%%%%%%%%%%%%%%%%%%%%%%%%%%%%%%%%%%%%%%%%%%%%%%%%%%%%%%%%%%%%%%%%%%%%%%%%%%%%%%%%%%%%%%%%%%%%%%%%%%%%%%%%%%%%%%%%%%%%%%%%%%%%%%%%%%%%%%%%%%%%%%%%%%%%%%%%%

\subsection{Forward-backward asymmetry, low-$q^2$ region}\label{sec:fbalowq2}
 The forward-backward asymmetry $A_\text{FB}$ and the related angular observable $H_A$ defined in eqs.~\eqref{eq:afb} and~\eqref{eq:projectionHI} are computed for the low-$q^2$ region. These observables have a zero-crossing at a position $q_0^2$ (in units of GeV$^2$) which we find to be
 \begin{align}\label{eq:zero}
 \dps (q_0^2)_{ee} =& \ 3.28\pm 0.11_{\text{scale}}\pm 0.001_{m_t}\pm 0.02_{C,m_c}\pm 0.05_{m_b} \nonumber \\
 & \hspace*{25pt} \pm 0.03_{\alpha _s}\pm 0.002_{\lambda _1}\pm 0.001_{\lambda _2}\pm 0.06_\text{resolved} =3.28\pm 0.14  \; , \\[0.5em]
 \dps (q_0^2)_{\mu\mu}=& \ 3.40\pm 0.12_{\text{scale}}\pm 0.001_{m_t}\pm 0.02_{C,m_c}\pm 0.05_{m_b} \nonumber \\
 & \hspace*{25pt} \pm 0.03_{\alpha _s}\pm 0.002_{\lambda _1}\pm 0.002_{\lambda _2}\pm 0.06_\text{resolved}=3.40\pm 0.15  \; .
 \end{align}
For the normalized forward-backward asymmetry it is natural to subdivide the low-$q^2$ region into two bins due to the zero-crossing, 
\begin{align}
\dps \overline{A}_{\rm FB}[1,3.5]_{ee} =& \ (-7.28\pm 0.67_{\text{scale}}\pm 0.01_{m_t}\pm 0.11_{C,m_c}\pm 0.23_{m_b}
\nonumber \\
& \hspace*{5pt}\pm 0.19_{\alpha _s} \pm  0.04_{\lambda _2} \pm 0.51_\text{resolved})\%=(-7.28\pm 0.90)\% \; , \\[0.5em]
\dps \overline{A}_{\rm FB}[3.5,6]_{ee} =& \ (8.57\pm 0.74_{\text{scale}}\pm 0.01_{m_t}\pm 0.13_{C,m_c}\pm 0.37_{m_b}
\nonumber \\
& \hspace*{5pt} \pm 0.18_{\alpha _s} \pm 0.11_{\lambda _2} \pm 0.60_\text{resolved})\%= (8.57\pm 1.05)\% \; ,\\[0.5em]
\dps \overline{A}_{\rm FB}[1,6]_{ee} =&\  (-0.18\pm 0.79_{\text{scale}}\pm 0.004_{m_t}\pm 0.13_{C,m_c}\pm 0.30_{m_b}
\nonumber \\
& \hspace*{5pt} \pm 0.20_{\alpha _s} \pm 0.02_{\lambda _2} \pm 0.01_\text{resolved})\%= (-0.18\pm 0.88)\% \; ,
\end{align}

\vspace*{-19pt}

\begin{align}
\dps \overline{A}_{\rm FB}[1,3.5]_{\mu\mu} =&\  (-8.16\pm 0.68_{\text{scale}}\pm 0.01_{m_t}\pm 0.11_{C,m_c}\pm 0.23_{m_b} \nonumber \\
& \hspace*{2pt} \pm 0.20_{\alpha _s} \pm 0.05_{\lambda _2} \pm 0.57_\text{resolved})\%= (-8.16\pm 0.95)\% \; , \\[0.5em]
\dps \overline{A}_{\rm FB}[3.5,6]_{\mu\mu} =& \ (8.13\pm 0.81_{\text{scale}}\pm 0.01_{m_t}\pm 0.13_{C,m_c}\pm 0.39_{m_b}\nonumber \\
& \hspace*{2pt} \pm 0.19_{\alpha _s} \pm 0.11_{\lambda _2} \pm 0.57_\text{resolved}) \%= (8.13\pm 1.09)\%  \; , \\[0.5em]
\dps \overline{A}_{\rm FB}[1,6]_{\mu\mu} =& \ (-0.77 \pm 0.84_{\text{scale}}\pm 0.004_{m_t}\pm 0.13_{C,m_c}\pm 0.31_{m_b}\nonumber \\
& \hspace*{2pt} \pm 0.21_{\alpha _s}  \pm 0.02_{\lambda _2} \pm 0.05_\text{resolved})\%= (-0.77\pm 0.93)\%  \; .
\end{align}

%%%%%%%%%%%%%%%%%%%%%%%%%%%%%%%%%%%%%%%%%%%%%%%%%%%%%%%%%%%%%%%%%%%%%%%%%%%%%%%%%%%%%%%%%%%%%%%%%%%%%%%%%%%%%%%%%%%%%%%%%%%%%%%%%%%%%%%%%%%%%%%%%%%%%%%%%%%%%%%%%%%%%%

%%%%%%%%%%%%%%%%%%%%%%%%%%%%%%%%%%%%%%%%%%%%%%%%%%%%%%%%%%%%%%%%%%%%%%%%%%%%%%%%%%%%%%%%%%%%%%%%%%%%%%%%%%%%%%%%%%%%%%%%%%%%%%%%%%%%%%%%%%%%%%%%%%%%%%%%%%%%%%%%%%%%%%
%%%%%%%%%%%%%%%%%%%%%%%%%%%%%%%%%%%%%%%%%%%%%%%%%%%%%%%%%%%%%%%%%%%%%%%%%%%%%%%%%%%%%%%%%%%%%%%%%%%%%%%%%%%%%%%%%%%%%%%%%%%%%%%%%%%%%%%%%%%%%%%%%%%%%%%%%%%%%%%%%%%%%%

\section{Treatment of collinear photons}
\label{sec:collinear}
In our calculation we include the effects of a single photon emission from the final state leptons. In the analytic expressions we derived in~\cite{Huber:2015sra}, the dilepton invariant mass is calculated without the inclusion of the photon, which is therefore considered to be part of the hadronic system. Contributions of photon radiation to the double differential branching ratio $d^2{\cal B}/dq^2 dz$ are calculated in the collinear approximation. One general result is that collinear radiation effects vanish once the differential rate is integrated over the entire phase space. Effects are only possible for low and high $q^2$ separately and tend to have opposite sign. The reason
is that the differential branching ratio $d{\cal B}/dq^2$ is not an infrared safe quantity with respect to collinear photon radiation off final state leptons. The integrated branching ratio, on the other hand, is infrared safe.

The typical size of the electromagnetic effects is expected to be small (i.e.~of the order of $\alpha_e/(4\pi) \log m_b^2/m_e^2 \sim 1\%$). There are, however, instances in which the net effect turns out to be disproportionately large. This is the case for $\mathcal{H}_T$ at low-$q^2$ for which collinear photon effects are $\mathcal{O}(1)$. We refer to section 7 of ref.~\cite{Huber:2015sra} for a complete discussion of this point.

In this section we discuss the effects of collinear radiation from the two narrow resonances $J/\psi$ and $\psi(2S)$. Compared to electrons, muons radiate much less due to their larger mass. Moreover, muons can be well separated from collinear photons in the detector, which is why we focus on the electron case in what follows.
It is easy to show that the emission of a real photon can only decrease the invariant mass of the dilepton: $(p_{e^+} + p_{e^-})^2 <  (p_{e^+} + p_{e^-}+ p_{\gamma,{\rm coll}})^2$. The net effect is a bin migration of the spectrum towards lower dilepton invariant mass: radiation from the resonances can only effect low-$q^2$ observables.

Unfortunately, it is impossible to produce a reliable estimate of collinear radiation from $J/\psi$ and $\psi(2S)$. In fact, while we are able to use the KS dispersive approach to achieve a complete description of resonances in the colour singlet channel, there is no accurate theoretical approach for the calculation of the colour  octet channel. Using the KS method, the colour singlet contributions to the branching ratios $\bar{B}\to X_s \psi \to X_s e^+ e^-$ are found to be $1.1\times 10^{-4}$ and $5.9\times 10^{-6}$ for the $J/\psi$ and $\psi(2S)$ cases; using the measured branching ratios for direct charmonium production~\cite{Tanabashi:2018oca} we find $4.7 \times 10^{-4}$ and $2.4 \times 10^{-5}$, respectively. The colour singlet channel accounts for only a quarter of the total resonance contribution: this result is well known. It can be taken into account by adding a corresponding multiplicative factor (also referred to ``fudge factors'' in the literature) of about 2 to the $\bar{B}\to (J/\psi X_s, \psi(2S) X_s) $ amplitudes. As we discussed at length in section 4 of ref.~\cite{Huber:2019iqf}, this problem becomes manageable at low-$q^2$ where the effects of the colour octet channel are included in the so-called resolved contributions which have been estimated to lead to a level below 5\%.

\begin{figure}[h!]
	\begin{center}
	\includegraphics[width=\textwidth]{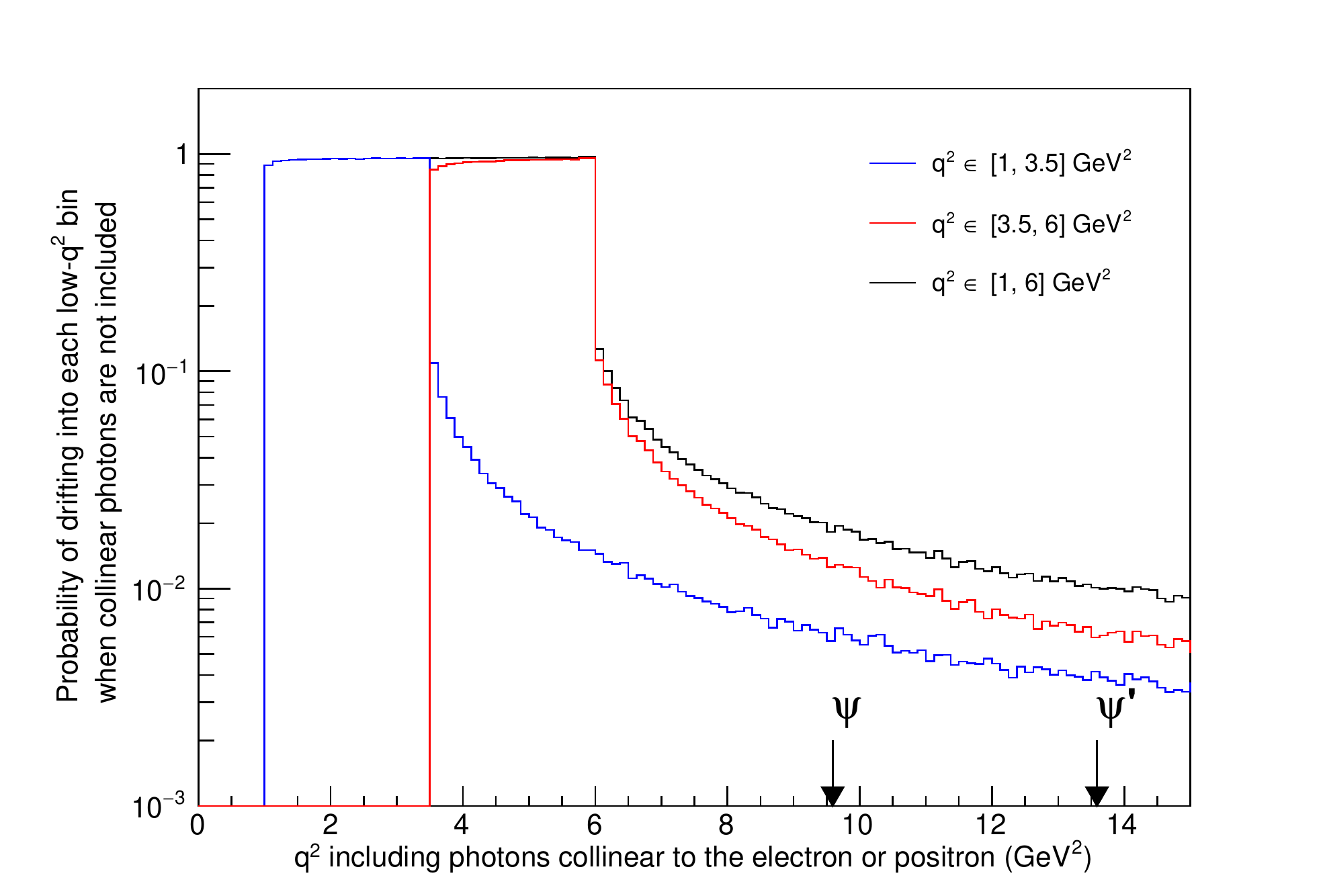}
	\caption{The probability that a $\bar{B} \to X_s \ell^+ \ell^-$ event with photons collinear to either outgoing lepton ($\theta<50\text{mrad}$ and $E_\gamma>20\text{MeV}$ in the Belle~II laboratory frame) will drift into the low-$q^2$ bin as a result of including only the charged particles in the definition of $q^2$, using events generated in Monte Carlo. The resonances $\psi$ and $\psi'$ are marked on the horizontal axis.}
	\label{fig:qed-drift}
	\end{center}
\end{figure}

Given our inability to calculate accurately the effects of the two narrow charmonium resonances, it is imperative to make sure that bin migration from the resonances does not pollute the low-$q^2$ branching ratio above the few percent level. Using Monte Carlo events generated using \texttt{EVTGEN}~\cite{Lange:2001uf}, \texttt{JETSET}~\cite{Sjostrand:1993yb} and \texttt{PHOTOS}~\cite{Golonka:2005pn} (see section 7 of ref.~\cite{Huber:2015sra} for a complete description of the event generation), it is straightforward to calculate the contribution of a given bin in $q^2$ to the integrated low-$q^2$ branching ratios. The results of this analysis are presented in figure~\ref{fig:qed-drift}, where the blue, red and black curves give the probability of migration into the $[1,3.5] \; {\rm GeV}^2$, $[3.5,6] \; {\rm GeV}^2$ and $[1,6] \; {\rm GeV}^2$ bins. Convoluting these results with the analytical expressions for resonant production (rescaled by the appropriate fudge factor to roughly take into account colour octet effects), we see that the contributions of the $J/\psi$ and $\psi(2S)$ to the low-$q^2$ branching ratios integrated in the three bins mentioned above can be roughly estimated as $(3,6,9)\times 10^{-6}$ and $(1,1.5,2.5)\times 10^{-7}$, respectively. In comparison with the results presented in section~\ref{sec:results}, we see that $J/\psi$ contamination is larger than the non-resonant contribution by almost an order of magnitude (the resonant contributions to the three bins are a factor of 3, 8 and 5 times larger than the non-resonant ones).

The problem discussed in the above paragraph is very well known and has been taken into account in existing experimental analyses. For instance, in the most recent Belle measurement of the low-$q^2$ branching ratio, the quantity $q^2_{ee\gamma} = m_{ee\gamma}^2$ was formed by including collinear photons (if any) with the leptons. Some of the events with $q^2_{ee\gamma}$ near the $J/\psi$ or $\psi(2S)$ resonances will have $q^2_{ee}$ in the $[1,6]\; {\rm GeV}^2$ range (as we mentioned above, drift is only possible  towards lower values of $q^2_{ee}$). Events with $q_{ee\gamma}^2$ in the ranges $[7.3,10.5]$ GeV$^2$ and $[12.5,14.3]$ GeV$^2$ were vetoed to suppress backgrounds from bin migration from $J/\psi$ and $\psi(2S)$ respectively.

We investigated the effect of this cut on all low-$q^2$ observables using events generated in Monte Carlo as follows: For each $\bar{B} \to X_s \ell^+ \ell^-$ event, photons with the ten highest energies in the lab frame were considered in addition to the two lepton momenta. For each photon, if the photon angle was within 50 mrad of $p_+ (p_-)$, it was added to a total photon vector $k_{+} (k_{-})$ (in case it was within both cones, there was an addition to the cone of the nearest lepton). If the energy of $k_+(k_-)$ exceeded a threshold of 20 MeV, then it was added to $p_+(p_-)$. The dilepton mass square and angular variable $z$ were then computed with the potentially modified lepton momenta. The results of this study are shown in the $``q^2 = q_{ee}^2$" section of table~\ref{tab:correction}. We also investigated the mild dependence of the cone angle and energy threshold.

Alternatively, the quantity $q_{ee\gamma}^2$ can be used in place of $q_{ee}^2$ to form histograms of observables, circumventing the need to correct for bin migration. However, including collinear photons in the definition of the dilepton momentum no longer corresponds to the definition used to make our theoretical predictions (recall that the photon is treated as part of the hadronic system in the theoretical predictions). In order to make bins in $q_{ee\gamma}^2$ in an experimental analysis and compare them to theoretical predictions, shifts need to be made and can be estimated in Monte Carlo in the same fashion as before (see the $``q^2 = q_{ee\gamma}^2$" section of table~\ref{tab:correction}). 

The shifts required for the latter analysis strategy are noticeably larger, in particular for the branching ratio in the high-$q^2$ region and for ${\cal H}_T$. This study suggests that the optimal strategy for dealing with collinear photons at Belle II is to treat all prompt photons as part of the hadronic system. After removing peaking backgrounds from the narrow resonances $J/\psi$ and $\psi(2S)$, the binned observables can be compared directly to our theoretical predictions after applying the appropriate ``$q^2 = q_{ee}^2$" correction terms presented in table~\ref{tab:correction}.

\begin{table}[t]
	\begin{center}
		\begin{tabular}{|r||rrr||rrr|}
			\hline
			\rule{0pt}{14pt}& & & $q^2 = q_{ee}^2$ & & & $ q^2 = q_{ee\gamma}^2$ \\
			& 50 mrad & 100 mrad & 50 mrad & 50 mrad & 100 mrad & 50 mrad \\
			& 20 MeV & 20 MeV & 80 MeV & 20 MeV & 20 MeV & 80 MeV \\
			\hline
			$\mathcal{B}[1, 3.5]$ & $-0.5~\%$ & $-0.6~\%$ & $-0.5~\%$ & $-1.9~\%$& $-2.2~\%$ & $-1.8~\%$\\
			$\mathcal{B}[3.5,6]$ & $-1.6~\%$ & $-1.9~\%$ & $-1.6~\%$ & $-1.3~\%$ & $-1.4~\%$ & $-1.2~\%$\\
			$\mathcal{B}[1,6]$ & $-1.0~\%$ & $-1.2~\%$ & $-1.0~\%$ & $-1.6~\%$ & $-1.9~\%$ & $-1.5~\%$\\
			\hline
			$\mathcal{H}_T[1, 3.5]$ & $-5.6~\%$ & $-6.7~\%$ & $-5.6~\%$ & $-13.8~\%$ & $-16.7~\%$ & $-13.5~\%$\\
			$\mathcal{H}_T[3.5, 6]$ & $-6.7~\%$ & $-8.0~\%$ & $-6.7~\%$  & $-12.9~\%$ & $-15.3~\%$ & $-12.5~\%$\\
			$\mathcal{H}_T[1,6]$ & $-6.1~\%$& $-7.3~\%$ & $-6.1~\%$ & $-13.4~\%$ & $-16.1~\%$ & $-13.0~\%$\\
			\hline
			$\mathcal{H}_L[1,3.5]$ & $1.1~\%$ & $1.3~\%$ & $1.1~\%$   & $2.0~\%$& $2.5~\%$& $2.0~\%$ \\
			$\mathcal{H}_L[3.5,6]$  & $0.2~\%$ & $0.2~\%$& $0.2~\%$  & $2.8~\%$& $3.4~\%$& $2.7~\%$\\
			$\mathcal{H}_L[1,6]$ & $0.7~\%$ & $0.9~\%$ & $0.7~\%$ & $2.3~\%$& $2.9~\%$& $2.3~\%$\\
			\hline
			$\mathcal{H}_A[1,3.5]$ & $0.2~\%$ & $0.2~\%$ & $0.2~\%$ & $2.1~\%$ & $2.6~\%$ & $2.0~\%$\\
			$\mathcal{H}_A[3.5,6]$  & $-4.7~\%$ & $-5.8~\%$ & $-4.7~\%$ & $-11.6~\%$ & $-13.8~\%$ & $-10.1~\%$\\
			$\mathcal{H}_A[1,6]$ & $2.6~\%$ & $3.3~\%$ & $2.6~\%$ & $9.0~\%$ & $10.9~\%$ & $8.2~\%$\\
			\hline
			$\mathcal{B}[>14.4]$ & $0~\%$ & $0~\%$ & $0~\%$ & $6.4~\%$ & $7.4~\%$ & $5.9~\%$\\
			\hline
  		\end{tabular}
	\end{center}
	\caption{Correction factors to the SM predictions presented in section~\ref{sec:results} required for a direct comparison with measurements performed using the two experimental strategies $q^2=q_{ee}^2$ and $q_{ee\gamma}^2$ as defined in the text.}
\label{tab:correction}
\end{table}

%%%%%%%%%%%%%%%%%%%%%%%%%%%%%%%%%%%%%%%%%%%%%%%%%%%%%%%%%%%%%%%%%%%%%%%%%%%%%%%%%%%%%%%%%%%%%%%%%%%%%%%%%%%%%%%%%%%%%%%%%%%%%%%%%%%%%%%%%%%%%%%%%%%%%%%%%%%%%%%%%%%%%%
%%%%%%%%%%%%%%%%%%%%%%%%%%%%%%%%%%%%%%%%%%%%%%%%%%%%%%%%%%%%%%%%%%%%%%%%%%%%%%%%%%%%%%%%%%%%%%%%%%%%%%%%%%%%%%%%%%%%%%%%%%%%%%%%%%%%%%%%%%%%%%%%%%%%%%%%%%%%%%%%%%%%%%

\section{New physics sensitivities}
\label{sec:newphysics}
In this section we discuss the existing constraints that Babar and Belle measurements impose on the Wilson coefficients and the projected sensitivity of Belle~II with $50\text{ab}^{-1}$ of integrated luminosity. We assume that the magnetic moment coefficients $C_7$ and $C_8$ do not receive appreciable new physics contributions and focus on the semileptonic operators. We express our results in terms of the new physics contributions to the Wilson coefficients evaluated at the matching scale $\mu_0 = 120$ GeV and adopt the parameterization
\begin{align}
C_{9,10}^\ell (\mu_0) &= C_{9,10}^{\ell,{\rm SM}} + C_{9,10}^{\ell{\rm NP}} \; , \label{eq:NPWC}
\end{align}
with $\ell=e,\mu$. Our operator basis is the same as in~\cite{Huber:2005ig}.

\begin{figure}
\begin{center}
\includegraphics[width=0.49\linewidth]{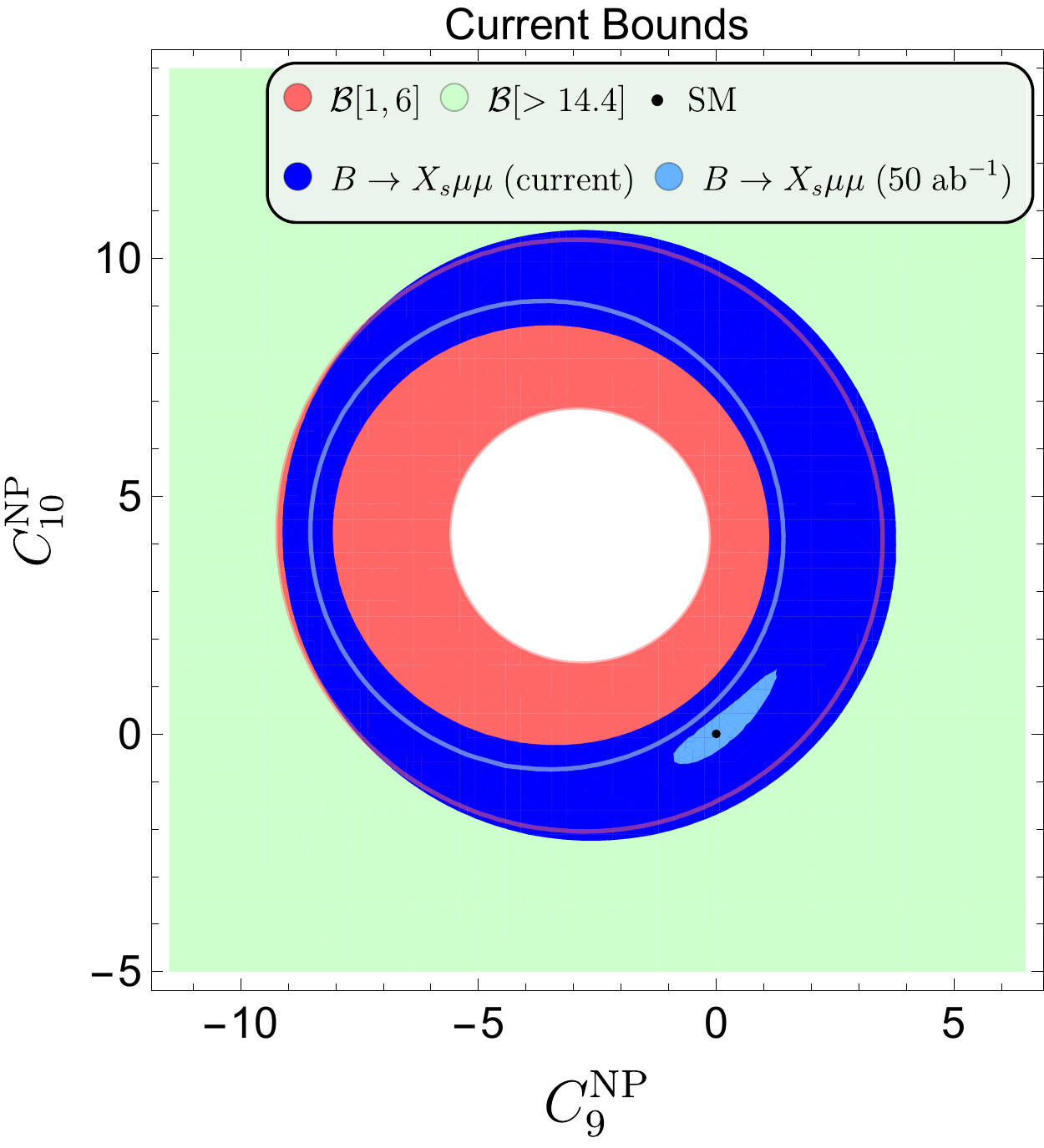}
\includegraphics[width=0.49\linewidth]{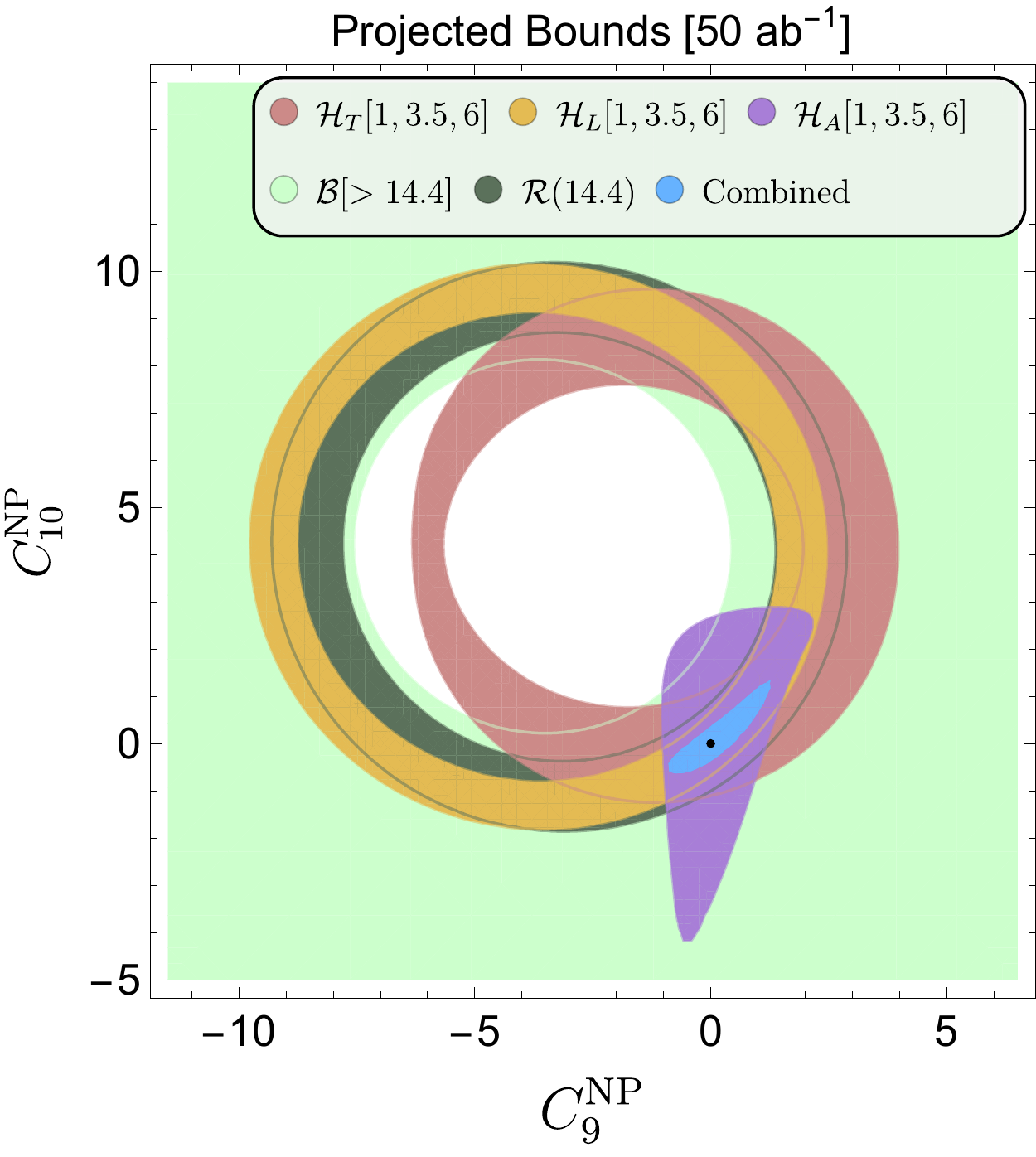}
\caption{Current (left panel) and expected bounds (right panel) on the new physics contributions to the Wilson coefficients $C_9 (\mu_0)$ and $C_{10} (\mu_0)$ with $\mu_0=120$ GeV. The coefficients $C_{7,8}$ are kept at their SM values. All regions are determined at 95\% C.L.. In
both panels we combine the electron and muon channels under the assumption $C_{9,10}^{e} = C_{9,10}^\mu$. In the 50 ab${}^{-1}$ extrapolation we combine separately projected measurements of $H_T$, $H_L$ and $H_A$ in the two low-$q^2$ bins.
}
\label{fig:bounds}
\end{center}
\end{figure}

We first consider the existing bounds which stem from branching ratio measurements at low- and high-$q^2$. The weighted average of the BaBar~\cite{Aubert:2004it, Lees:2013nxa} and Belle~\cite{Iwasaki:2005sy, Sato:2014pjr} experimental results are:
\begin{align}
{\cal B}[1,6]_{\ell\ell} &= (1.58 \pm 0.37 ) \times 10^{-6} \; , \\
{\cal B}[>14.4]_{\ell\ell} &= (0.48 \pm 0.10 ) \times 10^{-6} \; ,
\end{align}
where we have averaged over the electron and muon modes as well. We assume that the size of relative error in our theoretical predictions is independent of the Wilson coefficients $C_{9,10}(\mu_0)$. Using the numerical formulae presented in appendix~\ref{sec:formulae} we present the existing 95\% C.L. bounds on $C_{9,10}^{\rm NP}$ in the left panel of figure~\ref{fig:bounds} where we show separately the constraints from the low- and high-$q^2$ branching ratio measurements.

In order to determine the constraints that can be achieved with 50 ab${}^{-1}$, we assume SM central values and adopt projected experimental sensitivities obtained by combining the estimates for the branching ratio uncertainties presented in refs.~\cite{Kou:2018nap, Hurth:Cracow} with the method adopted in ref.~\cite{Huber:2015sra} for ${\cal H}_T$ and ${\cal H}_L$. In table~\ref{table:future-errors} we present the projected statistical uncertainties we use. The total uncertainties are obtained by adding a $5.8\%$ ($3.9\%$) systematic error to all low-$q^2$ (high-$q^2$) observables.

The projected uncertainty on the ratio ${\cal R}(14.4)$ requires an estimate of the expected experimental error on the semileptonic $\bar{B}\to X_u\ell\nu$ branching ratio measured with $q^2_{\ell\nu}> 14.4 \;{\rm GeV}^2$. We assess the latter by rescaling the expected experimental error on the extraction of $V_{ub}^{\rm incl}$ (see table 59 of ref.~\cite{Kou:2018nap}) by an estimate of the fraction of the semileptonic spectrum for  $q^2_{\ell\nu} > 14.4 \;{\rm GeV}^2$ which we obtained by a sample spectrum presented in ref.~\cite{Gambino:2007rp}. As a rough estimate of this projected uncertainty we find $[\delta {\cal R}(14.4)]^{\rm exp}_{50\; {\rm ab}^{-1}} = 7.3\%$.

\begin{figure}
\begin{center}
\includegraphics[width=0.49\linewidth]{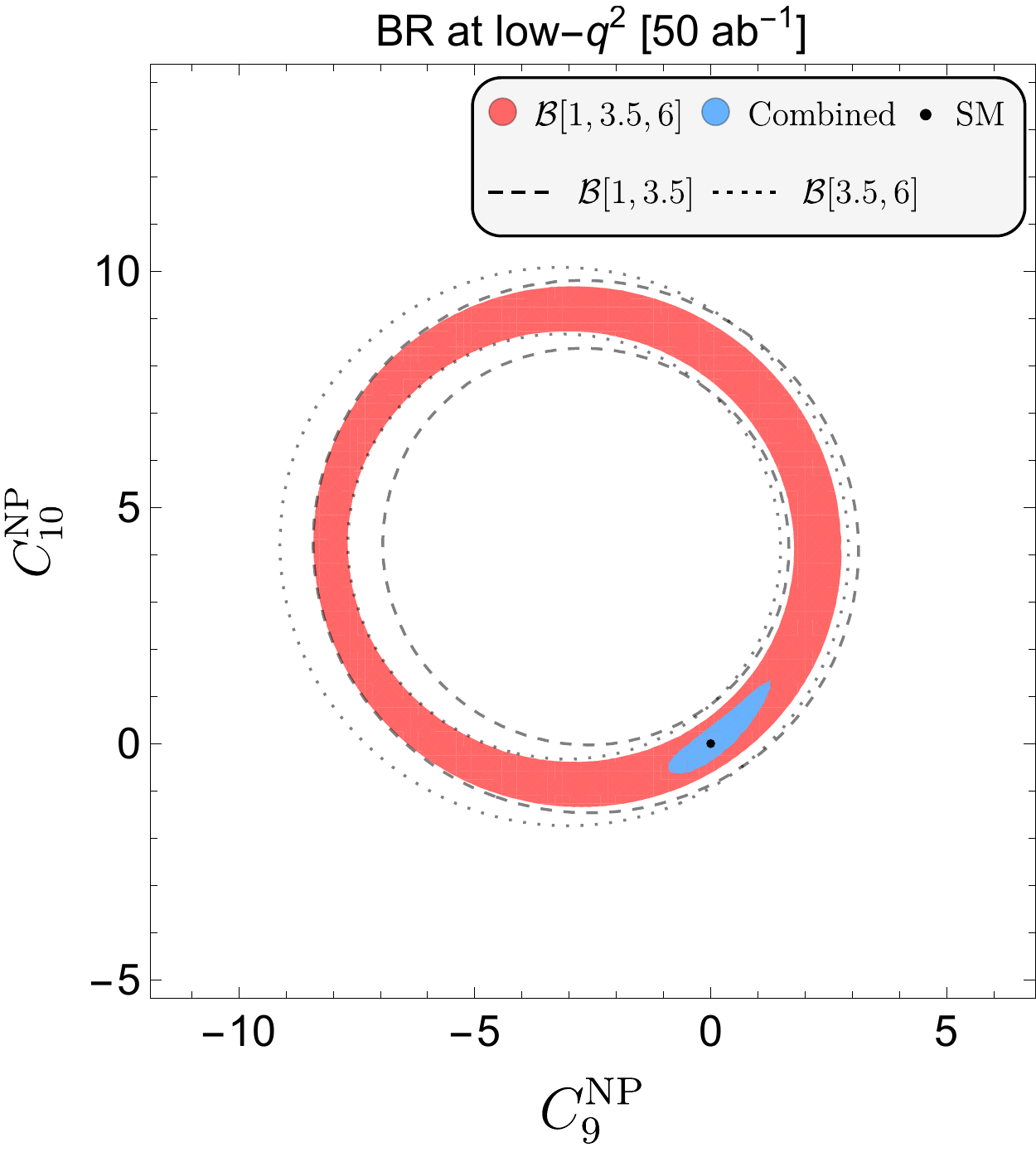}
\includegraphics[width=0.49\linewidth]{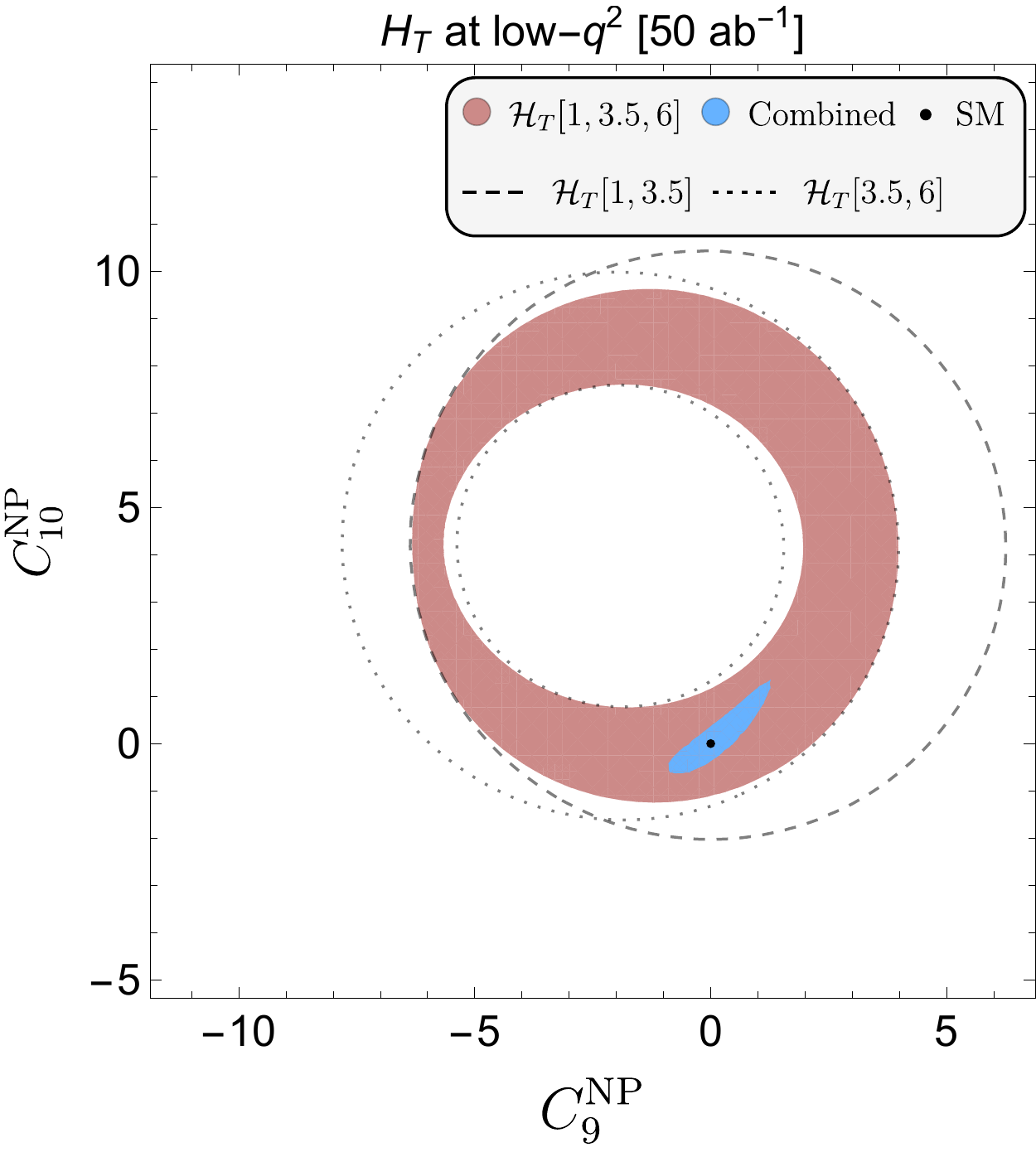}
\includegraphics[width=0.49\linewidth]{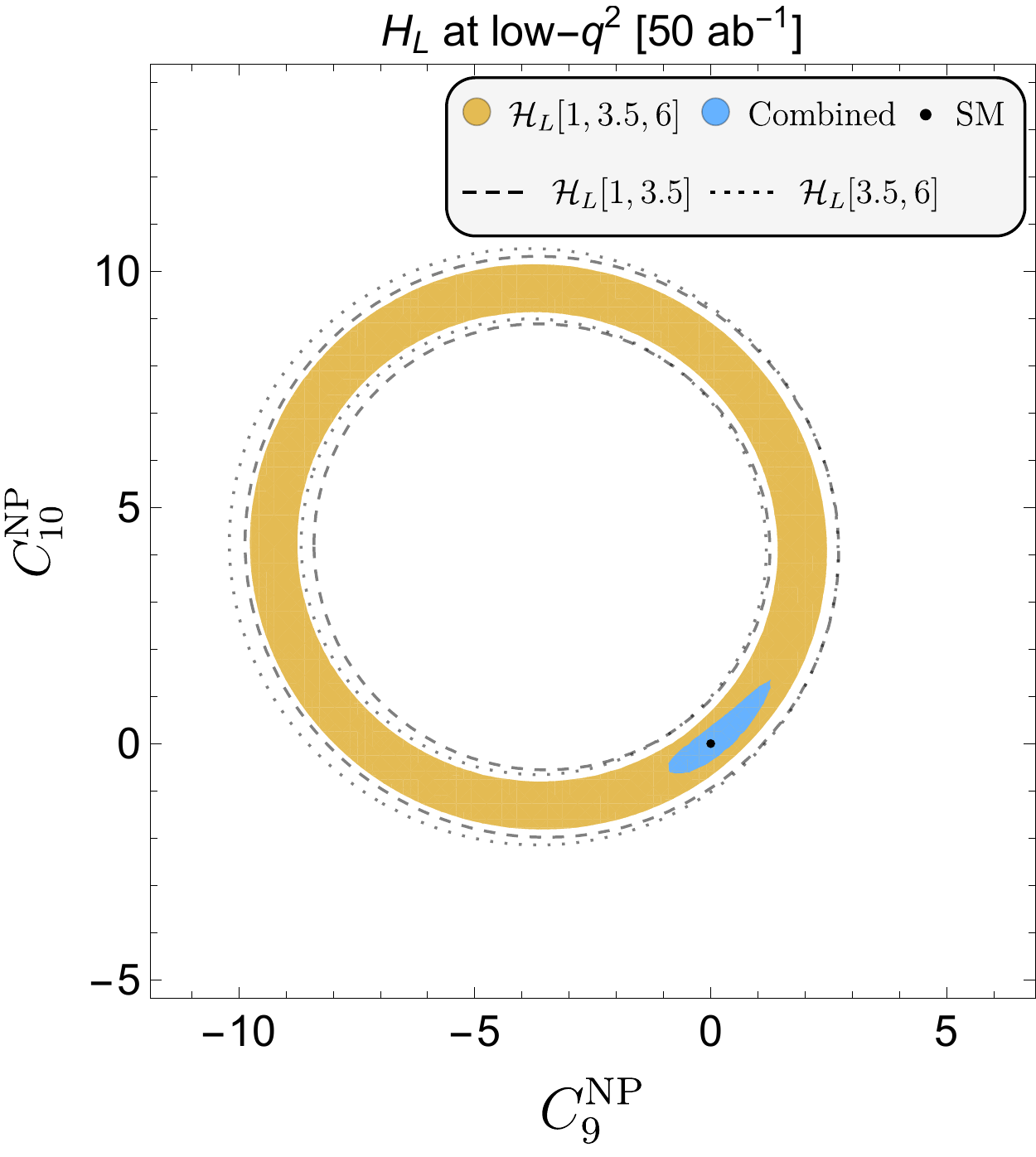}
\includegraphics[width=0.49\linewidth]{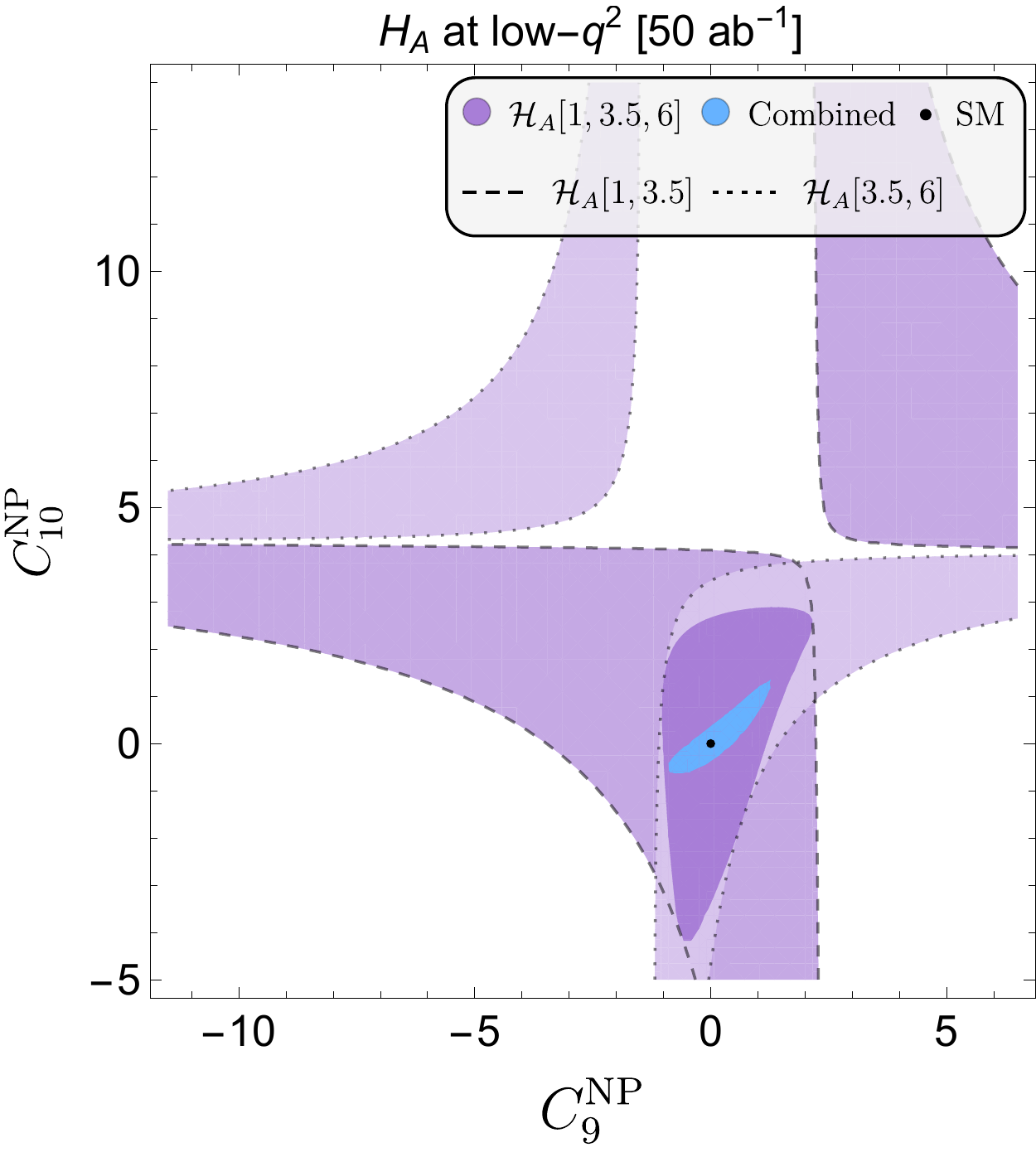}
\caption{Bounds from individual measurements of low-$q^2$ branching ratio, ${\cal H}_T$, ${\cal H}_L$ and ${\cal H}_A$. In each case we show the constraints from the two low-$q^2$ bins and from their combination. See the caption in figure~\ref{fig:bounds} for further details.}
\label{fig:bounds_breakdown_low}
\end{center}
\end{figure}

\begin{figure}
\begin{center}
\includegraphics[width=0.49\linewidth]{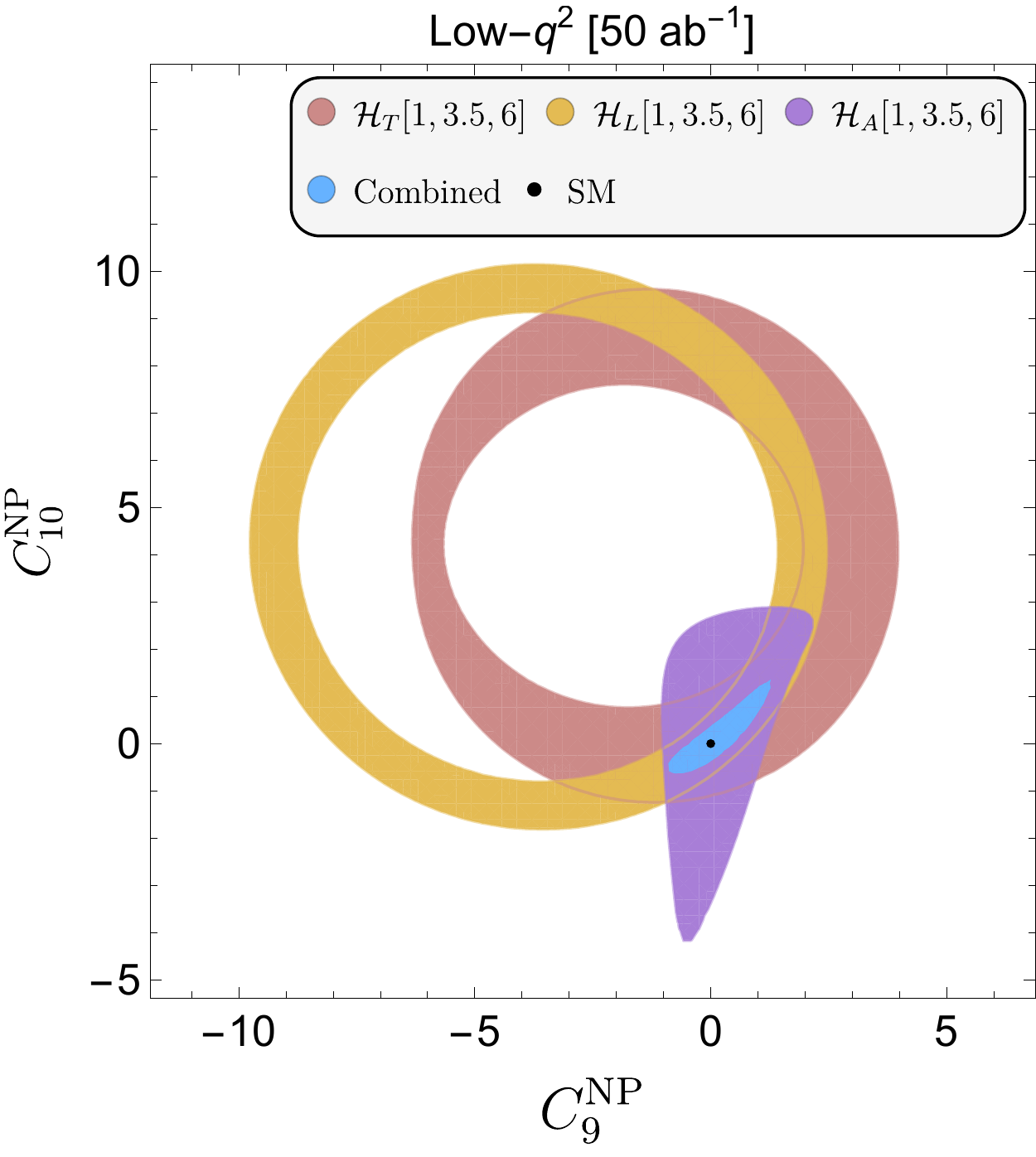}
\includegraphics[width=0.49\linewidth]{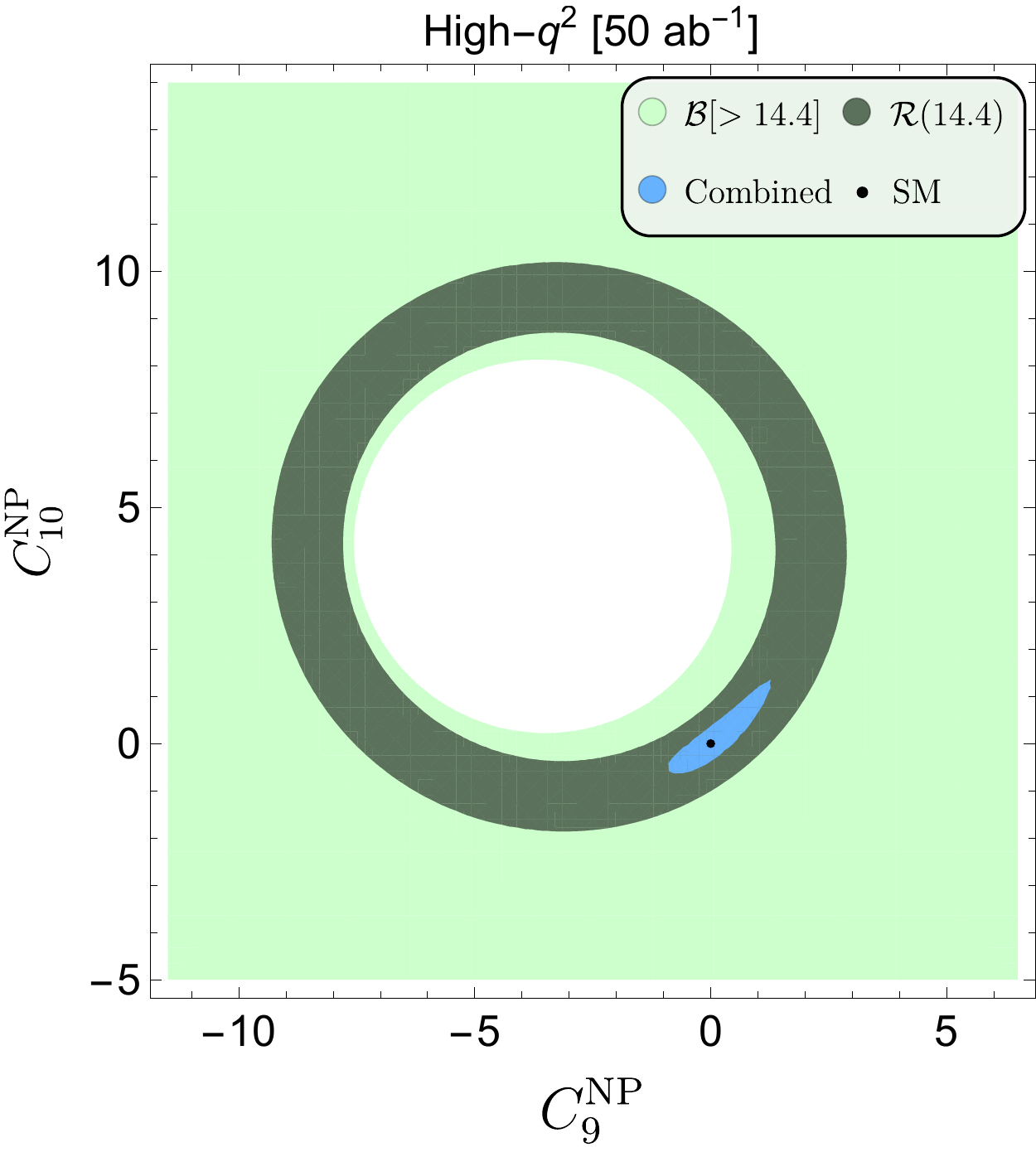}
\caption{Breakdown of the projected constraints at low- and high-$q^2$. See the caption in figure~\ref{fig:bounds} for further details.}
\label{fig:bounds_breakdown_lowVhigh}
\end{center}
\end{figure}

\begin{figure}
\begin{center}
\includegraphics[width=0.49\linewidth]{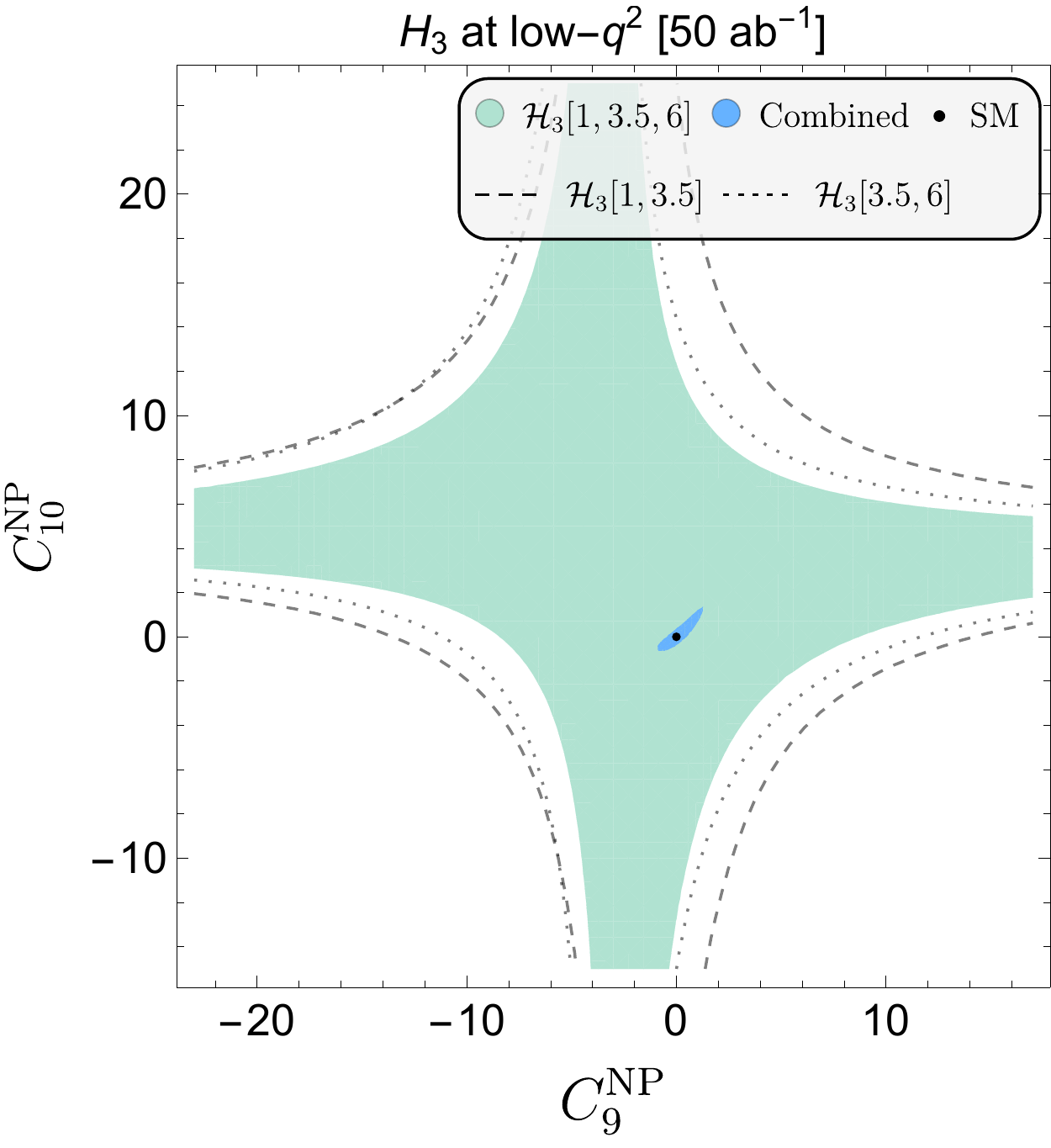}
\includegraphics[width=0.49\linewidth]{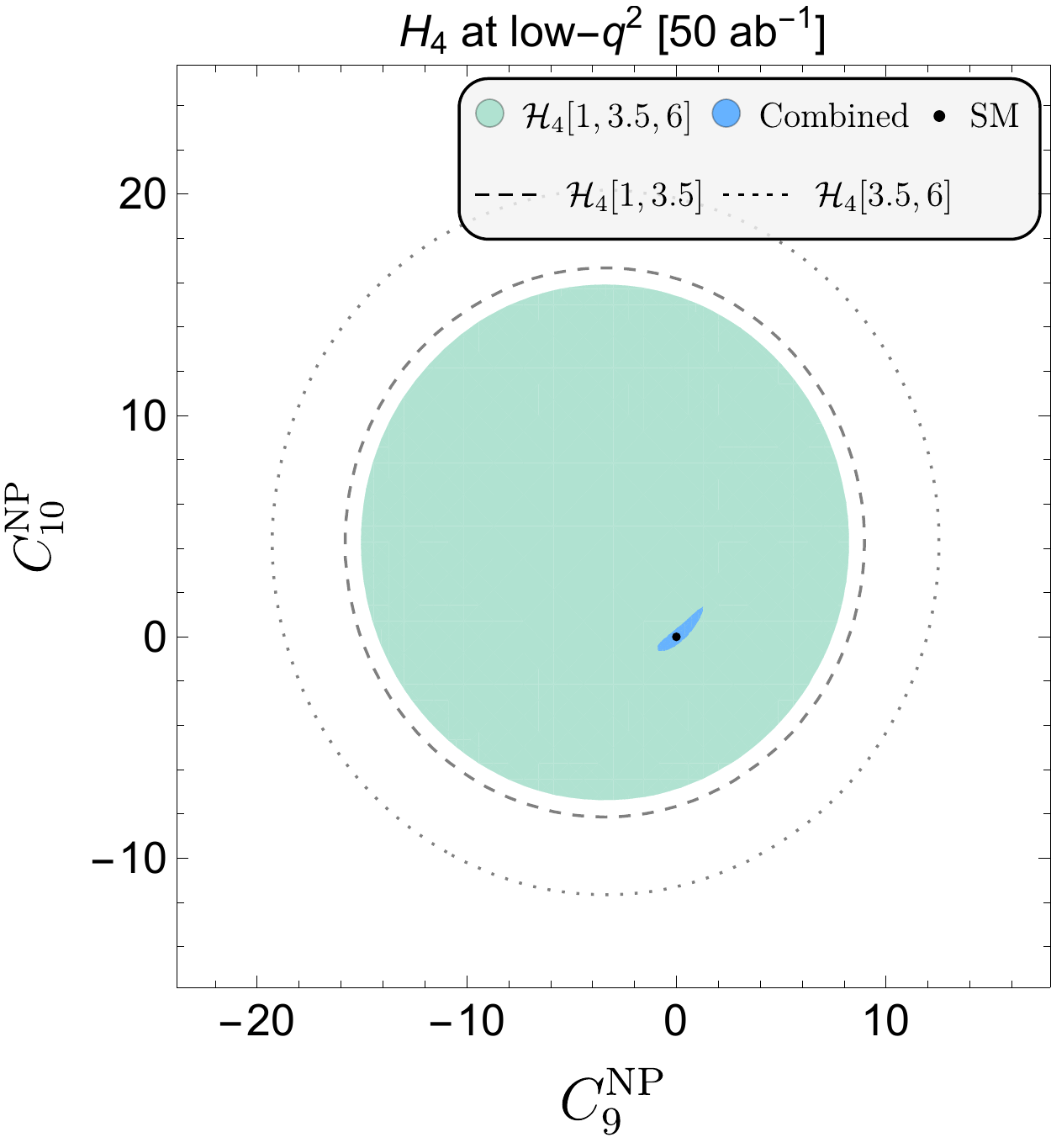}
\caption{Projected constraints from the QED observables ${\cal H}_3$ and ${\cal H}_4$ at low-$q^2$. See the caption in figure~\ref{fig:bounds} for further details.}
\label{fig:bounds_H34}
\end{center}
\end{figure}

The expected constraints obtained by considering separate measurements of ${\cal H}_{T,L,A}$ in the two low-$q^2$ bins, the high-$q^2$ branching ratio and the ratio ${\cal R}(14.4)$, are presented in  the right panel of figure~\ref{fig:bounds}. In figure~\ref{fig:bounds_breakdown_low} we show the breakdown of the low-$q^2$ constraints. In particular, we see that considering the two low-$q^2$ bins separately is mostly relevant for ${\cal H}_T$ and especially for ${\cal H}_A$. In the two panels of figure~\ref{fig:bounds_breakdown_lowVhigh} we show the relative contribution of low- and high-$q^2$ observables to the bounds expected. At high-$q^2$ it is imperative to consider the ratio ${\cal R}(14.4)$ in order to reduce exposure to large power corrections which stem from the breakdown of the OPE at the end-point of the spectrum. From
the SM results in eqs.~(\ref{eq:RZO1}),~(\ref{eq:RZO2}) we see that a large fraction of the uncertainty on ${\cal R}(14.4)$ is due to the direct determination of $|V_{ub}|$. In figure~\ref{fig:bounds_H34} we show the constraints from the QED observables ${\cal H}_{3,4}$. 

\begin{table}
\begin{center}
\begin{tabular}{|c|c c c c|}
\hline
 & $[1,3.5]$  & $[3.5, 6]$ & $[1,6]$ & $> 14.4$ \\ \hline
$\cal B$ & 3.1 \% & 2.6 \% & 2.0 \% & 2.6\%\\
${\cal H}_T$ & 24 \%  & 15 \%  & 13 \%  & - \\
${\cal H}_L$ & 5.5 \% & 5.0 \% & 3.7 \% & - \\
${\cal H}_A$ & 40 \%  & 33 \%  & -   \% & - \\
${\cal H}_3$ & 240 \% & 140 \% & 120 \% & - \\
${\cal H}_4$ & 140 \% & 270 \% & 120 \% & - \\
\hline
\end{tabular}
\caption{Projected statistical uncertainties that we expect at Belle~II with $50\; {\rm ab}^{-1}$ of integrated luminosity. The first row gives the considered $q^2$ bin in ${\rm GeV}^2$. The total projected error is obtained by adding a $5.8 (3.9) \%$ systematic uncertainty to all low-$q^2$ (high-$q^2$) observables. } 
\label{table:future-errors}
\end{center}
\end{table}

\begin{figure}
\begin{center}
\includegraphics[width=0.49\linewidth]{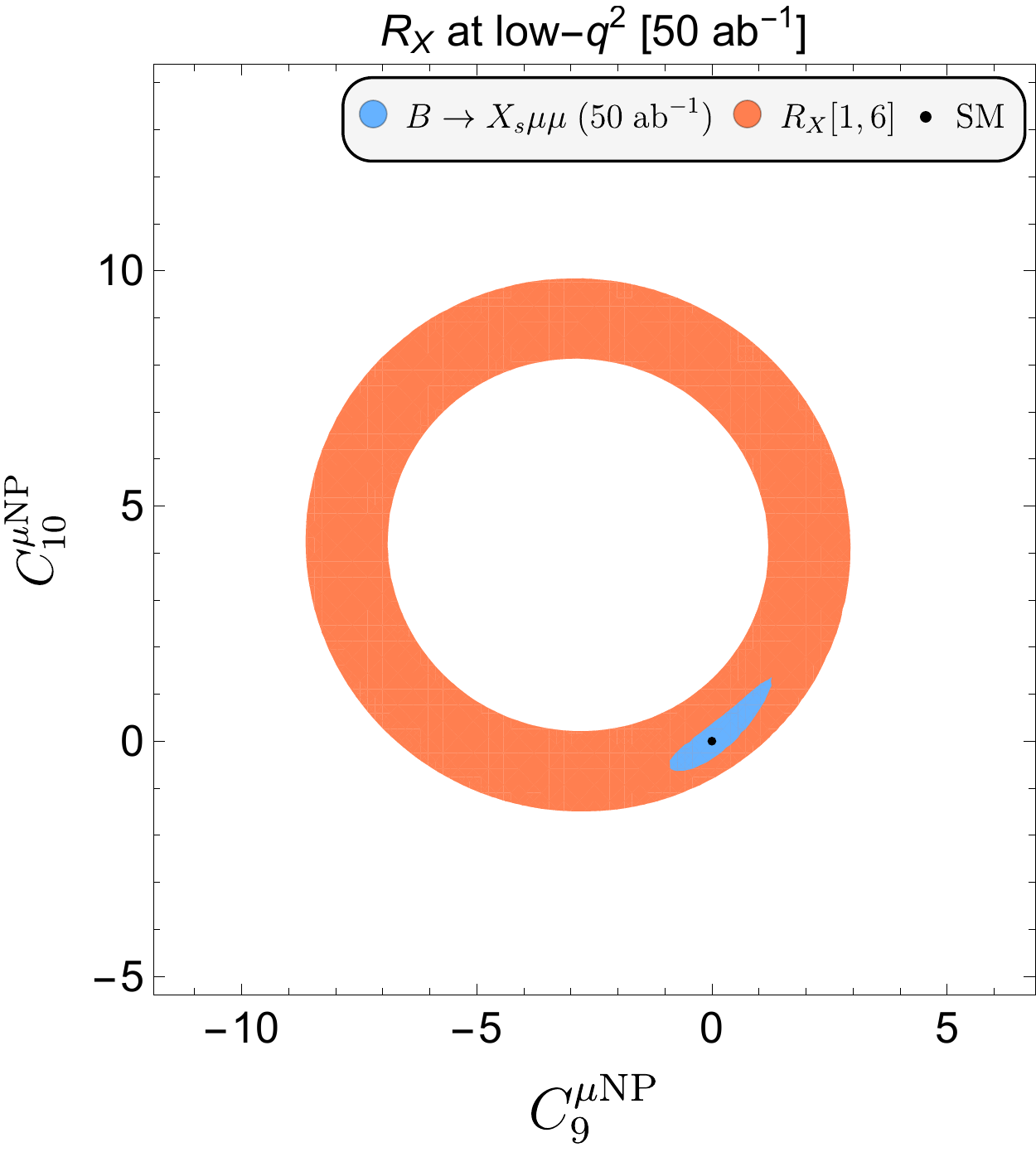}
\caption{The expected 50 ab${}^{-1}$ bounds under the assumption $C_{9,10}^{e} = C_{9,10}^{e{\rm SM}}$. The combined contour is slightly larger than the one in figure~\ref{fig:bounds} because it corresponds to the muon channel only. }
\label{fig:bounds_RX}
\end{center}
\end{figure}
\begin{figure}
\begin{center}
\includegraphics[width=0.49\linewidth]{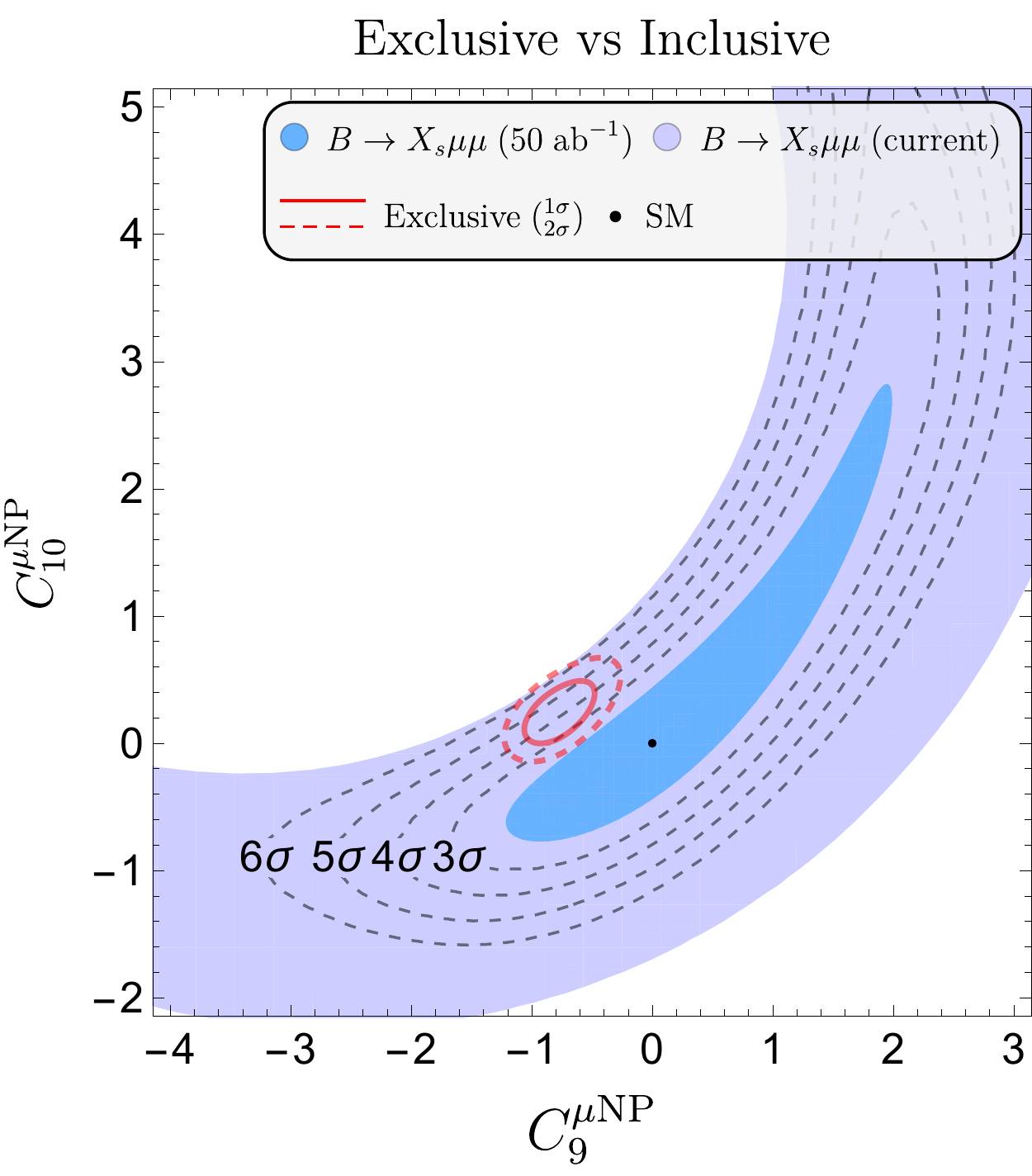}
\includegraphics[width=0.49\linewidth]{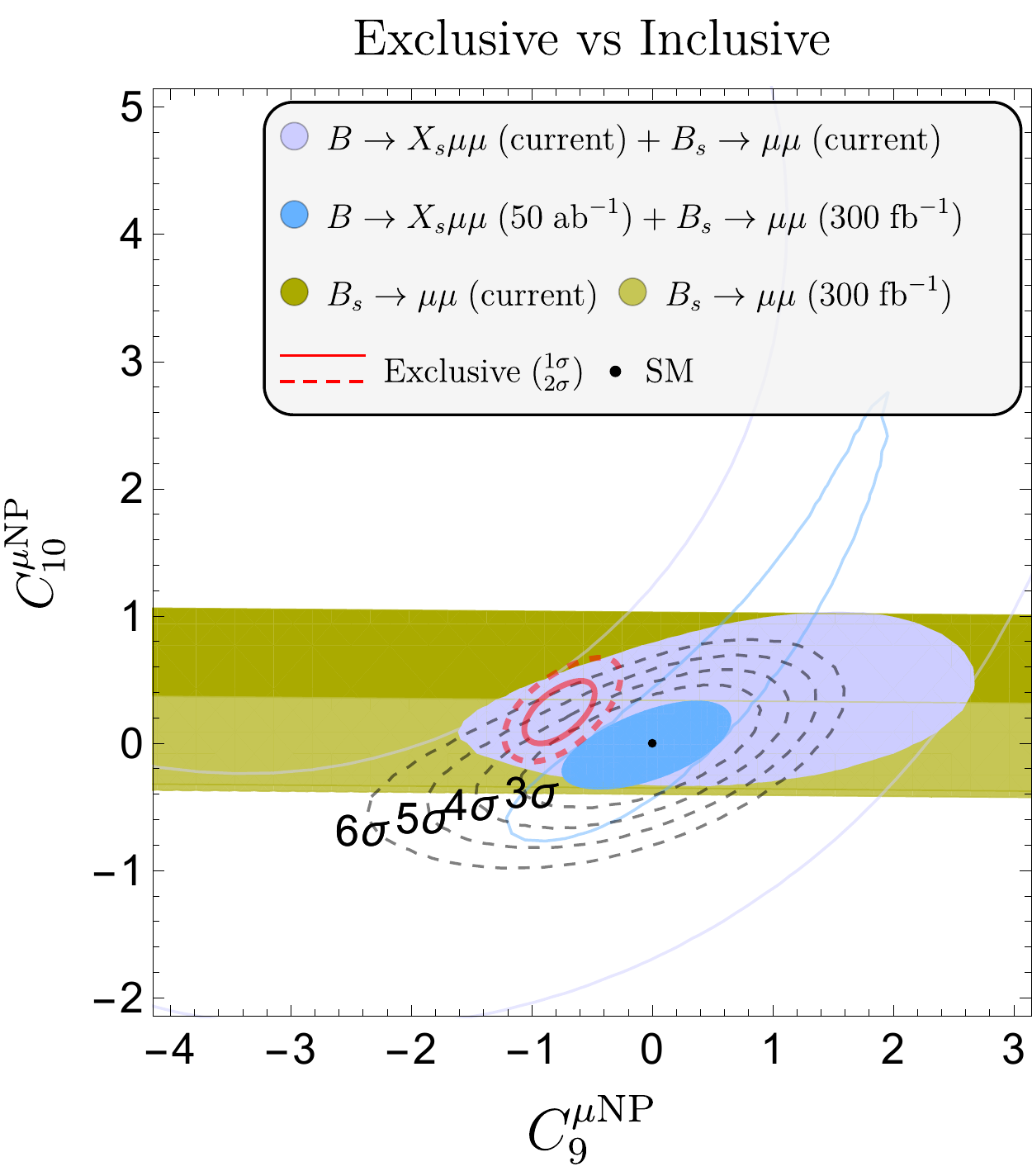}
\caption{Comparison between the expected bounds from inclusive measurements (up to six standard deviations) with the bounds from existing exclusive $b\to s\ell^+\ell^-$ measurements. The latter are derived using the Flavio~\cite{Straub:2018kue} and Smelli~\cite{Aebischer:2018iyb} packages and include constraints from branching ratio and asymmetries in $b\to s\mu^+\mu^-$ as well as from the lepton flavour universality violating ratios $R_{K^{(*)}}$. In the right panel we combine inclusive $\bar{B}\to X_s\mu^+\mu^-$ with the current determination of ${\cal B} (\bar{B}_s\to\mu^+\mu^-)$.}
\label{fig:bounds_excl_incl}
\end{center}
\end{figure}

\subsection{Interplay between inclusive and exclusive decays}
\label{sec:ratioRXSNP}
In this subsection we discuss the interplay between the experimental projections we discussed above and the existing anomalies in exclusive modes. Since some of the latter (such as $P_5^\prime$) are specific to the di-muon final state, and since modifying only the muonic Wilson coefficients can already accommodate the data, we present bounds in the $[C_9^{\mu {\rm  NP}},C_{10}^{\mu {\rm  NP}}]$ plane, assuming there are no new physics contributions to the coefficients $C_{9,10}^{e}$.

We begin by recalculating the expected constraints for the $\bar{B}\to X_s \mu^+\mu^-$ channel only (i.e.~the projected statistical experimental uncertainties increase by $\sqrt{2}$ because we loose the di-electron final state). The resulting projected Belle II reach is displayed in figure~\ref{fig:bounds_RX}, where we  also include the expected constraints from measurements of the ratio $R_{X_s}$ (which is essentially free of theoretical uncertainties, see the SM predictions given in section~\ref{sec:ratioRXs}). The constraints from $R_{X_s}$ are weaker than those from ${\cal H}_{T,L}$ mainly because of the much larger experimental statistical uncertainty: the ratio of the di-muon rate to the di-electron one has an expected statistical uncertainty which is twice as large than that for the combined electron and muon channel. Nevertheless, the absence of theoretical uncertainties makes this observable very interesting. 

In the left panel of figure~\ref{fig:bounds_excl_incl} we compare the expected constraints from inclusive di-muon modes with the existing bounds from exclusive $b\to s \mu^+ \mu^-$ observables. The exclusive contour has been calculated with the packages Flavio~\cite{Straub:2018kue} and Smelli~\cite{Aebischer:2018iyb} using the default likelihood but without the inclusion of $\bar{B}_s \to \mu^+\mu^-$. We see that if $C_{9,10}^{\mu{\rm NP}} = 0$, Belle II  results of the inclusive observables will exclude the current best-fit point of the exclusive  fits by slighly more than $4\sigma$. Moreover, we checked in a separate study that if the true values of $C_{9,10}^{\mu{\rm NP}}$ are at the current best fit point of the fit to the exclusive data, the SM point would be excluded with a similar significance.

In the right panel of figure~\ref{fig:bounds_excl_incl} we show the impact of $B_s \to \mu^+\mu^-$, which is essentially only dependent on the coefficient $C_{10}^{\mu}$. We choose to include the constraint from this purely leptonic decay in the inclusive semileptonic expected reach because both modes are considerably cleaner than the various exclusive semileptonic observables. The currently allowed region is obtained  by including the PDG average ${\cal B} (B_s \to \mu^+\mu^-) = (3.0 \pm 0.4) \times 10^{-9}$ and the theoretical description outlined in ref.~\cite{Bobeth:2013uxa}. The projected contour is obtained by assuming a $B_s \to \mu^+\mu^-$ measurement centered on the SM expectation~\cite{Bobeth:2013uxa} with an uncertainty corresponding to 300 fb${}^{-1}$ of LHCb data (which is the High-Luminosity LHC scenario considered in ref.~\cite{Cerri:2018ypt}). After including $B_s \to\mu^+\mu^-$, the reach in the $[C_9^{\mu{\rm NP}},C_{10}^{\mu{\rm NP}}]$ plane improves even further and the current exclusive best-fit point could be excluded with a significance close to $5\sigma$ if $C_{9,10}^{\mu{\rm NP}}=0$.

\subsection{Interplay with $b\to s\tau^+\tau^-$}

The $b\to s\tau^+\tau^-$ decays, both the exclusive and inclusive modes, are very challenging to measure in  experiments. The current experimental bounds on the decay rates are still far away from the corresponding SM
expectations \cite{TheBaBar:2016xwe,Aaij:2017xqt}. Alternatively, the $\tau^+\tau^-$ final state can be indirectly constrained by using the exclusive decay $B^+\to K^+\mu^+\mu^-$, which receives contributions from the $\tau^+\tau^-$ state via re-scattering~\cite{Cornella:2020aoq}.

Similar re-scattering also occurs in the inclusive channel, therefore $\bar{B}\to X_s \ell^+\ell^-$ measurements can be used to constrain the $b\to s\tau^+\tau^-$
amplitude. Defining, $C_9^\tau(\mu_0) \equiv R_9^\tau C_9^\text{SM}(\mu_0)$, we find 
\begin{align}
\mathcal{B}[1,6] =&\ \mathcal{B}_\text{SM}[1,6] - \left[ 2.9\times10^{-3}\; \mathcal{R}\left(R_9^\tau -1 \right) + 4.8\times10^{-5}\; \mathcal{I}\left(R_9^\tau \right) \right] \times 10^{-7} \; ,  \nnb \\
\mathcal{B}[1,3.5]  = &\ \mathcal{B}_\text{SM}[1,3.5] - \left[ 8.5\times10^{-4}\; \mathcal{R}\left(R_9^\tau -1 \right) + 1.8\times10^{-5}\; \mathcal{I}\left(R_9^\tau \right)\right] \times10^{-7} ,  \nnb\\
\mathcal{B}[3.5, 6] = &\ \mathcal{B}_\text{SM}[3.5, 6] - \left[ 2.1\times10^{-3}\; \mathcal{R}\left(R_9^\tau -1 \right) + 2.9\times10^{-5}\; \mathcal{I}\left(R_9^\tau \right)\right] \times10^{-7} ,  \nnb\\
\mathcal{B}[>14.4] = &\ \mathcal{B}_\text{SM}[>14.4] + \left[ 3.8\times10^{-3}\; \mathcal{R}\left(R_9^\tau -1 \right) - 3.6\times10^{-3}\; \mathcal{I}\left(R_9^\tau \right)\right] \times 10^{-7}.
\end{align}
We observe that the high-$q^2$ branching ratio $\mathcal{B}[>14.4]$ is most sensitive to $C_9^\tau$. For the sake of simplicity we assume that $C_9^\tau$ is real.  Assuming a projected uncertainty of 4.7\% on  $\mathcal{B}[>14.4]$ at Belle~II (see table~\ref{table:future-errors})~\cite{Kou:2018nap} leads to
\begin{align}
R_9^\tau \approx 0 \pm 230 \; , \qquad
\mathcal{B}(B^+ \to K^+ \tau^+\tau^-) \lsim \left\{ \begin{tabular}{ll} $2.5\times 10^{-3}$\;, & if $C_{10}^\tau = 0$\;, \\
 $8.1\times 10^{-3}$\;, & if $C_{10}^\tau = C_9^\tau$\;. \\ \end{tabular} \right.
\end{align}
This result is competitive to the current direct bound given by BaBar, $\mathcal{B}(B^+ \to K^+ \tau^+\tau^-)
< 2.25 \times 10^{-3}$ at 90\% CL \cite{TheBaBar:2016xwe}. Similar sensitivity can be obtained by considering 
$\mathcal{R}(14.4)$ which has a slightly larger projected experimental uncertainty  $[\delta {\cal R}(14.4)]^{\rm exp}_{50\; {\rm ab}^{-1}} = 7.3\%$ (as discussed in the previous section)
but a much smaller theoretical uncertainty than $\mathcal{B}[>14.4]$. We find 
\begin{align}
R_9^\tau \approx 0 \pm 66 \; , \qquad
\mathcal{B}(B^+ \to K^+ \tau^+\tau^-) \lsim \left\{ \begin{tabular}{ll} $2.1\times 10^{-4}$\;, & if $C_{10}^\tau = 0$\;, \\
 $6.7\times 10^{-4}$\;, & if $C_{10}^\tau = C_9^\tau$\;. \\ \end{tabular} \right.
\end{align}
This indirect constraint from the Belle II measurement of $\bar{B} \to X_s \mu^+\mu^-$ is comparable 
with the direct $B^+ \to K^+ \tau^+\tau^-$ measurement with the LHCb upgrade-II luminosity \cite{Cornella:2020aoq}.

%%%%%%%%%%%%%%%%%%%%%%%%%%%%%%%%%%%%%%%%%%%%%%%%%%%%%%%%%%%%%%%%%%%%%%%%%%%%%%%%%%%%%%%%%%%%%%%%%%%%%%%%%%%%%%%%%%%%%%%%%%%%%%%%%%%%%%%%%%%%%%%%%%%%%%%%%%%%%%%%%%%%%%
%%%%%%%%%%%%%%%%%%%%%%%%%%%%%%%%%%%%%%%%%%%%%%%%%%%%%%%%%%%%%%%%%%%%%%%%%%%%%%%%%%%%%%%%%%%%%%%%%%%%%%%%%%%%%%%%%%%%%%%%%%%%%%%%%%%%%%%%%%%%%%%%%%%%%%%%%%%%%%%%%%%%%%

\section{Conclusion}
\label{sec:conclusion}

In the absence of direct signals for physics beyond the SM, FCNC decays play a crucial role in searching for imprints of new physics in low-energy processes. With the experimental programs at LHCb, Belle~II and other experiments in operation, we are entering a new era of precision measurements of rare $B$ decays. One of the prime measurements which is expected to become available for the first time at Belle~II is a full angular analysis of inclusive $\bar{B}\to X_s \ell^+\ell^-$. This analysis is interesting on its own grounds, but also offers a unique opportunity to study the interplay with its exclusive $b\to s\ell^+\ell^-$ counterparts. In order to pave to road for precision phenomenology and extensive new-physics studies, a theoretical update of inclusive $\bar{B}\to X_s \ell^+\ell^-$ is mandatory.

In this paper we  therefore presented a comprehensive update of the SM theory predictions for the entire set of inclusive $\bar{B}\to X_s \ell^+\ell^-$ observables. As new observables we present predictions for the ratio $R_{X_s}$ (and 
similarly  for the angular parts). These are ratios of the inclusive $\bar{B}\to X_s \mu^+\mu^-$ versus $\bar{B}\to X_s e^+e^-$ transitions sensitive to lepton-flavour universality, in analogy to the exclusive ratios $R_{K^{(*)}}$.  Other main novelties in our analysis are updated input parameters, the implementation of the new and more sophisticated treatment of non-perturbative effects via the Kr\"uger-Sehgal mechanism~\cite{Huber:2019iqf}, and the inclusion of non-local power corrections via the resolved contributions~\cite{Hurth:2017xzf,Benzke:2017woq,Benzke:2020htm}. Along the lines of~\cite{Huber:2019iqf} we also implement the results of the updated study of power-suppressed effects in the high-$q^2$ region. Depending on the observable and the $q^2$-range, this leads to central values which differ by several percent from those in our previous analysis~\cite{Huber:2015sra}. For example, the low-$q^2$ integrated branching ratio for muons in units of $10^{-6}$ moves from $1.62 \pm 0.09$ to $1.73 \pm 0.13$, where the increase in uncertainty can be almost entirely attributed to the additional $5\%$ that we add to take into account the resolved contributions.

In addition, we investigated the effect of collinear photons in a detailed Monte Carlo study and gave a prescription for how to deal with these effects at Belle~II. An effect which has not been included in previous analysis is the bin migration from the charmonium resonances into the perturbative low-$q^2$ window. Table~\ref{tab:correction} contains a complete list of correction factors that have to be applied to compare our predictions for the electron channel (in which we always adopt the defintion $q^2 = (p_{e^+} + p_{e^-})^2$) to the Belle~II analysis which applies angular and energy cuts on collinear photons.

Finally, we presented an elaborate discussion on the new physics potential of inclusive $\bar{B}\to X_s \ell^+\ell^-$. First, we studied the bounds from current measurements, which are still rather loose. However, the projection to the final Belle~II data set and the inclusion of  all angular observables reveal that the inclusive channel has already power enough on its own
to tightly constrain $C_9^{\rm NP}$ and $C_{10}^{\rm NP}$. In combination with exclusive $b \to s \ell^+ \ell^-$ decays and the rare $\bar{B}_s \to \mu^+\mu^-$ decay, the full power of the synergy between inclusive and exclusive FCNC transitions becomes manifest. 
Should the true value of $C_9^{\rm NP}$ and $C_{10}^{\rm NP}$ be at either the SM point $C_9^{\rm NP} = C_{10}^{\rm NP} = 0$ or the current best-fit point of the exclusive fits, an analysis of inclusive $\bar{B}\to X_s\ell^+\ell^-$ at Belle~II with $50$ab$^{-1}$ of data will prefer that one with respect to the other one at the level of $\sim 5 \sigma$.
This again underlines the necessity of a full angular analysis of $\bar{B}\to X_s\ell^+\ell^-$ at Belle~II.

A point we addressed only marginally in the present article is that of a cut on the hadronic invariant mass $M_X$. While there is hope that a fully inclusive measurement using the recoil technique will become feasible towards the end of Belle~II, such a cut will remain necessary for a good portion of the Belle~II operation time. Despite the fact that there exists preliminary work on this topic~\cite{Lee:2005pk,Lee:2005pwa,Lee:2008xc}, better knowledge of sub-leading shape functions will certainly be required for more precise predictions. As for now, only the zero crossing of the forward backward asymmetry has been calculated in the presence of an $M_X$ cut~\cite{Bell:2010mg}. A study on the effect of a hadronic mass cut on the other observables will also build on~\cite{Hurth:2017xzf,Benzke:2017woq,Benzke:2020htm} and is left for future work.

%%%%%%%%%%%%%%%%%%%%%%%%%%%%%%%%%%%%%%%%%%%%%%%%%%%%%%%%%%%%%%%%%%%%%%%%%%%%%%%%%%%%%%%%%%%%%%%%%%%%%%%%%%%%%%%%%%%%%%%%%%%%%%%%%%%%%%%%%%%%%%%%%%%%%%%%%%%%%%%%%%%%%%
%%%%%%%%%%%%%%%%%%%%%%%%%%%%%%%%%%%%%%%%%%%%%%%%%%%%%%%%%%%%%%%%%%%%%%%%%%%%%%%%%%%%%%%%%%%%%%%%%%%%%%%%%%%%%%%%%%%%%%%%%%%%%%%%%%%%%%%%%%%%%%%%%%%%%%%%%%%%%%%%%%%%%%

\subsubsection*{Acknowledgements}

We would like to thank Akimasa Ishikawa,  Kevin Flood, Peter Stangl, Yo Sato and Wolfgang Altmannshofer for useful discussions and correspondence. The work of JJ and EL was supported in part by the U.S.\ Department of Energy under grant number DE-SC0010120. The work of T.~Huber was supported in part by the Deutsche Forschungsgemeinschaft (DFG, German Research Foundation) under grant  396021762 - TRR 257 ``Particle Physics Phenomenology after the Higgs Discovery''. The work of T.~Hurth was supported by the Cluster of Excellence `Precision Physics, Fundamental Interactions, and Structure of Matter' (PRISMA+ EXC 2118/1) funded by the German Research Foundation (DFG) within the German Excellence Strategy (Project ID 39083149), as well as by the BMBF Verbundprojekt 05H2018 - Belle II: Indirekte Suche nach neuer Physik bei Belle II. He also thanks the CERN theory group for its hospitality during his regular visits to CERN where part of this work was written. The work of KKV is supported by the DFG Sonderforschungsbereich/Transregio 110 Symmetries and the Emergence of Structure in QCD.

%%%%%%%%%%%%%%%%%%%%%%%%%%%%%%%%%%%%%%%%%%%%%%%%%%%%%%%%%%%%%%%%%%%%%%%%%%%%%%%%%%%%%%%%%%%%%%%%%%%%%%%%%%%%%%%%%%%%%%%%%%%%%%%%%%%%%%%%%%%%%%%%%%%%%%%%%%%%%%%%%%%%%%
%%%%%%%%%%%%%%%%%%%%%%%%%%%%%%%%%%%%%%%%%%%%%%%%%%%%%%%%%%%%%%%%%%%%%%%%%%%%%%%%%%%%%%%%%%%%%%%%%%%%%%%%%%%%%%%%%%%%%%%%%%%%%%%%%%%%%%%%%%%%%%%%%%%%%%%%%%%%%%%%%%%%%%

\begin{appendix}

%%%%%%%%%%%%%%%%%%%%%%%%%%%%%%%%%%%%%%%%%%%%%%%%%%%%%%%%%%%%%%%%%%%%%%%%%%%%%%%%%%%%%%%%%%%%%%%%%%%%%%%%%%%%%%%%%%%%%%%%%%%%%%%%%%%%%%%%%%%%%%%%%%%%%%%%%%%%%%%%%%%%%%
%%%%%%%%%%%%%%%%%%%%%%%%%%%%%%%%%%%%%%%%%%%%%%%%%%%%%%%%%%%%%%%%%%%%%%%%%%%%%%%%%%%%%%%%%%%%%%%%%%%%%%%%%%%%%%%%%%%%%%%%%%%%%%%%%%%%%%%%%%%%%%%%%%%%%%%%%%%%%%%%%%%%%%
\section{Phenomenological results}
\label{sec:pheno}

 \renewcommand{\arraystretch}{1.18}
\begin{table}
	\begin{center}
		\begin{tabular}{|c|c|c|c|}
			\hline
			\rule{0pt}{14pt}$q^2 \rm{~range} \;[{\rm GeV}^2]$	& $[1,6]$ & {$[1,3.5]$} & {$[3.5,6]$} \\
			\hline
			\rule{0pt}{14pt}$\cal B$ & $16.98\times 10^{-7}$& $9.23\times 10^{-7}$& $7.75\times 10^{-7}$\\
			${\cal H}_T$ & $3.13\times 10^{-7}$& $1.48\times 10^{-7}$& $1.64\times 10^{-7}$\\
			${\cal H}_L$ & $13.77\times 10^{-7}$& $7.69\times 10^{-7}$& $6.08\times 10^{-7}$\\
			${\cal H}_A$ & $-0.27\times 10^{-7}$ & $-1.08\times 10^{-7}$ & $0.81\times 10^{-7}$ \\
			\hline \hline
			\rule{0pt}{14pt}$q^2 \rm{~range} \;[{\rm GeV}^2]$ & \multicolumn{3}{c|}{$>14.4$ }\\ \hline
			\rule{0pt}{14pt}$\cal B$ & \multicolumn{3}{c|}{2.59 $\times 10^{-7}$}\\
			$\mathcal{R}(s_0)$ 		& \multicolumn{3}{c|}{27.71 $\times 10^{-4}$}\\ \hline
		\end{tabular}
		\caption{Phenomenological results without electromagnetic effects, i.e.\ log-enhanced QED corrections to the matrix elements at the scale $\mu_b$ are not included. The smaller effect of QED corrections in the matching and running is, however, taken into accout.}
		\label{tab:noem}
	\end{center}
\end{table}
\renewcommand{\arraystretch}{1}
In this appendix, we give the numerical results for the low-$q^2$ observables ${\cal H}_A, {\cal H}_T, {\cal H}_L, {\cal H}_3$ and ${\cal H}_4$ which we relegated from section~\ref{sec:results}. In table~\ref{tab:noem}, we list all observables without electromagnetic effects to also account for the case that electromagnetic radiation is taken care of entirely on the experimental side.

\subsection{${\cal H}_A$}
\begin{align}
\dps {\cal H}_\text{A}[1,3.5]_{ee} =& \ (-0.95\pm 0.05_{\text{scale}}\pm 0.008_{m_t}\pm 0.006_{C,m_c}\pm 0.02_{m_b} \pm 0.02_{\alpha _s} \pm 0.001_{\text{CKM}}
\nonumber \\
& \hspace*{14pt}\pm 0.01_{\text{BR}_{\text{sl}}}\pm 0.0002_{\lambda _2}\pm 0.05_{\text{resolved}})\cdot 10^{-7}=( -0.95\pm 0.08)\cdot 10^{-7}  \; , \nonumber\\[0.5em]
\dps {\cal H}_\text{A}[3.5,6]_{ee} =& \ (0.91\pm 0.13_{\text{scale}}\pm 0.009_{m_t}\pm 0.04_{C,m_c}\pm 0.05_{m_b}  \pm 0.03_{\alpha _s} \pm 0.001_{\text{CKM}}
\nonumber \\
& \hspace*{14pt}\pm 0.01_{\text{BR}_{\text{sl}}}\pm 0.005_{\lambda _2}\pm 0.05_{\text{resolved}})\cdot 10^{-7}=( 0.91\pm 0.16)\cdot 10^{-7}  \; , \nonumber\\[0.5em]
\dps {\cal H}_\text{A}[1,6]_{ee} =& \ (-0.04\pm 0.19_{\text{scale}}\pm 0.0004_{m_t}\pm 0.03_{C,m_c}\pm 0.07_{m_b} \pm 0.05_{\alpha _s} \pm 0.00004_{\text{CKM}}
\nonumber \\
& \hspace*{14pt}\pm 0.0006_{\text{BR}_{\text{sl}}}\pm 0.005_{\lambda _2}\pm 0.002_{\text{resolved}})\cdot 10^{-7}=( -0.04\pm 0.21)\cdot 10^{-7}  \; .
\end{align}
\begin{align}
\dps {\cal H}_\text{A}[1,3.5]_{\mu\mu} =& \ (-1.03\pm 0.05_{\text{scale}}\pm 0.009_{m_t}\pm 0.007_{C,m_c}\pm 0.02_{m_b}  \pm 0.02_{\alpha _s}
\pm 0.0009_{\text{CKM}}\nonumber \\
& \hspace*{14pt}\pm 0.02_{\text{BR}_{\text{sl}}}\pm 0.0006_{\lambda _2}\pm 0.05_{\text{resolved}})\cdot 10^{-7}=( -1.03\pm 0.08)\cdot 10^{-7}  \; , \nonumber\\[0.5em]
\dps {\cal H}_\text{A}[3.5,6]_{\mu\mu} =& \ (0.85\pm 0.13_{\text{scale}}\pm 0.008_{m_t}\pm 0.03_{C,m_c}\pm 0.05_{m_b} \pm 0.03_{\alpha _s}\pm 0.0008_{\text{CKM}}\nonumber \\
& \hspace*{14pt}\pm 0.01_{\text{BR}_{\text{sl}}}\pm 0.005_{\lambda _2}\pm 0.04_{\text{resolved}})\cdot 10^{-7}=( 0.85\pm 0.16)\cdot 10^{-7}  \; , \nonumber\\[0.5em]
\dps {\cal H}_\text{A}[1,6]_{\mu\mu} =& \ (-0.18\pm 0.19_{\text{scale}}\pm 0.0009_{m_t}\pm 0.03_{C,m_c}\pm 0.07_{m_b}  \pm 0.05_{\alpha _s}\pm 0.0002_{\text{CKM}}\nonumber \\
& \hspace*{14pt}\pm 0.003_{\text{BR}_{\text{sl}}}\pm 0.006_{\lambda _2}\pm 0.009_{\text{resolved}})\cdot 10^{-7}=( -0.18\pm 0.21)\cdot 10^{-7}  \; .
\end{align}
%

%%%%%%%%%%%%%%%%%%%%%%%%%%%%%%%%%%%%%%%%%%%%%%%%%%%%%%%%%%%%%%%%%%%%%%%%%%%%%%%%%%%%%%%%%%%%%%%%%%%%%%%%%%%%%%%%%%%%%%%%%%%%%%%%%%%%%%%%%%%%%%%%%%%%%%%%%%%%%%%%%%%%%%

\subsection{${\cal H}_T$ and   ${\cal H}_L$ }
\begin{align}
\dps {\cal H}_T[1,3.5]_{ee} =&\ ( 2.91 \pm 0.15_{\text{scale}} \pm 0.03_{m_t} \pm 0.05_{C,m_c} \pm 0.02_{m_b} \pm  0.005_{\alpha_s} \pm 0.003_{\text{CKM}} \nonumber \\
&\hspace*{0pt} \pm 0.04_{\text{BR}_{\text{sl}}}\pm 0.01_{\lambda _1}\pm 0.004_{\lambda _2}\pm 0.15_\text{resolved}) \cdot 10^{-7} = ( 2.91 \pm 0.22 ) \cdot 10^{-7} \; ,  \nonumber\\[0.5em]
\dps {\cal H}_T[3.5,6]_{ee} =&\ ( 2.51 \pm 0.18{\text{scale}} \pm 0.03_{m_t} \pm 0.06_{C,m_c} \pm 0.05_{m_b}  \pm  0.02_{\alpha_s} \pm 0.002_{\text{CKM}} \nonumber \\
&\hspace*{0pt} \pm 0.04_{\text{BR}_{\text{sl}}}\pm 0.02_{\lambda _1}\pm 0.003_{\lambda _2}\pm 0.13_\text{resolved}) \cdot 10^{-7} = ( 2.51 \pm 0.24 ) \cdot 10^{-7} \; , \nonumber\\[0.5em]
\dps {\cal H}_T[1,6]_{ee} =&\ ( 5.42 \pm 0.33_{\text{scale}} \pm 0.07_{m_t} \pm 0.11_{C,m_c} \pm 0.07_{m_b}  \pm    0.01_{\alpha_s} \pm 0.005_{\text{CKM}}\nonumber \\
&\hspace*{0pt}  \pm 0.08_{\text{BR}_{\text{sl}}}\pm 0.04_{\lambda _1}\pm 0.007_{\lambda _2}\pm 0.27_\text{resolved}) \cdot 10^{-7}  = ( 5.42 \pm 0.46  ) \cdot 10^{-7} \; .
\end{align}
\begin{align}
\dps {\cal H}_T[1,3.5]_{\mu\mu} =&\ ( 2.08 \pm 0.08_{\text{scale}} \pm 0.02_{m_t} \pm 0.03_{C,m_c} \pm 0.01_{m_b} \pm  0.009_{\alpha_s} \pm 0.002_{\text{CKM}}\nonumber \\
&\hspace*{0pt}  \pm 0.03_{\text{BR}_{\text{sl}}}\pm 0.01_{\lambda _1}\pm 0.0005_{\lambda _2}\pm 0.10_\text{resolved}) \cdot 10^{-7} = ( 2.08\pm 0.14 ) \cdot 10^{-7} \; ,  \nonumber\\[0.5em]
\dps {\cal H}_T[3.5,6]_{\mu\mu} =&\ ( 2.00 \pm 0.15_{\text{scale}} \pm 0.03_{m_t} \pm 0.05_{C,m_c} \pm 0.05_{m_b}\pm  0.01_{\alpha_s} \pm 0.002_{\text{CKM}} \nonumber \\
&\hspace*{0pt} \pm 0.03_{\text{BR}_{\text{sl}}}\pm 0.02_{\lambda _1}\pm 0.0007_{\lambda _2}\pm 0.10_\text{resolved}) \cdot 10^{-7} = ( 2.00 \pm 0.20 ) \cdot 10^{-7} \; , \nonumber\\[0.5em]
\dps {\cal H}_T[1,6]_{\mu\mu} =&\ ( 4.08 \pm 0.23_{\text{scale}} \pm 0.05_{m_t} \pm 0.08_{C,m_c} \pm 0.06_{m_b}  \pm    0.005_{\alpha_s} \pm 0.004_{\text{CKM}}\nonumber \\
&\hspace*{0pt}  \pm 0.06_{\text{BR}_{\text{sl}}}\pm 0.03_{\lambda _1}\pm 0.001_{\lambda _2}\pm 0.20_\text{resolved}) \cdot 10^{-7}  = ( 4.08 \pm 0.34  ) \cdot 10^{-7} \; .
\end{align}

\begin{align}
\dps {\cal H}_L[1,3.5]_{ee} =&\ ( 6.92 \pm 0.28_{\text{scale}} \pm 0.07_{m_t} \pm 0.16_{C,m_c} \pm 0.09_{m_b}  \pm  0.05_{\alpha_s} \pm 0.006_{\text{CKM}} \nonumber \\
&\hspace*{0pt} \pm 0.10_{\text{BR}_{\text{sl}}} \pm 0.01_{\lambda _1}\pm 0.06_{\lambda _2}\pm 0.35_\text{resolved} ) \cdot 10^{-7} = ( 6.92 \pm 0.50 ) \cdot 10^{-7} \; ,  \nonumber\\[0.5em]
\dps {\cal H}_L[3.5,6]_{ee} =&\ ( 5.43 \pm 0.29_{\text{scale}} \pm 0.06_{m_t} \pm 0.13_{C,m_c} \pm 0.04_{m_b} \pm  0.04_{\alpha_s} \pm 0.005_{\text{CKM}}  \nonumber \\
&\hspace*{0pt} \pm 0.08_{\text{BR}_{\text{sl}}} \pm 0.02_{\lambda _1}\pm 0.05_{\lambda _2}\pm 0.27_\text{resolved} ) \cdot 10^{-7} = ( 5.43 \pm 0.44 ) \cdot 10^{-7} \; , \nonumber\\[0.5em]
\dps {\cal H}_L[1,6]_{ee} =&\ ( 12.35 \pm 0.53_{\text{scale}} \pm 0.13_{m_t} \pm 0.29_{C,m_c} \pm 0.14_{m_b}  \pm    0.09_{\alpha_s} \pm 0.01_{\text{CKM}} \nonumber \\
&\hspace*{0pt} \pm 0.19_{\text{BR}_{\text{sl}}} \pm 0.03_{\lambda _1}\pm 0.11_{\lambda _2}\pm 0.62_\text{resolved} ) \cdot 10^{-7}  = (12.35 \pm 0.92  ) \cdot 10^{-7} \; .
\end{align}
\begin{align}
\dps {\cal H}_L[1,3.5]_{\mu\mu} =&\ ( 7.37 \pm 0.28_{\text{scale}} \pm 0.08_{m_t} \pm 0.17_{C,m_c} \pm 0.10_{m_b} \pm  0.05_{\alpha_s} \pm 0.007_{\text{CKM}} \nonumber \\
&\hspace*{0pt} \pm 0.11_{\text{BR}_{\text{sl}}} \pm 0.01_{\lambda _1}\pm 0.06_{\lambda _2}\pm 0.37_\text{resolved}) \cdot 10^{-7} = (7.37 \pm 0.52 ) \cdot 10^{-7} \; ,  \nonumber\\[0.5em]
\dps {\cal H}_L[3.5,6]_{\mu\mu} =&\ ( 5.81 \pm 0.31_{\text{scale}} \pm 0.06_{m_t} \pm 0.14_{C,m_c} \pm 0.05_{m_b}  \pm  0.04_{\alpha_s} \pm 0.005_{\text{CKM}} \nonumber \\
&\hspace*{0pt} \pm 0.09_{\text{BR}_{\text{sl}}} \pm 0.02_{\lambda _1}\pm 0.06_{\lambda _2}\pm 0.29_\text{resolved}) \cdot 10^{-7} = ( 5.81 \pm 0.47 ) \cdot 10^{-7} \; , \nonumber\\[0.5em]
\dps {\cal H}_L[1,6]_{\mu\mu} =&\ ( 13.18 \pm 0.53_{\text{scale}} \pm 0.14_{m_t} \pm 0.31_{C,m_c} \pm 0.15_{m_b}  \pm    0.09_{\alpha_s} \pm 0.01_{\text{CKM}}\nonumber \\
&\hspace*{0pt} \pm 0.20_{\text{BR}_{\text{sl}}}\pm 0.03_{\lambda _1}\pm 0.12_{\lambda _2}\pm 0.66_\text{resolved} ) \cdot 10^{-7}  = ( 13.18 \pm 0.96  ) \cdot 10^{-7} \;  .
\end{align}
%

%%%%%%%%%%%%%%%%%%%%%%%%%%%%%%%%%%%%%%%%%%%%%%%%%%%%%%%%%%%%%%%%%%%%%%%%%%%%%%%%%%%%%%%%%%%%%

\subsection{${\cal H}_3$ and   ${\cal H}_4$ }
\begin{align}
\dps {\cal H}_3[1,3.5]_{ee} =&\ ( 4.14 \pm 0.65_{\text{scale}} \pm 0.04_{m_t} \pm 0.09_{C,m_c} \pm 0.10_{m_b} \pm  0.05_{\alpha_s} \pm 0.004_{\text{CKM}} \nonumber \\
&\hspace*{0pt} \pm 0.06_{\text{BR}_{\text{sl}}}\pm 0.01_{\lambda _1}\pm 0.02_{\lambda _2}\pm 0.21_\text{resolved}) \cdot 10^{-9} = ( 4.14 \pm 0.70 ) \cdot 10^{-9} \; ,  \nonumber\\[0.5em]
\dps {\cal H}_3[3.5,6]_{ee} =&\ ( 5.00 \pm 0.51_{\text{scale}} \pm 0.05_{m_t} \pm 0.11_{C,m_c} \pm 0.07_{m_b} \pm  0.04_{\alpha_s} \pm 0.005_{\text{CKM}} \nonumber \\
&\hspace*{0pt} \pm 0.08_{\text{BR}_{\text{sl}}}\pm 0.01_{\lambda _1}\pm 0.02_{\lambda _2}\pm 0.25_\text{resolved}) \cdot 10^{-9} = ( 5.00 \pm 0.59 ) \cdot 10^{-9} \; ,  \nonumber\\[0.5em]
\dps {\cal H}_3[1,6]_{ee} =&\ ( 9.14 \pm 1.16_{\text{scale}} \pm 0.09_{m_t} \pm 0.19_{C,m_c} \pm 0.17_{m_b} \pm  0.09_{\alpha_s} \pm 0.008_{\text{CKM}} \nonumber \\
&\hspace*{0pt} \pm 0.14_{\text{BR}_{\text{sl}}}\pm 0.02_{\lambda _1}\pm 0.04_{\lambda _2}\pm 0.46_\text{resolved}) \cdot 10^{-9} = ( 9.14 \pm 1.29 ) \cdot 10^{-9} \;  .
\end{align}
\begin{align}
\dps {\cal H}_3[1,3.5]_{\mu\mu} =&\ ( 1.72 \pm 0.27_{\text{scale}} \pm 0.02_{m_t} \pm 0.04_{C,m_c} \pm 0.04_{m_b} \pm  0.02_{\alpha_s} \pm 0.002_{\text{CKM}} \nonumber \\
&\hspace*{0pt} \pm 0.03_{\text{BR}_{\text{sl}}}\pm 0.004_{\lambda _1}\pm 0.01_{\lambda _2}\pm 0.09_\text{resolved}) \cdot 10^{-9} = ( 1.72 \pm 0.29 ) \cdot 10^{-9} \; ,  \nonumber\\[0.5em]
\dps {\cal H}_3[3.5,6]_{\mu\mu} =&\ ( 2.08 \pm 0.21_{\text{scale}} \pm 0.02_{m_t} \pm 0.04_{C,m_c} \pm 0.03_{m_b} \pm  0.02_{\alpha_s} \pm 0.002_{\text{CKM}} \nonumber \\
&\hspace*{0pt} \pm 0.03_{\text{BR}_{\text{sl}}}\pm 0.005_{\lambda _1}\pm 0.01_{\lambda _2}\pm 0.10_\text{resolved}) \cdot 10^{-9} = ( 2.08 \pm 0.25 ) \cdot 10^{-9} \; , \nonumber\\[0.5em]
\dps {\cal H}_3[1,6]_{\mu\mu} =&\ ( 3.80 \pm 0.48_{\text{scale}} \pm 0.04_{m_t} \pm 0.08_{C,m_c} \pm 0.06_{m_b} \pm  0.04_{\alpha_s} \pm 0.003_{\text{CKM}} \nonumber \\
&\hspace*{0pt} \pm 0.06_{\text{BR}_{\text{sl}}}\pm 0.01_{\lambda _1}\pm 0.02_{\lambda _2}\pm 0.19_\text{resolved}) \cdot 10^{-9} = ( 3.80 \pm 0.53 ) \cdot 10^{-9} \; .
\end{align}
\begin{align}
\dps {\cal H}_4[1,3.5]_{ee} =&\ ( 6.37 \pm 0.56_{\text{scale}} \pm 0.07_{m_t} \pm 0.13_{C,m_c} \pm 0.01_{m_b} \pm  0.03_{\alpha_s} \pm 0.006_{\text{CKM}} \nonumber \\
&\hspace*{0pt} \pm 0.10_{\text{BR}_{\text{sl}}}\pm 0.01_{\lambda _1}\pm 0.03_{\lambda _2}\pm 0.32_\text{resolved}) \cdot 10^{-9} = ( 6.37 \pm 0.67 ) \cdot 10^{-9} \; ,  \nonumber\\[0.5em]
\dps {\cal H}_4[3.5,6]_{ee} =&\ ( 2.24 \pm 0.16_{\text{scale}} \pm 0.03_{m_t} \pm 0.05_{C,m_c} \pm 0.02_{m_b} \pm  0.01_{\alpha_s} \pm 0.002_{\text{CKM}} \nonumber \\
&\hspace*{0pt} \pm 0.03_{\text{BR}_{\text{sl}}}\pm 0.005_{\lambda _1}\pm 0.01_{\lambda _2}\pm 0.11_\text{resolved}) \cdot 10^{-9} = ( 2.24 \pm 0.21 ) \cdot 10^{-9} \; ,  \nonumber\\[0.5em]
\dps {\cal H}_4[1,6]_{ee} =&\ ( 8.60 \pm 0.73_{\text{scale}} \pm 0.09_{m_t} \pm 0.18_{C,m_c} \pm 0.02_{m_b} \pm  0.04_{\alpha_s} \pm 0.008_{\text{CKM}} \nonumber \\
&\hspace*{0pt} \pm 0.13_{\text{BR}_{\text{sl}}}\pm 0.02_{\lambda _1}\pm 0.04_{\lambda _2}\pm 0.43_\text{resolved}) \cdot 10^{-9} = ( 8.60 \pm 0.88 ) \cdot 10^{-9} \;  .
\end{align}
\begin{align}
\dps {\cal H}_4[1,3.5]_{\mu\mu} =&\ ( 2.65 \pm 0.23_{\text{scale}} \pm 0.03_{m_t} \pm 0.06_{C,m_c} \pm 0.01_{m_b} \pm  0.01_{\alpha_s} \pm 0.002_{\text{CKM}} \nonumber \\
&\hspace*{0pt} \pm 0.04_{\text{BR}_{\text{sl}}}\pm 0.01_{\lambda _1}\pm 0.01_{\lambda _2}\pm 0.13_\text{resolved}) \cdot 10^{-9} = ( 2.65 \pm 0.28 ) \cdot 10^{-9} \; ,  \nonumber\\[0.5em]
\dps {\cal H}_4[3.5,6]_{\mu\mu} =&\ ( 0.93 \pm 0.07_{\text{scale}} \pm 0.01_{m_t} \pm 0.02_{C,m_c} \pm 0.01_{m_b} \pm  0.004_{\alpha_s} \pm 0.001_{\text{CKM}} \nonumber \\
&\hspace*{0pt} \pm 0.01_{\text{BR}_{\text{sl}}}\pm 0.002_{\lambda _1}\pm 0.005_{\lambda _2}\pm 0.05_\text{resolved}) \cdot 10^{-9} = ( 0.93 \pm 0.09 ) \cdot 10^{-9} \; , \nonumber\\[0.5em]
\dps {\cal H}_4[1,6]_{\mu\mu} =&\ ( 3.58 \pm 0.30_{\text{scale}} \pm 0.04_{m_t} \pm 0.08_{C,m_c} \pm 0.02_{m_b} \pm  0.02_{\alpha_s} \pm 0.003_{\text{CKM}} \nonumber \\
&\hspace*{0pt} \pm 0.05_{\text{BR}_{\text{sl}}}\pm 0.01_{\lambda _1}\pm 0.02_{\lambda _2}\pm 0.18_\text{resolved}) \cdot 10^{-9} = ( 3.58 \pm 0.36 ) \cdot 10^{-9} \; .
\end{align}

%%%%%%%%%%%%%%%%%%%%%%%%%%%%%%%%%%%%%%%%%%%%%%%%%%%%%%%%%%%%%%%%%%%%%%%%%%%%%%%%%%%%%%%%%%%%%%%%%%%%%%%%%%%%%%%%%%%%%%%%%%%%%%%%%%%%%%%%%%%%%%%%%%%%%%%%%%%%
%%%%%%%%%%%%%%%%%%%%%%%%%%%%%%%%%%%%%%%%%%%%%%%%%%%%%%%%%%%%%%%%%%%%%%%%%%%%%%%%%%%%%%%%%%%%%%%%%%%%%%%%%%%%%%%%%%%%%%%%%%%%%%%%%%%%%%%%%%%%%%%%%%%%%%%%%%%%

\section{New Physics formulas}
\label{sec:formulae}

In this appendix we give the new-physics formulas of all observables in terms of the following ratios
\bea
R_{7,8} = \frac{C_{7,8}^{(00){\rm eff}} (\mu_0)}{C_{7,8}^{(00){\rm eff,SM}}(\mu_0)}  
\hspace{1cm} {\rm and} \hspace{1cm}
R_{9,10} = \frac{C_{9,10}^{(11)} (\mu_0)}{C_{9,10}^{(11){\rm SM}}(\mu_0)} \; .
\eea
The superscripts on the Wilson coefficients denote the order in the expansion in $\alpha_s$ and $\kappa = \alpha_e/\alpha_s$, see~\cite{Huber:2005ig,Huber:2015sra} for details. The connection to the new-physics part of the Wilson coefficients in eq.~\eqref{eq:NPWC} is straightforward. On the right-hand sides of all the equations below, $\mathcal{R}$ and $\mathcal{I}$ denote the real and imaginary part of the expression in parenthesis, respectively. The label 'no em' refers to leaving out log-enhanced  QED corrections as described in the caption of table~\ref{tab:noem}. The new-physics formulas  are provided electronically as ancillary files attached to the arXiv submission of the present paper.

\subsection{Branching ratio, low-$q^2$ region}
\begin{align}
\dps {\cal B}[1,3.5]_{ee} =& \Big[
0.216997 \left| R_7\right| {}^2 +0.00294962 \left| R_8\right| {}^2 +0.833492 \left| R_9\right| {}^2 \nnb \\
&+6.0782 \left| R_{10}\right| {}^2+0.0173276 \mathcal{I}\left(R_7 R_8^*\right)+0.00288963 \mathcal{I}\left(R_7 R_9^*\right) \nnb \\
&+0.0151859 \mathcal{I}\left(R_8 R_9^*\right)-0.000309907 \mathcal{I}\left(R_8 R_{10}^*\right)+0.0519547 \mathcal{R}\left(R_7 R_8^*\right) \nnb \\
& -0.519361 \mathcal{R}\left(R_7 R_9^*\right)+0.00893145 \mathcal{R}\left(R_7 R_{10}^*\right)-0.0597428 \mathcal{R}\left(R_8 R_9^*\right) \nnb\\
&+0.00109835 \mathcal{R}\left(R_8 R_{10}^*\right)-0.0573964 \mathcal{R}\left(R_9 R_{10}^*\right)+0.0183294 \mathcal{I}\left(R_7\right) \nnb \\
&-0.00310346 \mathcal{I}\left(R_8\right)+0.0477963 \mathcal{I}\left(R_9\right)-0.0020085 \mathcal{I}\left(R_{10}\right) \nnb \\
&+0.12521 \mathcal{R}\left(R_7\right)+0.00686405 \mathcal{R}\left(R_8\right)+1.66745 \mathcal{R}\left(R_9\right) \nnb \\
&-0.296469 \mathcal{R}\left(R_{10}\right)+1.76389 \, \Big] \, \times \, 10^{-7} \, , \\[0.5em]
%
%BR bin2 ee
%
\dps {\cal B}[3.5,6]_{ee} =& \Big[
0.0713305 \left| R_7\right| {}^2+0.000898232 \left| R_8\right| {}^2+0.731636 \left| R_9\right| {}^2 \nnb \\
&+5.28065 \left| R_{10}\right| {}^2+0.00587951 \mathcal{I}\left(R_7 R_8^*\right)+0.0021872 \mathcal{I}\left(R_7 R_9^*\right) \nnb \\
&+0.0114944 \mathcal{I}\left(R_8 R_9^*\right)-0.000234573 \mathcal{I}\left(R_8 R_{10}^*\right)+0.0170293 \mathcal{R}\left(R_7 R_8^*\right) \nnb \\
&-0.370929 \mathcal{R}\left(R_7 R_9^*\right)+0.00675653 \mathcal{R}\left(R_7 R_{10}^*\right)-0.0415991 \mathcal{R}\left(R_8 R_9^*\right) \nnb\\
&+0.000790901 \mathcal{R}\left(R_8 R_{10}^*\right)-0.0524327 \mathcal{R}\left(R_9 R_{10}^*\right)+0.0120584 \mathcal{I}\left(R_7\right) \nnb\\
&+0.00940699 \mathcal{I}\left(R_8\right)+0.000191673 \mathcal{I}\left(R_9\right)-0.000977984 \mathcal{I}\left(R_{10}\right) \nnb \\
&-0.370247 \mathcal{R}\left(R_7\right)-0.0404915 \mathcal{R}\left(R_8\right)+1.77487 \mathcal{R}\left(R_9\right) \nnb \\
&-0.274818 \mathcal{R}\left(R_{10}\right)+1.2456 \, \Big] \, \times \, 10^{-7} \, , \\[0.5em]
%
%BR low-s ee
%
\dps {\cal B}[1,6]_{ee} =& \Big[
0.288327 \left| R_7\right| {}^2+0.00384785 \left| R_8\right| {}^2+1.56513 \left| R_9\right| {}^2 \nonumber \\
&+11.3588 \left| R_{10}\right| {}^2+0.0232071 \mathcal{I}\left(R_7 R_8^*\right)+0.00507683 \mathcal{I}\left(R_7 R_9^*\right) \nonumber \\
&+0.0266804 \mathcal{I}\left(R_8 R_9^*\right)-0.00054448 \mathcal{I}\left(R_8 R_{10}^*\right)+0.068984 \mathcal{R}\left(R_7 R_8^*\right)\nonumber \\
&-0.89029 \mathcal{R}\left(R_7 R_9^*\right)+0.015688 \mathcal{R}\left(R_7 R_{10}^*\right)-0.101342 \mathcal{R}\left(R_8 R_9^*\right)\nonumber \\
&+0.00188925 \mathcal{R}\left(R_8 R_{10}^*\right)-0.109829 \mathcal{R}\left(R_9 R_{10}^*\right)+0.0303878 \mathcal{I}\left(R_7\right)\nonumber \\
&+0.00630353 \mathcal{I}\left(R_8\right)+0.047988 \mathcal{I}\left(R_9\right)-0.00298649 \mathcal{I}\left(R_{10}\right)\nonumber \\
&-0.245038 \mathcal{R}\left(R_7\right)-0.0336275 \mathcal{R}\left(R_8\right) +3.44232 \mathcal{R}\left(R_9\right) \nonumber\\
&-0.571287 \mathcal{R}\left(R_{10}\right)+3.00949 \, \Big] \, \times \, 10^{-7} \, , \\[0.5em]
%
%BR bin1 mumu
%
\dps {\cal B}[1,3.5]_{\mu\mu} =& \Big[
0.221569 \left| R_7\right| {}^2+0.00297589 \left| R_8\right| {}^2+0.800733 \left| R_9\right| {}^2 \nnb \\
&+5.85121 \left| R_{10}\right| {}^2+0.0173276 \mathcal{I}\left(R_7 R_8^*\right)+0.00288963 \mathcal{I}\left(R_7 R_9^*\right) \nnb \\
&+0.0151859 \mathcal{I}\left(R_8 R_9^*\right)-0.000309907 \mathcal{I}\left(R_8 R_{10}^*\right)+0.0526479 \mathcal{R}\left(R_7 R_8^*\right) \nnb\\
&-0.503222 \mathcal{R}\left(R_7 R_9^*\right)+0.00893145 \mathcal{R}\left(R_7 R_{10}^*\right)-0.0585195 \mathcal{R}\left(R_8 R_9^*\right) \nnb\\
&+0.00109835 \mathcal{R}\left(R_8 R_{10}^*\right)-0.0573964 \mathcal{R}\left(R_9 R_{10}^*\right)+0.018682 \mathcal{I}\left(R_7\right) \nnb\\
&-0.00307673 \mathcal{I}\left(R_8\right)+0.0461066 \mathcal{I}\left(R_9\right)-0.0020085 \mathcal{I}\left(R_{10}\right) \nnb\\
&+0.170386 \mathcal{R}\left(R_7\right)+0.0103486 \mathcal{R}\left(R_8\right)+1.56162 \mathcal{R}\left(R_9\right) \nnb\\
&-0.296469 \mathcal{R}\left(R_{10}\right)+1.67348 \, \Big] \, \times \, 10^{-7} \, , \\[0.5em]
%
%BR bin2 mumu
%
\dps {\cal B}[3.5,6]_{\mu\mu} =& \Big[
0.0745453 \left| R_7\right| {}^2+0.000916702 \left| R_8\right| {}^2+0.724126 \left| R_9\right| {}^2 \nnb\\
&+5.22861 \left| R_{10}\right| {}^2+0.00587951 \mathcal{I}\left(R_7 R_8^*\right)+0.0021872 \mathcal{I}\left(R_7 R_9^*\right) \nnb\\
&+0.0114944 \mathcal{I}\left(R_8 R_9^*\right)-0.000234573 \mathcal{I}\left(R_8 R_{10}^*\right)+0.0175166 \mathcal{R}\left(R_7 R_8^*\right) \nnb\\
&-0.371854 \mathcal{R}\left(R_7 R_9^*\right)+0.00675653 \mathcal{R}\left(R_7 R_{10}^*\right)-0.0416692 \mathcal{R}\left(R_8 R_9^*\right) \nnb\\
&+0.000790901 \mathcal{R}\left(R_8 R_{10}^*\right)-0.0524327 \mathcal{R}\left(R_9 R_{10}^*\right)+0.0125784 \mathcal{I}\left(R_7\right) \nnb\\
&+0.00944641 \mathcal{I}\left(R_8\right)-0.00227977 \mathcal{I}\left(R_9\right)-0.000977984 \mathcal{I}\left(R_{10}\right) \nnb\\
&-0.360151 \mathcal{R}\left(R_7\right)-0.0397297 \mathcal{R}\left(R_8\right)+1.73717 \mathcal{R}\left(R_9\right) \nnb\\
&-0.274818 \mathcal{R}\left(R_{10}\right)+1.19972 \, \Big] \, \times \, 10^{-7} \, , \\[0.5em]
%
%BR low-s mumu
%
\dps {\cal B}[1,6]_{\mu\mu} =& \Big[
0.296114 \left| R_7\right| {}^2+0.00389259 \left| R_8\right| {}^2+1.52486 \left| R_9\right| {}^2 \nnb \\
&+11.0798 \left| R_{10}\right| {}^2+0.0232071 \mathcal{I}\left(R_7 R_8^*\right)+0.00507683 \mathcal{I}\left(R_7 R_9^*\right) \nnb\\
&+0.0266804 \mathcal{I}\left(R_8 R_9^*\right)-0.00054448 \mathcal{I}\left(R_8 R_{10}^*\right)+0.0701645 \mathcal{R}\left(R_7 R_8^*\right) \nnb\\
&-0.875076 \mathcal{R}\left(R_7 R_9^*\right)+0.015688 \mathcal{R}\left(R_7 R_{10}^*\right)-0.100189 \mathcal{R}\left(R_8 R_9^*\right) \nnb\\
&+0.00188925 \mathcal{R}\left(R_8 R_{10}^*\right)-0.109829 \mathcal{R}\left(R_9 R_{10}^*\right)+0.0312604 \mathcal{I}\left(R_7\right) \nnb\\
&+0.00636967 \mathcal{I}\left(R_8\right)+0.0438268 \mathcal{I}\left(R_9\right)-0.00298649 \mathcal{I}\left(R_{10}\right) \nnb\\
&-0.189764 \mathcal{R}\left(R_7\right)-0.0293811 \mathcal{R}\left(R_8\right)+3.29879 \mathcal{R}\left(R_9\right) \nnb\\
&-0.571287 \mathcal{R}\left(R_{10}\right)+2.8732 \, \Big] \, \times \, 10^{-7} \, , \\[0.5em]
%
%BR bin1 no em
%
\dps {\cal B}[1,3.5]_\text{no em} =& \Big[
0.224822 \left| R_7\right| {}^2+0.00299458 \left| R_8\right| {}^2+0.785074 \left| R_9\right| {}^2 \nnb\\
&+5.68972 \left| R_{10}\right| {}^2+0.0173276 \mathcal{I}\left(R_7 R_8^*\right)+0.00288963 \mathcal{I}\left(R_7 R_9^*\right) \nnb\\
&+0.0151859 \mathcal{I}\left(R_8 R_9^*\right)-0.000309907 \mathcal{I}\left(R_8 R_{10}^*\right)+0.053141 \mathcal{R}\left(R_7 R_8^*\right) \nnb\\
&-0.49174 \mathcal{R}\left(R_7 R_9^*\right)+0.00893145 \mathcal{R}\left(R_7 R_{10}^*\right)-0.0576492 \mathcal{R}\left(R_8 R_9^*\right) \nnb\\
&+0.00109835 \mathcal{R}\left(R_8 R_{10}^*\right)-0.0573964 \mathcal{R}\left(R_9 R_{10}^*\right)+0.0189329 \mathcal{I}\left(R_7\right) \nnb\\
&-0.00305772 \mathcal{I}\left(R_8\right)+0.0449044 \mathcal{I}\left(R_9\right)-0.0020085 \mathcal{I}\left(R_{10}\right) \nnb\\
&+0.202569 \mathcal{R}\left(R_7\right)+0.0128309 \mathcal{R}\left(R_8\right)+1.51378 \mathcal{R}\left(R_9\right) \nnb\\
&-0.296469 \mathcal{R}\left(R_{10}\right)+1.63384 \, \Big] \, \times \, 10^{-7} \, , \\[0.5em]
%
%BR bin2 no em
%
\dps {\cal B}[3.5,6]_\text{no em} =& \Big[
0.0768325 \left| R_7\right| {}^2+0.000929842 \left| R_8\right| {}^2+0.717985 \left| R_9\right| {}^2 \nnb\\
&+5.19159 \left| R_{10}\right| {}^2+0.00587951 \mathcal{I}\left(R_7 R_8^*\right)+0.0021872 \mathcal{I}\left(R_7 R_9^*\right) \nnb \\
&+0.0114944 \mathcal{I}\left(R_8 R_9^*\right)-0.000234573 \mathcal{I}\left(R_8 R_{10}^*\right)+0.0178634 \mathcal{R}\left(R_7 R_8^*\right) \nnb\\
&-0.372512 \mathcal{R}\left(R_7 R_9^*\right)+0.00675653 \mathcal{R}\left(R_7 R_{10}^*\right)-0.0417191 \mathcal{R}\left(R_8 R_9^*\right) \nnb\\
&+0.000790901 \mathcal{R}\left(R_8 R_{10}^*\right)-0.0524327 \mathcal{R}\left(R_9 R_{10}^*\right)+0.0129484 \mathcal{I}\left(R_7\right) \nnb\\
&+0.00947445 \mathcal{I}\left(R_8\right)-0.00403808 \mathcal{I}\left(R_9\right)-0.000977984 \mathcal{I}\left(R_{10}\right) \nnb\\
&-0.35297 \mathcal{R}\left(R_7\right)-0.0391879 \mathcal{R}\left(R_8\right)+1.70797 \mathcal{R}\left(R_9\right) \nnb\\
&-0.274818 \mathcal{R}\left(R_{10}\right)+1.16546 \, \Big] \, \times \, 10^{-7} \, , \\[0.5em]
%
%BR low-s no em
%
\dps {\cal B}[1,6]_\text{no em} =& \Big[
0.301655 \left| R_7\right| {}^2+0.00392442 \left| R_8\right| {}^2+1.50306 \left| R_9\right| {}^2 \nnb\\
&+10.8813 \left| R_{10}\right| {}^2+0.0232071 \mathcal{I}\left(R_7 R_8^*\right)+0.00507683 \mathcal{I}\left(R_7 R_9^*\right) \nnb\\
&+0.0266804 \mathcal{I}\left(R_8 R_9^*\right)-0.00054448 \mathcal{I}\left(R_8 R_{10}^*\right)+0.0710044 \mathcal{R}\left(R_7 R_8^*\right)\nnb\\
&-0.864251 \mathcal{R}\left(R_7 R_9^*\right)+0.015688 \mathcal{R}\left(R_7 R_{10}^*\right)-0.0993683 \mathcal{R}\left(R_8 R_9^*\right) \nnb\\
&+0.00188925 \mathcal{R}\left(R_8 R_{10}^*\right)-0.109829 \mathcal{R}\left(R_9 R_{10}^*\right)+0.0318813 \mathcal{I}\left(R_7\right) \nnb\\
&+0.00641673 \mathcal{I}\left(R_8\right)+0.0408664 \mathcal{I}\left(R_9\right)-0.00298649 \mathcal{I}\left(R_{10}\right) \nnb\\
&-0.150401 \mathcal{R}\left(R_7\right)-0.026357 \mathcal{R}\left(R_8\right)+3.22175 \mathcal{R}\left(R_9\right) \nnb\\
&-0.571287 \mathcal{R}\left(R_{10}\right)+2.79931 \, \Big] \, \times \, 10^{-7} \, .
\end{align}

%%%%%%%%%%%%%%%%%%%%%%%%%%%%%%%%%%%%%%%%%%%%%%%%%%%%%%%%%%%%%%%%%%%%%%%%%%%%%%%%%%%%%%%%%%%%%%%%%%%%%%%%%%%%%%%%%%%%%%%%%%%%%%%%%%%%%%%%%%%%%%%%%%%%%%%%%%%%%%%%%%%%%%

\subsection{Branching ratio, high-$q^2$ region}
%%%%%%%%%%%%%%%%%%%%%%%%%%%%%%%%%%%%%%%%%%%%%%%%%%%%%%%%%%%%%%%%%%%%%%%%%%%%%%%%%%%%%%%%%%%%%%%%%%%%%%%%%%%%%%%%%%%%%%%%%%%%%%%%%%%%%%%%%%%%%%%%%%%%%%%%%%%%%%%%%%%%%%

\begin{align}
\dps {\cal B}[>14.4]_{ee} =& \Big[
0.000257481 \mathcal{I}\left(R_7 R_8^*\right)+0.000385537 \mathcal{I}\left(R_7 R_9^*\right)-0.0000314797 \mathcal{I}\left(R_8 R_{10}^*\right) \nnb \\
&+0.00154984 \mathcal{I}\left(R_8 R_9^*\right)+0.000626787 \mathcal{R}\left(R_7 R_8^*\right)-0.0448958 \mathcal{R}\left(R_7 R_9^*\right) \nnb \\
&-0.00499875 \mathcal{R}\left(R_8 R_9^*\right)+0.00106976 \mathcal{R}\left(R_7 R_{10}^*\right)-0.0163127 \mathcal{R}\left(R_9 R_{10}^*\right) \nnb \\
&+0.000114054 \mathcal{R}\left(R_8 R_{10}^*\right)+0.00237824 \left| R_7\right| {}^2+0.0000338564 \left| R_8\right| {}^2 \nnb \\
&+0.190037 \left| R_9\right| {}^2+1.3514 \left| R_{10}\right| {}^2+0.00458151 \mathcal{I}\left(R_7\right) \nnb \\
&+0.00208694 \mathcal{I}\left(R_8\right)+0.00740457 \mathcal{I}\left(R_9\right)-0.000683152 \mathcal{I}\left(R_{10}\right) \nnb \\
&-0.0704095 \mathcal{R}\left(R_7\right)-0.00781914 \mathcal{R}\left(R_8\right)+0.497853 \mathcal{R}\left(R_9\right) \nnb \\
&-0.0766318 \mathcal{R}\left(R_{10}\right)+0.216631 \, \Big] \, \times \, 10^{-7} \, , \\[0.5em]
%
%BR high-s mumu
%
\dps {\cal B}[>14.4]_{\mu\mu} =& \Big[
0.000257481 \mathcal{I}\left(R_7 R_8^*\right)+0.000385537 \mathcal{I}\left(R_7 R_9^*\right)-0.0000314797 \mathcal{I}\left(R_8 R_{10}^*\right) \nnb \\
&+0.00154984 \mathcal{I}\left(R_8 R_9^*\right)+0.000756249 \mathcal{R}\left(R_7 R_8^*\right)-0.0537067 \mathcal{R}\left(R_7 R_9^*\right) \nnb \\
&-0.00566659 \mathcal{R}\left(R_8 R_9^*\right)+0.00106976 \mathcal{R}\left(R_7 R_{10}^*\right)-0.0163127 \mathcal{R}\left(R_9 R_{10}^*\right) \nnb \\
&+0.000114054 \mathcal{R}\left(R_8 R_{10}^*\right)+0.00323224 \left| R_7\right| {}^2+0.0000387628 \left| R_8\right| {}^2 \nnb \\
&+0.213448 \left| R_9\right| {}^2+1.51361 \left| R_{10}\right| {}^2+0.00272213 \mathcal{I}\left(R_7\right) \nnb \\
&+0.001946 \mathcal{I}\left(R_8\right)+0.0173032 \mathcal{I}\left(R_9\right)-0.000683152 \mathcal{I}\left(R_{10}\right) \nnb \\
&-0.0857515 \mathcal{R}\left(R_7\right)-0.00901494 \mathcal{R}\left(R_8\right)+0.58311 \mathcal{R}\left(R_9\right) \nnb \\
&-0.0766318 \mathcal{R}\left(R_{10}\right)+0.311364 \, \Big] \, \times \, 10^{-7} \, , \\[0.5em]
%
%BR high-s mumu
%
\dps {\cal B}[>14.4]_{\rm no \; em} =& \Big[
0.000257481 \mathcal{I}\left(R_7 R_8^*\right)+0.000385537 \mathcal{I}\left(R_7 R_9^*\right)-0.0000314797 \mathcal{I}\left(R_8 R_{10}^*\right) \nnb \\
&+0.00154984 \mathcal{I}\left(R_8 R_9^*\right)+0.000848356 \mathcal{R}\left(R_7 R_8^*\right)-0.0599752 \mathcal{R}\left(R_7 R_9^*\right) \nnb \\
&-0.00614173 \mathcal{R}\left(R_8 R_9^*\right)+0.00106976 \mathcal{R}\left(R_7 R_{10}^*\right)-0.0163127 \mathcal{R}\left(R_9 R_{10}^*\right) \nnb \\
&+0.000114054 \mathcal{R}\left(R_8 R_{10}^*\right)+0.00383983 \left| R_7\right| {}^2+0.0000422535 \left| R_8\right| {}^2 \nnb \\
&+0.22634 \left| R_9\right| {}^2+1.62902 \left| R_{10}\right| {}^2+0.00139926 \mathcal{I}\left(R_7\right) \nnb \\
&+0.00184573 \mathcal{I}\left(R_8\right)+0.0243456 \mathcal{I}\left(R_9\right)-0.000683152 \mathcal{I}\left(R_{10}\right) \nnb \\
&-0.0966894 \mathcal{R}\left(R_7\right)-0.00986742 \mathcal{R}\left(R_8\right)+0.629448 \mathcal{R}\left(R_9\right) \nnb \\
&-0.0766318 \mathcal{R}\left(R_{10}\right)+0.364954 \, \Big] \, \times \, 10^{-7} \, .
\end{align}

\subsection{The ratio ${\cal R}(s_0)$}

\begin{align}
\dps {\cal R}(14.4)_{ee} =& \Big[
0.00345454 \mathcal{I}\left(R_7 R_8^*\right)+0.00518561 \mathcal{I}\left(R_7 R_9^*\right)+0.0204459 \mathcal{I}\left(R_8 R_9^*\right)\nnb\\
&-0.000422075 \mathcal{I}\left(R_8 R_{10}^*\right)+0.00769598 \mathcal{R}\left(R_7 R_8^*\right)-0.509477 \mathcal{R}\left(R_7 R_9^*\right)\nnb\\
&-0.0594276 \mathcal{R}\left(R_8 R_9^*\right)+0.0116279 \mathcal{R}\left(R_7 R_{10}^*\right)+0.00131861 \mathcal{R}\left(R_8 R_{10}^*\right)\nnb\\
&-0.178208 \mathcal{R}\left(R_9 R_{10}^*\right)+0.0274262 \left| R_7\right| {}^2+0.000425936 \left| R_8\right| {}^2\nnb\\
&+2.12663 \left| R_9\right| {}^2+15.1222 \left| R_{10}\right| {}^2+0.0610087 \mathcal{I}\left(R_7\right)\nnb\\
&+0.0240118 \mathcal{I}\left(R_8\right)+0.0219189 \mathcal{I}\left(R_9\right)-0.00627745 \mathcal{I}\left(R_{10}\right)\nnb\\
&-0.691986 \mathcal{R}\left(R_7\right)-0.0813958 \mathcal{R}\left(R_8\right)+4.65255 \mathcal{R}\left(R_9\right)\nnb\\
&-0.898911 \mathcal{R}\left(R_{10}\right)+1.99568 \, \Big] \, \times \, 10^{-4} \, , \\[0.5em]
%
%R0 high-s mumu
%
\dps {\cal R}(14.4)_{\mu\mu} =& \Big[
0.00345454 \mathcal{I}\left(R_7 R_8^*\right)+0.00518561 \mathcal{I}\left(R_7 R_9^*\right)+0.0204459 \mathcal{I}\left(R_8 R_9^*\right)\nnb\\
&-0.000422075 \mathcal{I}\left(R_8 R_{10}^*\right)+0.00925535 \mathcal{R}\left(R_7 R_8^*\right)-0.615603 \mathcal{R}\left(R_7 R_9^*\right)\nnb\\
&-0.0674718 \mathcal{R}\left(R_8 R_9^*\right)+0.0116279 \mathcal{R}\left(R_7 R_{10}^*\right)+0.00131861 \mathcal{R}\left(R_8 R_{10}^*\right)\nnb\\
&-0.178208 \mathcal{R}\left(R_9 R_{10}^*\right)+0.0377127 \left| R_7\right| {}^2+0.000485034 \left| R_8\right| {}^2\nnb\\
&+2.40861 \left| R_9\right| {}^2+17.0761 \left| R_{10}\right| {}^2+0.0386124 \mathcal{I}\left(R_7\right)\nnb\\
&+0.0223142 \mathcal{I}\left(R_8\right)+0.141148 \mathcal{I}\left(R_9\right)-0.00627745 \mathcal{I}\left(R_{10}\right)\nnb\\
&-0.852337 \mathcal{R}\left(R_7\right)-0.0939464 \mathcal{R}\left(R_8\right)+5.54958 \mathcal{R}\left(R_9\right)\nnb\\
&-0.898911 \mathcal{R}\left(R_{10}\right)+2.93907 \, \Big] \, \times \, 10^{-4} \, .
\\[0.5em]
%
%R0 high-s no em
%
\dps {\cal R}(14.4)_{\rm no\; em} =& \Big[
0.00345454 \mathcal{I}\left(R_7 R_8^*\right)+0.00518561 \mathcal{I}\left(R_7 R_9^*\right)+0.0204459 \mathcal{I}\left(R_8 R_9^*\right)\nnb\\
&-0.000422075 \mathcal{I}\left(R_8 R_{10}^*\right)+0.0103648 \mathcal{R}\left(R_7 R_8^*\right)-0.691108 \mathcal{R}\left(R_7 R_9^*\right)\nnb\\
&-0.0731948 \mathcal{R}\left(R_8 R_9^*\right)+0.0116279 \mathcal{R}\left(R_7 R_{10}^*\right)+0.00131861 \mathcal{R}\left(R_8 R_{10}^*\right)\nnb\\
&-0.178208 \mathcal{R}\left(R_9 R_{10}^*\right)+0.045031 \left| R_7\right| {}^2+0.00052708 \left| R_8\right| {}^2\nnb\\
&+2.5639 \left| R_9\right| {}^2+18.4662 \left| R_{10}\right| {}^2+0.0226784 \mathcal{I}\left(R_7\right)\nnb\\
&+0.0211065 \mathcal{I}\left(R_8\right)+0.225975 \mathcal{I}\left(R_9\right)-0.00627745 \mathcal{I}\left(R_{10}\right)\nnb\\
&-0.966693 \mathcal{R}\left(R_7\right)-0.102896 \mathcal{R}\left(R_8\right)+6.03618 \mathcal{R}\left(R_9\right)\nnb\\
&-0.898911 \mathcal{R}\left(R_{10}\right)+3.47739 \, \Big] \, \times \, 10^{-4} \, .
\end{align}

%%%%%%%%%%%%%%%%%%%%%%%%%%%%%%%%%%%%%%%%%%%%%%%%%%%%%%%%%%%%%%%%%%%%%%%%%%%%%%%%%%%%%%%%%%%%%%%%%%%%%%%%%%%%%%%%%%%%%%%%%%%%%%%%%%%%%%%%%%%%%%%%%%%%%%%%%%%%%%%%%%%%%%

\subsection{Forward-backward asymmetry, low-$q^2$ region}

\begin{align}
\dps {\cal H}_A[1,3.5]_{ee} =& \Big[
-0.00177919 \left| R_9\right| {}^2 -0.0123279 \left| R_{10}\right| {}^2 -0.0000776477 \mathcal{I}\left(R_8 R_9^*\right) \nnb\\
&+0.0263923 \mathcal{I}\left(R_8 R_{10}^*\right) +0.00325843 \mathcal{I}\left(R_9 R_{10}^*\right)+0.00220437 \mathcal{R}\left(R_7 R_9^*\right) \nnb\\
&-0.870054 \mathcal{R}\left(R_7 R_{10}^*\right)+0.000227347 \mathcal{R}\left(R_8 R_9^*\right)-0.0864769 \mathcal{R}\left(R_8 R_{10}^*\right) \nnb\\
&+0.73759 \mathcal{R}\left(R_9 R_{10}^*\right)-0.0000857354 \mathcal{I}\left(R_8\right)+0.000225495 \mathcal{I}\left(R_9\right) \nnb\\
&-0.10319 \mathcal{I}\left(R_{10}\right)+0.023259 \mathcal{R}\left(R_7\right)+0.00182876 \mathcal{R}\left(R_8\right) \nnb\\
&-0.0136603 \mathcal{R}\left(R_9\right)-0.744674 \mathcal{R}\left(R_{10}\right)+0.0102674 \, \Big] \, \times \, 10^{-7} \, , \\[0.5em]
%
%BR bin2 ee
%
\dps {\cal H}_A[3.5,6]_{ee} =& \Big[
-0.0028283 \left| R_9\right| {}^2-0.0195971 \left| R_{10}\right| {}^2-0.0000582814 \mathcal{I}\left(R_8 R_9^*\right) \nnb\\
&+0.0198099 \mathcal{I}\left(R_8 R_{10}^*\right)+0.00520128 \mathcal{I}\left(R_9 R_{10}^*\right)+0.00164717 \mathcal{R}\left(R_7 R_9^*\right) \nnb\\
&-0.620306 \mathcal{R}\left(R_7 R_{10}^*\right)+0.000173633 \mathcal{R}\left(R_8 R_9^*\right)-0.0636331 \mathcal{R}\left(R_8 R_{10}^*\right)\nnb\\
&+1.10604 \mathcal{R}\left(R_9 R_{10}^*\right)-0.0000642074 \mathcal{I}\left(R_8\right)+0.000138567 \mathcal{I}\left(R_9\right) \nnb\\
&-0.0924194 \mathcal{I}\left(R_{10}\right)+0.0174663 \mathcal{R}\left(R_7\right)+0.00137799 \mathcal{R}\left(R_8\right) \nnb\\
&-0.0266284 \mathcal{R}\left(R_9\right)+0.533272 \mathcal{R}\left(R_{10}\right)-0.0149706 \, \Big] \, \times \, 10^{-7} \, , \\[0.5em]
%
%BR low-s ee
%
\dps {\cal H}_A[1,6]_{ee} =& \Big[
-0.00460749 \left| R_9\right| {}^2-0.031925 \left| R_{10}\right| {}^2-0.000135929 \mathcal{I}\left(R_8 R_9^*\right) \nnb\\
&+0.0462021 \mathcal{I}\left(R_8 R_{10}^*\right)+0.00845971 \mathcal{I}\left(R_9 R_{10}^*\right)+0.00385155 \mathcal{R}\left(R_7 R_9^*\right) \nnb\\
&-1.49036 \mathcal{R}\left(R_7 R_{10}^*\right)+0.00040098 \mathcal{R}\left(R_8 R_9^*\right)-0.15011 \mathcal{R}\left(R_8 R_{10}^*\right) \nnb\\
&+1.84363 \mathcal{R}\left(R_9 R_{10}^*\right)-0.000149943 \mathcal{I}\left(R_8\right)+0.000364062 \mathcal{I}\left(R_9\right) \nnb\\
&-0.19561 \mathcal{I}\left(R_{10}\right)+0.0407254 \mathcal{R}\left(R_7\right)+0.00320675 \mathcal{R}\left(R_8\right) \nnb\\
&-0.0402886 \mathcal{R}\left(R_9\right)-0.211402 \mathcal{R}\left(R_{10}\right)-0.0047032 \, \Big] \, \times \, 10^{-7} \, , \\[0.5em]
%
%BR bin1 mumu
%
\dps {\cal H}_A[1,3.5]_{\mu\mu} =& \Big[
-0.00177919 \left| R_9\right| {}^2-0.0123279 \left| R_{10}\right| {}^2-0.0000776477 \mathcal{I}\left(R_8 R_9^*\right) \nnb\\
&+0.0263923 \mathcal{I}\left(R_8 R_{10}^*\right)+0.00325843 \mathcal{I}\left(R_9 R_{10}^*\right)+0.00220437 \mathcal{R}\left(R_7 R_9^*\right)\nnb\\
&-0.86614 \mathcal{R}\left(R_7 R_{10}^*\right)+0.000227347 \mathcal{R}\left(R_8 R_9^*\right)-0.0861802 \mathcal{R}\left(R_8 R_{10}^*\right)\nnb\\
&+0.708627 \mathcal{R}\left(R_9 R_{10}^*\right)-0.0000857354 \mathcal{I}\left(R_8\right)+0.000225495 \mathcal{I}\left(R_9\right) \nnb\\
&-0.10319 \mathcal{I}\left(R_{10}\right)+0.023259 \mathcal{R}\left(R_7\right)+0.00182876 \mathcal{R}\left(R_8\right) \nnb\\
&-0.0136603 \mathcal{R}\left(R_9\right)-0.79481 \mathcal{R}\left(R_{10}\right)+0.0112923 \, \Big] \, \times \, 10^{-7} \, , \\[0.5em]
%
%BR bin2 mumu
%
\dps {\cal H}_A[3.5,6]_{\mu\mu} =& \Big[
-0.0028283 \left| R_9\right| {}^2-0.0195971 \left| R_{10}\right| {}^2-0.0000582814 \mathcal{I}\left(R_8 R_9^*\right) \nnb\\
&+0.0198099 \mathcal{I}\left(R_8 R_{10}^*\right)+0.00520128 \mathcal{I}\left(R_9 R_{10}^*\right)+0.00164717 \mathcal{R}\left(R_7 R_9^*\right) \nnb\\
&-0.635041 \mathcal{R}\left(R_7 R_{10}^*\right)+0.000173633 \mathcal{R}\left(R_8 R_9^*\right)-0.0647501 \mathcal{R}\left(R_8 R_{10}^*\right) \nnb\\
&+1.09972 \mathcal{R}\left(R_9 R_{10}^*\right)-0.0000642074 \mathcal{I}\left(R_8\right)+0.000138567 \mathcal{I}\left(R_9\right) \nnb\\
&-0.092495 \mathcal{I}\left(R_{10}\right)+0.0174663 \mathcal{R}\left(R_7\right)+0.00137799 \mathcal{R}\left(R_8\right) \nnb\\
&-0.0266284 \mathcal{R}\left(R_9\right)+0.494194 \mathcal{R}\left(R_{10}\right)-0.0147469 \, \Big] \, \times \, 10^{-7} \, , \\[0.5em]
%
%BR low-s mumu
%
\dps {\cal H}_A[1,6]_{\mu\mu} =& \Big[
-0.00460749 \left| R_9\right| {}^2-0.031925 \left| R_{10}\right| {}^2-0.000135929 \mathcal{I}\left(R_8 R_9^*\right) \nnb\\
&+0.0462021 \mathcal{I}\left(R_8 R_{10}^*\right)+0.00845971 \mathcal{I}\left(R_9 R_{10}^*\right)+0.00385155 \mathcal{R}\left(R_7 R_9^*\right) \nnb\\
&-1.50118 \mathcal{R}\left(R_7 R_{10}^*\right)+0.00040098 \mathcal{R}\left(R_8 R_9^*\right)-0.15093 \mathcal{R}\left(R_8 R_{10}^*\right) \nnb\\
&+1.80835 \mathcal{R}\left(R_9 R_{10}^*\right)-0.000149943 \mathcal{I}\left(R_8\right)+0.000364062 \mathcal{I}\left(R_9\right) \nnb\\
&-0.195685 \mathcal{I}\left(R_{10}\right)+0.0407254 \mathcal{R}\left(R_7\right)+0.00320675 \mathcal{R}\left(R_8\right) \nnb\\
&-0.0402886 \mathcal{R}\left(R_9\right)-0.300616 \mathcal{R}\left(R_{10}\right)-0.00345465 \, \Big] \, \times \, 10^{-7} \, , \\[0.5em]
%
%BR bin1 no em
%
\dps {\cal H}_A[1,3.5]_\text{no em} =& \Big[
-0.00177919 \left| R_9\right| {}^2-0.0123279 \left| R_{10}\right| {}^2-0.0000776477 \mathcal{I}\left(R_8 R_9^*\right) \nnb\\
&+0.0263923 \mathcal{I}\left(R_8 R_{10}^*\right)+0.00325843 \mathcal{I}\left(R_9 R_{10}^*\right)+0.00220437 \mathcal{R}\left(R_7 R_9^*\right) \nnb\\
&-0.863355 \mathcal{R}\left(R_7 R_{10}^*\right)+0.000227347 \mathcal{R}\left(R_8 R_9^*\right)-0.0859692 \mathcal{R}\left(R_8 R_{10}^*\right)\nnb\\
&+0.688022 \mathcal{R}\left(R_9 R_{10}^*\right)-0.0000857354 \mathcal{I}\left(R_8\right)+0.000225495 \mathcal{I}\left(R_9\right) \nnb\\
&-0.10319 \mathcal{I}\left(R_{10}\right)+0.023259 \mathcal{R}\left(R_7\right)+0.00182876 \mathcal{R}\left(R_8\right) \nnb\\
&-0.0136603 \mathcal{R}\left(R_9\right)-0.830553 \mathcal{R}\left(R_{10}\right)+0.0120214 \, \Big] \, \times \, 10^{-7} \, , \\[0.5em]
%
%BR bin2 no em
%
\dps {\cal H}_A[3.5,6]_\text{no em} =& \Big[
-0.0028283 \left| R_9\right| {}^2-0.0195971 \left| R_{10}\right| {}^2-0.0000582814 \mathcal{I}\left(R_8 R_9^*\right) \nnb\\
&+0.0198099 \mathcal{I}\left(R_8 R_{10}^*\right)+0.00520128 \mathcal{I}\left(R_9 R_{10}^*\right)+0.00164717 \mathcal{R}\left(R_7 R_9^*\right) \nnb\\
&-0.645525 \mathcal{R}\left(R_7 R_{10}^*\right)+0.000173633 \mathcal{R}\left(R_8 R_9^*\right)-0.0655447 \mathcal{R}\left(R_8 R_{10}^*\right)\nnb\\
&+1.09522 \mathcal{R}\left(R_9 R_{10}^*\right)-0.0000642074 \mathcal{I}\left(R_8\right)+0.000138567 \mathcal{I}\left(R_9\right) \nnb\\
&-0.0925487 \mathcal{I}\left(R_{10}\right)+0.0174663 \mathcal{R}\left(R_7\right)+0.00137799 \mathcal{R}\left(R_8\right) \nnb\\
&-0.0266284 \mathcal{R}\left(R_9\right)+0.466375 \mathcal{R}\left(R_{10}\right)-0.0145878 \, \Big] \, \times \, 10^{-7} \, , \\[0.5em]
%
%BR low-s no em
%
\dps {\cal H}_A[1,6]_\text{no em} =& \Big[
-0.00460749 \left| R_9\right| {}^2-0.031925 \left| R_{10}\right| {}^2-0.000135929 \mathcal{I}\left(R_8 R_9^*\right) \nnb\\
&+0.0462021 \mathcal{I}\left(R_8 R_{10}^*\right)+0.00845971 \mathcal{I}\left(R_9 R_{10}^*\right)+0.00385155 \mathcal{R}\left(R_7 R_9^*\right) \nnb\\
&-1.50888 \mathcal{R}\left(R_7 R_{10}^*\right)+0.00040098 \mathcal{R}\left(R_8 R_9^*\right)-0.151514 \mathcal{R}\left(R_8 R_{10}^*\right) \nnb\\
&+1.78325 \mathcal{R}\left(R_9 R_{10}^*\right)-0.000149943 \mathcal{I}\left(R_8\right)+0.000364062 \mathcal{I}\left(R_9\right) \nnb\\
&-0.195739 \mathcal{I}\left(R_{10}\right)+0.0407254 \mathcal{R}\left(R_7\right)+0.00320675 \mathcal{R}\left(R_8\right) \nnb\\
&-0.0402886 \mathcal{R}\left(R_9\right)-0.364178 \mathcal{R}\left(R_{10}\right)-0.00256636 \, \Big] \, \times \, 10^{-7} \, .
\end{align}

\subsection{$H_L$ and $H_T$, low-$q^2$ region}

\begin{align}
\dps {\cal H}_L[1,3.5]_{ee} =& \Big[
0.000747411 \mathcal{I}\left(R_7 R_8^*\right)+0.000977063 \mathcal{I}\left(R_7 R_9^*\right)+0.00513472 \mathcal{I}\left(R_8 R_9^*\right) \nnb\\
& -0.000104788 \mathcal{I}\left(R_8 R_{10}^*\right)+0.00214866 \mathcal{R}\left(R_7 R_8^*\right)-0.140096 \mathcal{R}\left(R_7 R_9^*\right)\nnb\\
&-0.0191872 \mathcal{R}\left(R_8 R_9^*\right)+0.00308685 \mathcal{R}\left(R_7 R_{10}^*\right)-0.047482 \mathcal{R}\left(R_9 R_{10}^*\right)\nnb\\
&+0.000411202 \mathcal{R}\left(R_8 R_{10}^*\right)+0.00586189 \left| R_7\right| {}^2+0.00012925 \left| R_8\right| {}^2 \nnb\\
&+0.591901 \left| R_9\right| {}^2+4.30913 \left| R_{10}\right| {}^2-0.00448177 \mathcal{I}\left(R_7\right) \nnb\\
&+0.00578567 \mathcal{I}\left(R_8\right)+0.0501124 \mathcal{I}\left(R_9\right)-0.00204632 \mathcal{I}\left(R_{10}\right) \nnb\\
&-0.218574 \mathcal{R}\left(R_7\right)-0.0279545 \mathcal{R}\left(R_8\right)+1.59319 \mathcal{R}\left(R_9\right) \nnb\\
&-0.257169 \mathcal{R}\left(R_{10}\right)+1.12382 \, \Big] \, \times \, 10^{-7} \, , \\[0.5em]
%
%BR bin2 ee
%
\dps {\cal H}_L[3.5,6]_{ee} =& \Big[
0.000564413 \mathcal{I}\left(R_7 R_8^*\right)+0.000742411 \mathcal{I}\left(R_7 R_9^*\right)+0.00390158 \mathcal{I}\left(R_8 R_9^*\right)\nnb\\
&-0.0000796221 \mathcal{I}\left(R_8 R_{10}^*\right)+0.00176752 \mathcal{R}\left(R_7 R_8^*\right)-0.110436 \mathcal{R}\left(R_7 R_9^*\right)\nnb\\
&-0.0139287 \mathcal{R}\left(R_8 R_9^*\right)+0.00234522 \mathcal{R}\left(R_7 R_{10}^*\right)+0.000292701 \mathcal{R}\left(R_8 R_{10}^*\right)\nnb\\
&-0.0365245 \mathcal{R}\left(R_9 R_{10}^*\right)+0.00629886 \left| R_7\right| {}^2+0.0000978272 \left| R_8\right| {}^2\nnb\\
&+0.453059 \left| R_9\right| {}^2+3.29103 \left| R_{10}\right| {}^2+0.00231621 \mathcal{I}\left(R_7\right) \nnb\\
&+0.00530567 \mathcal{I}\left(R_8\right)-0.0014697 \mathcal{I}\left(R_9\right)-0.000854216 \mathcal{I}\left(R_{10}\right) \nnb\\
&-0.184255 \mathcal{R}\left(R_7\right)-0.0217067 \mathcal{R}\left(R_8\right)+1.31804 \mathcal{R}\left(R_9\right) \nnb\\
&-0.198876 \mathcal{R}\left(R_{10}\right)+0.92649 \, \Big] \, \times \, 10^{-7} \, , \\[0.5em]
%
%BR low-s ee
%
\dps {\cal H}_L[1,6]_{ee} =& \Big[
0.00131182 \mathcal{I}\left(R_7 R_8^*\right)+0.00171947 \mathcal{I}\left(R_7 R_9^*\right)+0.0090363 \mathcal{I}\left(R_8 R_9^*\right) \nnb\\
&-0.00018441 \mathcal{I}\left(R_8 R_{10}^*\right)+0.00391618 \mathcal{R}\left(R_7 R_8^*\right)-0.250532 \mathcal{R}\left(R_7 R_9^*\right) \nnb\\
&-0.0331159 \mathcal{R}\left(R_8 R_9^*\right)+0.00543208 \mathcal{R}\left(R_7 R_{10}^*\right)+0.000703903 \mathcal{R}\left(R_8 R_{10}^*\right)\nnb\\
&-0.0840066 \mathcal{R}\left(R_9 R_{10}^*\right)+0.0121607 \left| R_7\right| {}^2+0.000227077 \left| R_8\right| {}^2 \nnb\\
&+1.04496 \left| R_9\right| {}^2+7.60015 \left| R_{10}\right| {}^2-0.00216555 \mathcal{I}\left(R_7\right) \nnb\\
&+0.0110913 \mathcal{I}\left(R_8\right)+0.0486427 \mathcal{I}\left(R_9\right)-0.00290053 \mathcal{I}\left(R_{10}\right) \nnb\\
&-0.402829 \mathcal{R}\left(R_7\right)-0.0496612 \mathcal{R}\left(R_8\right)+2.91123 \mathcal{R}\left(R_9\right) \nnb\\
&-0.456045 \mathcal{R}\left(R_{10}\right)+2.05031 \, \Big] \, \times \, 10^{-7} \, , \\[0.5em]
%
%BR bin1 mumu
%
\dps {\cal H}_L[1,3.5]_{\mu\mu} =& \Big[
0.000747411 \mathcal{I}\left(R_7 R_8^*\right)+0.000977063 \mathcal{I}\left(R_7 R_9^*\right)+0.00513472 \mathcal{I}\left(R_8 R_9^*\right)\nnb\\
&-0.000104788 \mathcal{I}\left(R_8 R_{10}^*\right)+0.00262758 \mathcal{R}\left(R_7 R_8^*\right)-0.15727 \mathcal{R}\left(R_7 R_9^*\right)\nnb\\
&-0.020489 \mathcal{R}\left(R_8 R_9^*\right)+0.00308685 \mathcal{R}\left(R_7 R_{10}^*\right)+0.000411202 \mathcal{R}\left(R_8 R_{10}^*\right)\nnb\\
&-0.047482 \mathcal{R}\left(R_9 R_{10}^*\right)+0.0090211 \left| R_7\right| {}^2+0.0001474 \left| R_8\right| {}^2 \nnb\\
&+0.625902 \left| R_9\right| {}^2+4.54472 \left| R_{10}\right| {}^2-0.00541775 \mathcal{I}\left(R_7\right) \nnb\\
&+0.00571472 \mathcal{I}\left(R_8\right)+0.0546139 \mathcal{I}\left(R_9\right)-0.00204632 \mathcal{I}\left(R_{10}\right) \nnb\\
&-0.241612 \mathcal{R}\left(R_7\right)-0.0297648 \mathcal{R}\left(R_8\right)+1.70159 \mathcal{R}\left(R_9\right) \nnb\\
&-0.257169 \mathcal{R}\left(R_{10}\right)+1.23483 \, \Big] \, \times \, 10^{-7} \, , \\[0.5em]
%
%BR bin2 mumu
%
\dps {\cal H}_L[3.5,6]_{\mu\mu} =& \Big[
0.000564413 \mathcal{I}\left(R_7 R_8^*\right)+0.000742411 \mathcal{I}\left(R_7 R_9^*\right)+0.00390158 \mathcal{I}\left(R_8 R_9^*\right)\nnb\\
&-0.0000796221 \mathcal{I}\left(R_8 R_{10}^*\right)+0.00195014 \mathcal{R}\left(R_7 R_8^*\right)-0.121246 \mathcal{R}\left(R_7 R_9^*\right)\nnb\\
&-0.0147481 \mathcal{R}\left(R_8 R_9^*\right)+0.00234522 \mathcal{R}\left(R_7 R_{10}^*\right)+0.000292701 \mathcal{R}\left(R_8 R_{10}^*\right)\nnb\\
&-0.0365245 \mathcal{R}\left(R_9 R_{10}^*\right)+0.00750349 \left| R_7\right| {}^2+0.000104748 \left| R_8\right| {}^2\nnb\\
&+0.480992 \left| R_9\right| {}^2+3.48458 \left| R_{10}\right| {}^2+0.00137075 \mathcal{I}\left(R_7\right) \nnb\\
&+0.005234 \mathcal{I}\left(R_8\right)+0.00314813 \mathcal{I}\left(R_9\right)-0.000854216 \mathcal{I}\left(R_{10}\right) \nnb\\
&-0.200992 \mathcal{R}\left(R_7\right)-0.0230157 \mathcal{R}\left(R_8\right)+1.41053 \mathcal{R}\left(R_9\right) \nnb\\
&-0.198876 \mathcal{R}\left(R_{10}\right)+1.01906 \, \Big] \, \times \, 10^{-7} \, , \\[0.5em]
%
%BR low-s mumu
%
\dps {\cal H}_L[1,6]_{\mu\mu} =& \Big[
0.00131182 \mathcal{I}\left(R_7 R_8^*\right)+0.00171947 \mathcal{I}\left(R_7 R_9^*\right)+0.0090363 \mathcal{I}\left(R_8 R_9^*\right) \nnb\\
&-0.00018441 \mathcal{I}\left(R_8 R_{10}^*\right)+0.00457772 \mathcal{R}\left(R_7 R_8^*\right)-0.278516 \mathcal{R}\left(R_7 R_9^*\right) \nnb\\
&-0.0352371 \mathcal{R}\left(R_8 R_9^*\right)+0.00543208 \mathcal{R}\left(R_7 R_{10}^*\right)+0.000703903 \mathcal{R}\left(R_8 R_{10}^*\right)\nnb\\
&-0.0840066 \mathcal{R}\left(R_9 R_{10}^*\right)+0.0165246 \left| R_7\right| {}^2+0.000252149 \left| R_8\right| {}^2 \nnb\\
&+1.10689 \left| R_9\right| {}^2+8.02929 \left| R_{10}\right| {}^2-0.004047 \mathcal{I}\left(R_7\right) \nnb\\
&+0.0109487 \mathcal{I}\left(R_8\right)+0.0577621 \mathcal{I}\left(R_9\right)-0.00290053 \mathcal{I}\left(R_{10}\right) \nnb\\
&-0.442604 \mathcal{R}\left(R_7\right)-0.0527806 \mathcal{R}\left(R_8\right)+3.11211 \mathcal{R}\left(R_9\right) \nnb\\
&-0.456045 \mathcal{R}\left(R_{10}\right)+2.25389 \, \Big] \, \times \, 10^{-7} \, , \\[0.5em]
%
%BR bin1 no em
%
\dps {\cal H}_L[1,3.5]_\text{no em} =& \Big[
0.000747411 \mathcal{I}\left(R_7 R_8^*\right)+0.000977063 \mathcal{I}\left(R_7 R_9^*\right)+0.00513472 \mathcal{I}\left(R_8 R_9^*\right) \nnb\\
&-0.000104788 \mathcal{I}\left(R_8 R_{10}^*\right)+0.00296831 \mathcal{R}\left(R_7 R_8^*\right)-0.169489 \mathcal{R}\left(R_7 R_9^*\right) \nnb\\
&-0.0214151 \mathcal{R}\left(R_8 R_9^*\right)+0.00308685 \mathcal{R}\left(R_7 R_{10}^*\right)+0.000411202 \mathcal{R}\left(R_8 R_{10}^*\right)\nnb\\
&-0.047482 \mathcal{R}\left(R_9 R_{10}^*\right)+0.0112687 \left| R_7\right| {}^2+0.000160314 \left| R_8\right| {}^2 \nnb\\
&+0.650092 \left| R_9\right| {}^2+4.71233 \left| R_{10}\right| {}^2-0.00608367 \mathcal{I}\left(R_7\right) \nnb\\
&+0.00566425 \mathcal{I}\left(R_8\right)+0.0578166 \mathcal{I}\left(R_9\right)-0.00204632 \mathcal{I}\left(R_{10}\right) \nnb\\
&-0.258047 \mathcal{R}\left(R_7\right)-0.0310562 \mathcal{R}\left(R_8\right)+1.77889 \mathcal{R}\left(R_9\right) \nnb\\
&-0.257169 \mathcal{R}\left(R_{10}\right)+1.31413 \, \Big] \, \times \, 10^{-7} \, , \\[0.5em]
%
%BR bin2 no em
%
\dps {\cal H}_L[3.5,6]_\text{no em} =& \Big[
0.000564413 \mathcal{I}\left(R_7 R_8^*\right)+0.000742411 \mathcal{I}\left(R_7 R_9^*\right)+0.00390158 \mathcal{I}\left(R_8 R_9^*\right)\nnb\\
&-0.0000796221 \mathcal{I}\left(R_8 R_{10}^*\right)+0.00208006 \mathcal{R}\left(R_7 R_8^*\right)-0.128937 \mathcal{R}\left(R_7 R_9^*\right)\nnb\\
&-0.015331 \mathcal{R}\left(R_8 R_9^*\right)+0.00234522 \mathcal{R}\left(R_7 R_{10}^*\right)+0.000292701 \mathcal{R}\left(R_8 R_{10}^*\right)\nnb\\
&-0.0365245 \mathcal{R}\left(R_9 R_{10}^*\right)+0.00836054 \left| R_7\right| {}^2+0.000109672 \left| R_8\right| {}^2\nnb\\
&+0.500866 \left| R_9\right| {}^2+3.62228 \left| R_{10}\right| {}^2+0.000698096 \mathcal{I}\left(R_7\right) \nnb\\
&+0.00518302 \mathcal{I}\left(R_8\right)+0.0064335 \mathcal{I}\left(R_9\right)-0.000854216 \mathcal{I}\left(R_{10}\right) \nnb\\
&-0.212928 \mathcal{R}\left(R_7\right)-0.0239492 \mathcal{R}\left(R_8\right)+1.47647 \mathcal{R}\left(R_9\right) \nnb\\
&-0.198876 \mathcal{R}\left(R_{10}\right)+1.08519 \, \Big] \, \times \, 10^{-7} \, , \\[0.5em]
%
%BR low-s no em
%
\dps {\cal H}_L[1,6]_\text{no em} =& \Big[
0.00131182 \mathcal{I}\left(R_7 R_8^*\right)+0.00171947 \mathcal{I}\left(R_7 R_9^*\right)+0.0090363 \mathcal{I}\left(R_8 R_9^*\right) \nnb\\
&-0.00018441 \mathcal{I}\left(R_8 R_{10}^*\right)+0.00504837 \mathcal{R}\left(R_7 R_8^*\right)-0.298426 \mathcal{R}\left(R_7 R_9^*\right) \nnb\\
&-0.0367462 \mathcal{R}\left(R_8 R_9^*\right)+0.00543208 \mathcal{R}\left(R_7 R_{10}^*\right)+0.000703903 \mathcal{R}\left(R_8 R_{10}^*\right) \nnb\\
&-0.0840066 \mathcal{R}\left(R_9 R_{10}^*\right)+0.0196293 \left| R_7\right| {}^2+0.000269986 \left| R_8\right| {}^2 \nnb\\
&+1.15096 \left| R_9\right| {}^2+8.33461 \left| R_{10}\right| {}^2-0.00538557 \mathcal{I}\left(R_7\right) \nnb\\
&+0.0108473 \mathcal{I}\left(R_8\right)+0.0642501 \mathcal{I}\left(R_9\right)-0.00290053 \mathcal{I}\left(R_{10}\right) \nnb\\
&-0.470975 \mathcal{R}\left(R_7\right)-0.0550053 \mathcal{R}\left(R_8\right)+3.25535 \mathcal{R}\left(R_9\right) \nnb\\
&-0.456045 \mathcal{R}\left(R_{10}\right)+2.39933 \, \Big] \, \times \, 10^{-7} \, .
\end{align}

\begin{align}
\dps {\cal H}_T[1,3.5]_{ee} =& \Big[
0.0165802 \mathcal{I}\left(R_7 R_8^*\right)+0.00191257 \mathcal{I}\left(R_7 R_9^*\right)+0.0100512 \mathcal{I}\left(R_8 R_9^*\right) \nnb\\
&-0.000205119 \mathcal{I}\left(R_8 R_{10}^*\right)+0.0469885 \mathcal{R}\left(R_7 R_8^*\right)-0.379261 \mathcal{R}\left(R_7 R_9^*\right)\nnb\\
&-0.0377252 \mathcal{R}\left(R_8 R_9^*\right)+0.0058446 \mathcal{R}\left(R_7 R_{10}^*\right)+0.000628542 \mathcal{R}\left(R_8 R_{10}^*\right) \nnb\\
&-0.00991432 \mathcal{R}\left(R_9 R_{10}^*\right)+0.211135 \left| R_7\right| {}^2+0.00234936 \left| R_8\right| {}^2 \nnb\\
&+0.249236 \left| R_9\right| {}^2+1.76904 \left| R_{10}\right| {}^2+0.0362042 \mathcal{I}\left(R_7\right) \nnb\\
&-0.00833705 \mathcal{I}\left(R_8\right)-0.0146216 \mathcal{I}\left(R_9\right)+0.000290922 \mathcal{I}\left(R_{10}\right) \nnb\\
&+0.352468 \mathcal{R}\left(R_7\right)+0.03572 \mathcal{R}\left(R_8\right)+0.0901749 \mathcal{R}\left(R_9\right) \nnb\\
&-0.0390521 \mathcal{R}\left(R_{10}\right)+0.614659 \, \Big] \, \times \, 10^{-7} \, , \\[0.5em]
%
%BR bin2 ee
%
\dps {\cal H}_T[3.5,6]_{ee} =& \Big[
0.0053151 \mathcal{I}\left(R_7 R_8^*\right)+0.00144479 \mathcal{I}\left(R_7 R_9^*\right)+0.00759287 \mathcal{I}\left(R_8 R_9^*\right) \nnb\\
&-0.00015495 \mathcal{I}\left(R_8 R_{10}^*\right)+0.0147212 \mathcal{R}\left(R_7 R_8^*\right)-0.260494 \mathcal{R}\left(R_7 R_9^*\right) \nnb\\
&-0.0266142 \mathcal{R}\left(R_8 R_9^*\right)+0.0044113 \mathcal{R}\left(R_7 R_{10}^*\right)+0.000476328 \mathcal{R}\left(R_8 R_{10}^*\right)\nnb\\
&-0.0159082 \mathcal{R}\left(R_9 R_{10}^*\right)+0.0650316 \left| R_7\right| {}^2+0.00074246 \left| R_8\right| {}^2 \nnb\\
&+0.27778 \left| R_9\right| {}^2+1.98963 \left| R_{10}\right| {}^2+0.0155002 \mathcal{I}\left(R_7\right) \nnb\\
&+0.00455726 \mathcal{I}\left(R_8\right)-0.00974966 \mathcal{I}\left(R_9\right)+0.000110893 \mathcal{I}\left(R_{10}\right) \nnb\\
&-0.181332 \mathcal{R}\left(R_7\right)-0.0172841 \mathcal{R}\left(R_8\right)+0.443444 \mathcal{R}\left(R_9\right) \nnb\\
&-0.0757004 \mathcal{R}\left(R_{10}\right)+0.292097 \, \Big] \, \times \, 10^{-7} \, , \\[0.5em]
%
%BR low-s ee
%
\dps {\cal H}_T[1,6]_{ee} =& \Big[
0.0218953 \mathcal{I}\left(R_7 R_8^*\right)+0.00335736 \mathcal{I}\left(R_7 R_9^*\right)+0.017644 \mathcal{I}\left(R_8 R_9^*\right) \nnb\\
&-0.00036007 \mathcal{I}\left(R_8 R_{10}^*\right)+0.0617098 \mathcal{R}\left(R_7 R_8^*\right)-0.639755 \mathcal{R}\left(R_7 R_9^*\right)\nnb\\
&-0.0643394 \mathcal{R}\left(R_8 R_9^*\right)+0.0102559 \mathcal{R}\left(R_7 R_{10}^*\right)+0.00110487 \mathcal{R}\left(R_8 R_{10}^*\right) \nnb\\
&-0.0258225 \mathcal{R}\left(R_9 R_{10}^*\right)+0.276167 \left| R_7\right| {}^2+0.00309182 \left| R_8\right| {}^2 \nnb\\
&+0.527016 \left| R_9\right| {}^2+3.75867 \left| R_{10}\right| {}^2+0.0517044 \mathcal{I}\left(R_7\right) \nnb\\
&-0.00377979 \mathcal{I}\left(R_8\right)-0.0243713 \mathcal{I}\left(R_9\right)+0.000401815 \mathcal{I}\left(R_{10}\right) \nnb\\
&+0.171135 \mathcal{R}\left(R_7\right)+0.018436 \mathcal{R}\left(R_8\right)+0.533619 \mathcal{R}\left(R_9\right) \nnb\\
&-0.114753 \mathcal{R}\left(R_{10}\right)+0.906757 \, \Big] \, \times \, 10^{-7} \, , \\[0.5em]
%
%BR bin1 mumu
%
\dps {\cal H}_T[1,3.5]_{\mu\mu} =& \Big[
0.0165802 \mathcal{I}\left(R_7 R_8^*\right)+0.00191257 \mathcal{I}\left(R_7 R_9^*\right)+0.0100512 \mathcal{I}\left(R_8 R_9^*\right) \nnb\\
&-0.000205119 \mathcal{I}\left(R_8 R_{10}^*\right)+0.0472027 \mathcal{R}\left(R_7 R_8^*\right)-0.34595 \mathcal{R}\left(R_7 R_9^*\right)\nnb\\
&-0.0352003 \mathcal{R}\left(R_8 R_9^*\right)+0.0058446 \mathcal{R}\left(R_7 R_{10}^*\right)+0.000628542 \mathcal{R}\left(R_8 R_{10}^*\right) \nnb\\
&-0.00991432 \mathcal{R}\left(R_9 R_{10}^*\right)+0.212548 \left| R_7\right| {}^2+0.00235748 \left| R_8\right| {}^2 \nnb\\
&+0.182478 \left| R_9\right| {}^2+1.30648 \left| R_{10}\right| {}^2+0.0374908 \mathcal{I}\left(R_7\right) \nnb\\
&-0.00823953 \mathcal{I}\left(R_8\right)-0.0208026 \mathcal{I}\left(R_9\right)+0.000290922 \mathcal{I}\left(R_{10}\right) \nnb\\
&+0.420669 \mathcal{R}\left(R_7\right)+0.0410139 \mathcal{R}\left(R_8\right)-0.12403 \mathcal{R}\left(R_9\right) \nnb\\
&-0.0390521 \mathcal{R}\left(R_{10}\right)+0.413317 \, \Big] \, \times \, 10^{-7} \, , \\[0.5em]
%
%BR bin2 mumu
%
\dps {\cal H}_T[3.5,6]_{\mu\mu} =& \Big[
0.0053151 \mathcal{I}\left(R_7 R_8^*\right)+0.00144479 \mathcal{I}\left(R_7 R_9^*\right)+0.00759287 \mathcal{I}\left(R_8 R_9^*\right) \nnb\\
&-0.00015495 \mathcal{I}\left(R_8 R_{10}^*\right)+0.015026 \mathcal{R}\left(R_7 R_8^*\right)-0.250608 \mathcal{R}\left(R_7 R_9^*\right)\nnb\\
&-0.0258649 \mathcal{R}\left(R_8 R_9^*\right)+0.0044113 \mathcal{R}\left(R_7 R_{10}^*\right)+0.000476328 \mathcal{R}\left(R_8 R_{10}^*\right)\nnb\\
&-0.0159082 \mathcal{R}\left(R_9 R_{10}^*\right)+0.0670418 \left| R_7\right| {}^2+0.000754009 \left| R_8\right| {}^2 \nnb\\
&+0.242336 \left| R_9\right| {}^2+1.74404 \left| R_{10}\right| {}^2+0.0169642 \mathcal{I}\left(R_7\right) \nnb\\
&+0.00466823 \mathcal{I}\left(R_8\right)-0.0168279 \mathcal{I}\left(R_9\right)+0.000110893 \mathcal{I}\left(R_{10}\right) \nnb\\
&-0.154501 \mathcal{R}\left(R_7\right)-0.0152134 \mathcal{R}\left(R_8\right)+0.313295 \mathcal{R}\left(R_9\right) \nnb\\
&-0.0757004 \mathcal{R}\left(R_{10}\right)+0.153748 \, \Big] \, \times \, 10^{-7} \, , \\[0.5em]
%
%BR low-s mumu
%
\dps {\cal H}_T[1,6]_{\mu\mu} =& \Big[
0.0218953 \mathcal{I}\left(R_7 R_8^*\right)+0.00335736 \mathcal{I}\left(R_7 R_9^*\right)+0.017644 \mathcal{I}\left(R_8 R_9^*\right) \nnb\\
&-0.00036007 \mathcal{I}\left(R_8 R_{10}^*\right)+0.0622287 \mathcal{R}\left(R_7 R_8^*\right)-0.596558 \mathcal{R}\left(R_7 R_9^*\right) \nnb\\
&-0.0610652 \mathcal{R}\left(R_8 R_9^*\right)+0.0102559 \mathcal{R}\left(R_7 R_{10}^*\right)+0.00110487 \mathcal{R}\left(R_8 R_{10}^*\right)\nnb\\
&-0.0258225 \mathcal{R}\left(R_9 R_{10}^*\right)+0.27959 \left| R_7\right| {}^2+0.00311149 \left| R_8\right| {}^2 \nnb\\
&+0.424814 \left| R_9\right| {}^2+3.05052 \left| R_{10}\right| {}^2+0.054455 \mathcal{I}\left(R_7\right) \nnb\\
&-0.0035713 \mathcal{I}\left(R_8\right)-0.0376305 \mathcal{I}\left(R_9\right)+0.000401815 \mathcal{I}\left(R_{10}\right) \nnb\\
&+0.266168 \mathcal{R}\left(R_7\right)+0.0258005 \mathcal{R}\left(R_8\right)+0.189265 \mathcal{R}\left(R_9\right) \nnb\\
&-0.114753 \mathcal{R}\left(R_{10}\right)+0.567065 \, \Big] \, \times \, 10^{-7} \, , \\[0.5em]
%
%BR bin1 no em
%
\dps {\cal H}_T[1,3.5]_\text{no em} =& \Big[
0.0165802 \mathcal{I}\left(R_7 R_8^*\right)+0.00191257 \mathcal{I}\left(R_7 R_9^*\right)+0.0100512 \mathcal{I}\left(R_8 R_9^*\right) \nnb\\
&-0.000205119 \mathcal{I}\left(R_8 R_{10}^*\right)+0.0473551 \mathcal{R}\left(R_7 R_8^*\right)-0.322251 \mathcal{R}\left(R_7 R_9^*\right)\nnb\\
&-0.0334039 \mathcal{R}\left(R_8 R_9^*\right)+0.0058446 \mathcal{R}\left(R_7 R_{10}^*\right)+0.000628542 \mathcal{R}\left(R_8 R_{10}^*\right)\nnb\\
&-0.00991432 \mathcal{R}\left(R_9 R_{10}^*\right)+0.213553 \left| R_7\right| {}^2+0.00236325 \left| R_8\right| {}^2\nnb\\
&+0.134982 \left| R_9\right| {}^2+0.977388 \left| R_{10}\right| {}^2+0.0384061 \mathcal{I}\left(R_7\right) \nnb\\
&-0.00817015 \mathcal{I}\left(R_8\right)-0.0252 \mathcal{I}\left(R_9\right)+0.000290922 \mathcal{I}\left(R_{10}\right) \nnb\\
&+0.469276 \mathcal{R}\left(R_7\right)+0.0447868 \mathcal{R}\left(R_8\right)-0.276771 \mathcal{R}\left(R_9\right) \nnb\\
&-0.0390521 \mathcal{R}\left(R_{10}\right)+0.269424 \, \Big] \, \times \, 10^{-7} \, , \\[0.5em]
%
%BR bin2 no em
%
\dps {\cal H}_T[3.5,6]_\text{no em} =& \Big[
0.0053151 \mathcal{I}\left(R_7 R_8^*\right)+0.00144479 \mathcal{I}\left(R_7 R_9^*\right)+0.00759287 \mathcal{I}\left(R_8 R_9^*\right) \nnb\\
&-0.00015495 \mathcal{I}\left(R_8 R_{10}^*\right)+0.0152428 \mathcal{R}\left(R_7 R_8^*\right)-0.243575 \mathcal{R}\left(R_7 R_9^*\right)\nnb\\
&-0.0253318 \mathcal{R}\left(R_8 R_9^*\right)+0.0044113 \mathcal{R}\left(R_7 R_{10}^*\right)+0.000476328 \mathcal{R}\left(R_8 R_{10}^*\right)\nnb\\
&-0.0159082 \mathcal{R}\left(R_9 R_{10}^*\right)+0.0684719 \left| R_7\right| {}^2+0.000762226 \left| R_8\right| {}^2 \nnb\\
&+0.217119 \left| R_9\right| {}^2+1.56931 \left| R_{10}\right| {}^2+0.0180057 \mathcal{I}\left(R_7\right) \nnb\\
&+0.00474718 \mathcal{I}\left(R_8\right)-0.0218638 \mathcal{I}\left(R_9\right)+0.000110893 \mathcal{I}\left(R_{10}\right) \nnb\\
&-0.135386 \mathcal{R}\left(R_7\right)-0.0137382 \mathcal{R}\left(R_8\right)+0.220517 \mathcal{R}\left(R_9\right) \nnb\\
&-0.0757004 \mathcal{R}\left(R_{10}\right)+0.0549339 \, \Big] \, \times \, 10^{-7} \, , \\[0.5em]
%
%BR low-s no em
%
\dps {\cal H}_T[1,6]_\text{no em} =& \Big[
0.0218953 \mathcal{I}\left(R_7 R_8^*\right)+0.00335736 \mathcal{I}\left(R_7 R_9^*\right)+0.017644 \mathcal{I}\left(R_8 R_9^*\right) \nnb\\
&-0.00036007 \mathcal{I}\left(R_8 R_{10}^*\right)+0.0625979 \mathcal{R}\left(R_7 R_8^*\right)-0.565826 \mathcal{R}\left(R_7 R_9^*\right)\nnb\\
&-0.0587357 \mathcal{R}\left(R_8 R_9^*\right)+0.0102559 \mathcal{R}\left(R_7 R_{10}^*\right)+0.00110487 \mathcal{R}\left(R_8 R_{10}^*\right)\nnb\\
&-0.0258225 \mathcal{R}\left(R_9 R_{10}^*\right)+0.282025 \left| R_7\right| {}^2+0.00312548 \left| R_8\right| {}^2 \nnb\\
&+0.352102 \left| R_9\right| {}^2+2.5467 \left| R_{10}\right| {}^2+0.0564119 \mathcal{I}\left(R_7\right) \nnb\\
&-0.00342297 \mathcal{I}\left(R_8\right)-0.0470638 \mathcal{I}\left(R_9\right)+0.000401815 \mathcal{I}\left(R_{10}\right) \nnb\\
&+0.33389 \mathcal{R}\left(R_7\right)+0.0310485 \mathcal{R}\left(R_8\right)-0.0562543 \mathcal{R}\left(R_9\right) \nnb\\
&-0.114753 \mathcal{R}\left(R_{10}\right)+0.324358 \, \Big] \, \times \, 10^{-7} \, .
\end{align}

\subsection{${\cal H}_3$ and  ${\cal H}_4$}

	\begin{align}
	\dps {\cal H}_3[1,3.5]_{ee} =& \Big[
	-1.77941 \mathcal{R}\left(R_7 R_{10}^*\right)-0.134875 \mathcal{R}\left(R_8 R_{10}^*\right)+3.00277 \mathcal{R}\left(R_9 R_{10}^*\right) \nnb \\
	&+0.0272335 \mathcal{I}\left(R_{10}\right)+3.15458 \mathcal{R}\left(R_{10}\right)-0.106259   \,
	\Big] \, \times \, 10^{-9} \, , \\[0.5em]
	%
	%H3 bin2 ee
	%
	\dps {\cal H}_3[3.5,6]_{ee} =& \Big[
	-0.933808 \mathcal{R}\left(R_7 R_{10}^*\right)-0.0707802 \mathcal{R}\left(R_8 R_{10}^*\right)+2.48483 \mathcal{R}\left(R_9 R_{10}^*\right) \nnb \\
	&+0.136987 \mathcal{I}\left(R_{10}\right)+3.6078 \mathcal{R}\left(R_{10}\right)-0.0879303
	\, \Big] \, \times \, 10^{-9} \, , \\[0.5em]
	%
	%H3 low-s ee
	%
	\dps {\cal H}_3[1,6]_{ee} =& \Big[
	-2.71322 \mathcal{R}\left(R_7 R_{10}^*\right)-0.205655 \mathcal{R}\left(R_8 R_{10}^*\right)+5.48761 \mathcal{R}\left(R_9 R_{10}^*\right) \nnb \\
	&+0.164221 \mathcal{I}\left(R_{10}\right)+6.76238 \mathcal{R}\left(R_{10}\right)-0.194189
	\, \Big] \, \times \, 10^{-9} \, , \\[0.5em]
	%
	%H3 bin1 mumu
	%
	\dps {\cal H}_3[1,3.5]_{\mu\mu} =& \Big[
	-0.739705 \mathcal{R}\left(R_7 R_{10}^*\right)-0.0560677 \mathcal{R}\left(R_8 R_{10}^*\right)+1.24826 \mathcal{R}\left(R_9 R_{10}^*\right) \nnb \\
	&+0.011321 \mathcal{I}\left(R_{10}\right)+1.31401 \mathcal{R}\left(R_{10}\right)-0.0441719
	\, \Big] \, \times \, 10^{-9} \, , \\[0.5em]
	%
	%H3 bin2 mumu
	%
	\dps {\cal H}_3[3.5,6]_{\mu\mu} =& \Big[
	-0.388185 \mathcal{R}\left(R_7 R_{10}^*\right)-0.0294234 \mathcal{R}\left(R_8 R_{10}^*\right)+1.03295 \mathcal{R}\left(R_9 R_{10}^*\right) \nnb \\
	&+0.0569458 \mathcal{I}\left(R_{10}\right)+1.50196 \mathcal{R}\left(R_{10}\right)-0.0365528
	\, \Big] \, \times \, 10^{-9} \, , \\[0.5em]
	%
	%H3 low-s mumu
	%
	\dps {\cal H}_3[1,6]_{\mu\mu} =& \Big[
	-1.12789 \mathcal{R}\left(R_7 R_{10}^*\right)-0.0854912 \mathcal{R}\left(R_8 R_{10}^*\right)+2.28121 \mathcal{R}\left(R_9 R_{10}^*\right) \nnb \\
	&+0.0682668 \mathcal{I}\left(R_{10}\right)+2.81597 \mathcal{R}\left(R_{10}\right)-0.0807247
	\, \Big] \, \times \, 10^{-9} \, , \\[0.5em]
	%
	%%%%%%%%%%%%%%%%%%%%%%%%%%%%%%%%%%%%%%%%%%%%%%%%%%%%%%%%%%%%%%%%%%%%%%%%%%%%%%%%%%%%%%%%%%%%%%%%%%%%%%%%%%%%%
	%%%%%%%%%%%%%%%%%%%%%%%%%%%%%%%%%%%%%%%%%%%%%%%%%%%%%%%%%%%%%%%%%%%%%%%%%%%%%%%%%%%%%%%%%%%%%%%%%%%%%%%%%%%%%
	%
	%H4 bin1 ee
	%
	\dps {\cal H}_4[1,3.5]_{ee} =& \Big[
	0.013885 \mathcal{R}\left(R_7 R_8^*\right)-0.36974 \mathcal{R}\left(R_7 R_9^*\right)-0.0280253 \mathcal{R}\left(R_8 R_9^*\right)  \nnb \\
	&+0.0915928 \left| R_7\right| {}^2+0.000526223 \left| R_8\right| {}^2+0.493116 \left| R_9\right| {}^2  \nnb \\
	&+3.41677 \left| R_{10}\right| {}^2-0.0427247 \mathcal{I}\left(R_7\right)-0.00323842 \mathcal{I}\left(R_8\right)  \nnb \\
	&+0.206797 \mathcal{I}\left(R_9\right)-0.44318 \mathcal{R}\left(R_7\right)-0.0349731 \mathcal{R}\left(R_8\right)  \nnb \\
	&+1.50448 \mathcal{R}\left(R_9\right)+1.72231
	\, \Big] \, \times \, 10^{-9} \, , \\[0.5em]
	%
	%H4 bin2 ee
	%
	\dps {\cal H}_4[3.5,6]_{ee} =& \Big[
	0.00356372 \mathcal{R}\left(R_7 R_8^*\right)-0.130439 \mathcal{R}\left(R_7 R_9^*\right)-0.00988691 \mathcal{R}\left(R_8 R_9^*\right) \nnb \\
	&+0.0235082 \left| R_7\right| {}^2+0.00013506 \left| R_8\right| {}^2+0.183801 \left| R_9\right| {}^2  \nnb \\
	&+1.27355 \left| R_{10}\right| {}^2-0.026613 \mathcal{I}\left(R_7\right)-0.00201719 \mathcal{I}\left(R_8\right)  \nnb \\
	&+0.13151 \mathcal{I}\left(R_9\right)-0.181082 \mathcal{R}\left(R_7\right)-0.0142128 \mathcal{R}\left(R_8\right) \nnb \\
	&+0.541921 \mathcal{R}\left(R_9\right)+0.544544
	\, \Big] \, \times \, 10^{-9} \, , \\[0.5em]
	%
	%H4 low-s ee
	%
	\dps {\cal H}_4[1,6]_{ee} =& \Big[ 0.0174487 \mathcal{R}\left(R_7 R_8^*\right)-0.500179 \mathcal{R}\left(R_7 R_9^*\right)-0.0379122 \mathcal{R}\left(R_8 R_9^*\right) \nnb \\
	&+0.115101 \left| R_7\right| {}^2+0.000661283 \left| R_8\right| {}^2+0.676917 \left| R_9\right| {}^2 \nnb \\
	&+4.69032 \left| R_{10}\right| {}^2-0.0693376 \mathcal{I}\left(R_7\right)-0.00525561 \mathcal{I}\left(R_8\right)  \nnb \\
	&+0.338308 \mathcal{I}\left(R_9\right)-0.624262 \mathcal{R}\left(R_7\right)-0.0491858 \mathcal{R}\left(R_8\right) \nnb \\
	&+2.0464 \mathcal{R}\left(R_9\right)+2.26685   \, \Big] \, \times \, 10^{-9} \, , \\[0.5em]
	%
	%H4 bin1 mumu
	%
	\dps {\cal H}_4[1,3.5]_{\mu\mu} =& \Big[
	0.00577201 \mathcal{R}\left(R_7 R_8^*\right)-0.153702 \mathcal{R}\left(R_7 R_9^*\right)-0.0116502 \mathcal{R}\left(R_8 R_9^*\right) \nnb \\
	&+0.0380753 \left| R_7\right| {}^2+0.000218752 \left| R_8\right| {}^2+0.204989 \left| R_9\right| {}^2 \nnb \\
	&+1.42036 \left| R_{10}\right| {}^2-0.0177607 \mathcal{I}\left(R_7\right)-0.00134622 \mathcal{I}\left(R_8\right) \nnb \\
	&+0.085966 \mathcal{I}\left(R_9\right)-0.184556 \mathcal{R}\left(R_7\right)-0.014563 \mathcal{R}\left(R_8\right) \nnb \\
	&+0.626285 \mathcal{R}\left(R_9\right)+0.717523   \, \Big] \, \times \, 10^{-9} \, , \\[0.5em]
	%
	%H4 bin2 mumu
	%
	\dps {\cal H}_4[3.5,6]_{\mu\mu} =& \Big[
	0.00148144 \mathcal{R}\left(R_7 R_8^*\right)-0.0542236 \mathcal{R}\left(R_7 R_9^*\right)-0.00411001 \mathcal{R}\left(R_8 R_9^*\right) \nnb \\
	&+0.00977239 \left| R_7\right| {}^2+0.0000561448 \left| R_8\right| {}^2+0.0764064 \left| R_9\right| {}^2 \nnb \\
	&+0.529415 \left| R_{10}\right| {}^2-0.0110631 \mathcal{I}\left(R_7\right)-0.000838551 \mathcal{I}\left(R_8\right) \nnb \\
	&+0.0546691 \mathcal{I}\left(R_9\right)-0.0753908 \mathcal{R}\left(R_7\right)-0.00591698 \mathcal{R}\left(R_8\right) \nnb \\
	&+0.225601 \mathcal{R}\left(R_9\right)+0.22693
	\, \Big] \, \times \, 10^{-9} \, , \\[0.5em]
	%
	%H4 low-s mumu
	%
	\dps {\cal H}_4[1,6]_{\mu\mu} =& \Big[
	0.00725346 \mathcal{R}\left(R_7 R_8^*\right)-0.207925 \mathcal{R}\left(R_7 R_9^*\right)-0.0157602 \mathcal{R}\left(R_8 R_9^*\right) \nnb \\
	&+0.0478477 \left| R_7\right| {}^2+0.000274897 \left| R_8\right| {}^2+0.281396 \left| R_9\right| {}^2 \nnb \\
	&+1.94977 \left| R_{10}\right| {}^2-0.0288238 \mathcal{I}\left(R_7\right)-0.00218477 \mathcal{I}\left(R_8\right) \nnb \\
	&+0.140635 \mathcal{I}\left(R_9\right)-0.259947 \mathcal{R}\left(R_7\right)-0.02048 \mathcal{R}\left(R_8\right) \nnb \\
	&+0.851886 \mathcal{R}\left(R_9\right)+0.944453
	\, \Big] \, \times \, 10^{-9} \, ,
	\end{align}
	%%%%%%%%%%%%%%%%%%%%%%%%%%%%%%%%%%%%%%%%%%%%%%%%%%%%%%%%%%%%%%%%%%%%%%%%%%%%%%%%%%%%%%%%%%%%%%%%%%%%%%%%%%

\end{appendix}

%%%%%%%%%%%%%%%%%%%%%%%%%%%%%%%%%%%%%%%%%%%%%%%%%%%%%%%%%%%%%%%%%%%%%%%%%%%%%%%%%%%%%%%%%%%%%%%%%%%%%%%%%%%%%%%%%%%%%%%%%%%%%%%%%%%%%%%%%%%%%%%%%%%%%%%%%%%%%%%%%%%%%%
%%%%%%%%%%%%%%%%%%%%%%%%%%%%%%%%%%%%%%%%%%%%%%%%%%%%%%%%%%%%%%%%%%%%%%%%%%%%%%%%%%%%%%%%%%%%%%%%%%%%%%%%%%%%%%%%%%%%%%%%%%%%%%%%%%%%%%%%%%%%%%%%%%%%%%%%%%%%%%%%%%%%%%

\bibliography{references}{}
\bibliographystyle{JHEP} % bst file

\end{document}